\documentclass[12pt]{article}
\usepackage{epsfig}
\usepackage{pstricks,rotating, graphicx}
\usepackage{a4}
\usepackage{latexsym}
\usepackage{cite}

\usepackage{color}
\usepackage{colordvi}
\usepackage[hang]{subfigure}

\graphicspath{{../Pics/}}
\DeclareGraphicsExtensions{.eps,.ps}

\usepackage[width=16.5cm,left=2.2cm,top=2.5cm,height=24.cm]{geometry}

\usepackage{pslatex}
\usepackage{txfonts}
\usepackage[latin1]{inputenc}
\usepackage[T1]{fontenc}

\newcommand{\GeV}{\ensuremath{\,\mathrm{GeV}}}

\def\mus{{\mu^{\,2}}}
\def\muf{{\mu^{}_f}}
\def\mufs{{\mu^{\,2}_f}}
\def\mur{{\mu^{}_r}}
\def\murs{{\mu^{\,2}_r}}
\def\MSbar{{$\overline{\mbox{MS}}\,$}}

\newcommand{\gsim}{\raisebox{-0.07cm}{$\:\:\stackrel{>}{{\scriptstyle \sim}}\:\: $} }

\begin{document}

\begin{titlepage}
\noindent
DESY 12-023\\
DO-TH 11/31\\
LPN 12-033 \\
SFB/CPP-12-08\\
February 2012 \\
\vspace{1.75cm}

\begin{center}
  {\bf
    \Large
    Parton distribution functions and benchmark cross sections \\ at NNLO\\
  }
  \vspace{1.5cm}
  {\large
    S.~Alekhin$^{\, a,b,}$\footnote{{\bf e-mail}: sergey.alekhin@ihep.ru},
    J.~Bl\"umlein$^{\, a,}$\footnote{{\bf e-mail}: johannes.bluemlein@desy.de},
    and S.~Moch$^{\, a,}$\footnote{{\bf e-mail}: sven-olaf.moch@desy.de} \\
  }
  \vspace{1.2cm}
  {\it
    $^a$Deutsches Elektronensynchrotron DESY \\
    Platanenallee 6, D--15738 Zeuthen, Germany \\
    \vspace{0.2cm}
    $^b$Institute for High Energy Physics \\
    142281 Protvino, Moscow region, Russia\\
  }
  \vspace{1.4cm}
  \large {\bf Abstract}
  \vspace{-0.2cm}
\end{center}
We present a determination of parton distribution functions (ABM11) 
and the strong coupling constant $\alpha_s$ 
at next-to-leading order and next-to-next-to-leading order (NNLO) in QCD 
based on world data for deep-inelastic scattering and fixed-target data 
for the Drell-Yan process.
The analysis is performed in the fixed-flavor number scheme for $n_f=3,4,5$ 
and uses the \MSbar-scheme for $\alpha_s$ and the heavy-quark masses.
At NNLO we obtain the value $\alpha_s(M_Z) = 0.1134 \pm 0.0011$.
The fit results are used to compute benchmark cross sections at hadron 
colliders to NNLO accuracy and to compare to data from the LHC. 
\end{titlepage}

%\newpage
%
%\tableofcontents

\newpage
\setcounter{footnote}{0}
%%
%% ---------------------------------------------------------------------------
%%
\section{Introduction}
\label{sec:intro}

Parton distribution functions (PDFs) in the nucleon are an
indispensable ingredient of modern collider phenomenology and their
study has a long history. In the perturbative approach to the gauge
theory of the strong interactions, Quantum Chromodynamics (QCD),
factorization allows for the computation of the hard parton
scattering processes as a power series in the strong coupling
constant $\alpha_s$ and, typically, to leading power $1/Q^2$ dominating for 
large momentum transfer $Q^2$. Predictions for physical cross
sections involving initial hadrons, however, do require further
non-perturbative information, that is knowledge of the PDFs in the
nucleon as well as the value of $\alpha_s(Q)$ and of the masses of
heavy quarks. Since PDFs cannot be calculated in perturbative QCD,
they need to be extracted from a comparison of theory predictions to
available experimental precision data on deep-inelastic scattering (DIS), on
the production of lepton-pairs (Drell-Yan process) or jets in hadron
collisions or any other suitable hard scattering reaction.

The accuracy of PDF determinations in such analyses has steadily
improved over the years, both due to more accurate experimental input
and due to refined theory predictions for the hard parton
scattering reactions including higher orders in perturbation theory.
As of now, complete next-to-next-to-leading order (NNLO)
calculations in perturbative QCD form the backbone of this endeavor.
These allow for the computation of many important benchmark cross
sections, e.g. in the proton-proton collisions at the LHC, with an
unprecedented precision. The determination of PDFs to NNLO accuracy
in QCD was pioneered more than a decade
 ago in~\cite{Alekhin:2001ih} and builds in particular on the known corrections for the PDF
evolution~\cite{Moch:2004pa,Vogt:2004mw} as well as on the hard
scattering corrections for
DIS~\cite{vanNeerven:1991nn,Zijlstra:1991qc,Zijlstra:1992qd,Zijlstra:1992kj,Moch:1999eb,Moch:2004xu,Vermaseren:2005qc},
and hadronic $W$- and $Z$-gauge-boson production, both at an
inclusive~\cite{Hamberg:1990np,Harlander:2002wh} and a differential
level~\cite{Anastasiou:2003yy,Anastasiou:2003ds,Catani:2009sm}.

Presently, NNLO PDFs have been obtained by a number of different
groups. In detail, these are
ABKM09~\cite{Alekhin:2009ni,Alekhin:2010iu},
HERAPDF1.5~\cite{herapdf:2009wt,herapdfgrid:2011},
JR09~\cite{JimenezDelgado:2008hf,JimenezDelgado:2009tv},
MSTW~\cite{Martin:2009iq} and NN21~\cite{Ball:2011uy}, while
CT10~\cite{Lai:2010vv} still remains at next-to-leading order (NLO)
accuracy only. There exist, of course, differences between these PDF
sets. These arise from variations in the choice of the parameters,
e.g., the value of $\alpha_s(M_Z)$, but also from a different theoretical
footing for the data analysis. In the latter case, this comprises
for instance, the treatment of the heavy-quark contributions in DIS,
the corrections for nuclear effects, the inclusion of higher twist (HT)
terms and so on. The implications for precision predictions at
TeV-scale hadron colliders can be profound, though, as benchmark
cross sections at NNLO in QCD for the production of $W$- and
$Z$-gauge-bosons or the Higgs boson through gluon-gluon-fusion (ggF)
show, see e.g., the recent discussion
in~\cite{Baglio:2010um,Alekhin:2010dd,Alekhin:2011ey,Thorne:2011kq,Watt:2011kp,Ball:2011we}.

In this article we present the PDF set ABM11, which is an updated version of the PDF analyses of
ABKM09~\cite{Alekhin:2009ni} and ABM10~\cite{Alekhin:2010iu} in the
3-, 4-, and 5-flavor scheme at NNLO in QCD. 
These PDFs are obtained from an analysis of the world DIS data combined with fixed-target
data for the Drell-Yan (DY) process and for di-muon production in neutrino-nucleon DIS. 
In the ABM11 fit we are now using the final version of the DIS inclusive data collected
by the HERA experiments in run~I~\cite{herapdf:2009wt} together
with new data of the H1 collaboration from the HERA low-energy
run~\cite{Aaron:2010ry}. Moreover, our update is based on theoretical
improvements. For instance, the treatment of the heavy-quark
contributions in DIS now employs the running-mass definition in the
\MSbar-scheme for the heavy quarks~\cite{Alekhin:2010sv}.

The strong coupling constant $\alpha_s(M_Z)$ or, respectively, the QCD scale $\Lambda_{\rm QCD}$,  
is a mandatory parameter to be fitted in DIS analyses of world data, 
its correct value being of paramount importance for many processes 
in DIS and at hadron colliders, in particular for Higgs boson production in ggF~\cite{Alekhin:2010dd}.
An essential criterion for the selection of additional precision data on top of the world DIS data, 
e.g. those for the DY process or for hadronic weak-boson and jet-production cross sections, 
in the measurement of the QCD scale is the compatibility of these data sets 
with respect to the experimental systematics in the different measurements. 
Strictly speaking, combined analyses require a theoretical description at the same perturbative order.
Because of these reasons, the combination of different data sets needs great care
if performed with the goal of a precision measurement of $\alpha_s(M_Z)$. 
Combinations of a wide range of hard-scattering data sets of differing quality, 
as sometimes used in more global fits, are useful only, 
if they indeed lead to a statistically and systematically improved value of $\alpha_s(M_Z)$.
Of course, a careful check is always required when new data sets are added.
As a result of our new analysis we determine in the ABM11 fit the strong coupling constant at NNLO in the
\MSbar-scheme and present a detailed discussion of the 
uncertainties and of the impact of individual experiments, 
showing the great stability in the obtained value of $\alpha_s(M_Z)$.

The papers is organized as follows. In Sec.~\ref{sec:theory} we
describe the theoretical framework of our analysis, in
Sec.~\ref{sec:pQCD} in particular the perturbative QCD input
including the framework for heavy-quark DIS.
Secs.~\ref{sec:power-cor} and \ref{sec:nucl-cor} are concerned with a
detailed account of the non-perturbative corrections and nuclear
corrections, which have already been applied in previous PDF
analyses~\cite{Alekhin:2009ni,Alekhin:2010iu}. 
Sec.~\ref{sec:data} features in detail the data analysed with an emphasis on the
systematic and normalization uncertainties. 
This comprises data on inclusive DIS from the HERA collider and fixed target
experiments in Sec.~\ref{sec:incldis}, on the DY process in Sec.~\ref{sec:drell-yan}
and on di-muon production in neutrino-nucleon DIS in Sec.~\ref{sec:di-muon}.

The main results of the present work are contained in Sec.~\ref{sec:results},
where we present all PDF parameters along with illustrations of the shapes of PDFs.
The numerous checks include studies of the pulls and the statistical quality 
for all individual experiments as well as a detailed assessment 
of the power corrections induced by the higher twist terms.
As we work in a scheme with a fixed number $n_f$ of light quark flavors,
a detailed discussion is also devoted to the generation of heavy-quark PDFs.
Our determination of the strong coupling constant $\alpha_s$ 
at NNLO in QCD leads to the value $\alpha_s(M_Z) = 0.1134 \pm 0.0011$.
We show the impact of the individual data sets on $\alpha_s(M_Z)$ and 
compare with the determinations from other PDF fits 
and other measurements included in the current world average.
Finally, Sec.~\ref{sec:results} is complemented with a comparison of 
moments of PDFs with recent lattice results.

The consequences of the new PDF set ABM11 on standard candle cross section
benchmarks are illustrated in Sec.~\ref{sec:crs}. 
We provide cross section values for $W$- and $Z$-boson production in schemes
with $n_f=4$ and $n_f=5$ flavors and we address the accuracy of theory
predictions for all dominant Higgs boson search channels at the LHC.
The PDF uncertainties for top-quark pair-production are also illustrated
highlighting the combined uncertainty in the gluon PDF, $\alpha_s$ and the
top-quark mass $m_t$. 
Sec.~\ref{sec:crs} finishes with comments on the issue of hadronic jet
production, especially from the Tevatron, and the impact of its data on PDF fits.
We conclude in Sec.~\ref{sec:conclusions} and summarize our approach for the handling
of the correlated systematic and normalization uncertainties along with
the explicit tables for the covariance matrix of the ABM11 fit in App.~\ref{sec:appA}.

%%
%% ---------------------------------------------------------------------------
%%
\renewcommand{\theequation}{\thesection.\arabic{equation}}
\setcounter{equation}{0}
\renewcommand{\thefigure}{\thesection.\arabic{figure}}
\setcounter{figure}{0}
\section{Theoretical framework}
\label{sec:theory}

Here, we briefly recall the theoretical basis of our PDF analysis,
which is conducted in the so-called fixed-flavor number scheme
(FFNS) for $n_f$ light (massless) quarks. 
That is to say, we consider QCD with $n_f$ light quarks 
in the PDF evolution, while heavy (massive) quarks only appear in the final state.
As far as QCD perturbation theory is concerned, we specifically focus on
aspects relevant to NNLO accuracy. 
For completeness our treatment of power corrections
and also of nuclear corrections as needed e.g.,  for DIS from
fixed-target experiments, is documented in Sec.~\ref{sec:power-cor}
and \ref{sec:nucl-cor}. The latter have already been used in our
previous PDF determinations~\cite{Alekhin:2009ni,Alekhin:2010iu}.

%%
%% ---------------------------------------------------------------------
%%
\subsection{Perturbative QCD}
\label{sec:pQCD}

The ability to make quantitative predictions in QCD which is a
strongly coupled gauge theory, rests entirely on its factorization property. 
A cross section for the production of some final state $X$ from scattering of initial state
hadrons can be expressed in lepton-nucleon ($ep$) DIS as,
\begin{eqnarray}
\label{eq:factorization-ep}
  \displaystyle
  \sigma_{ep \to lX} &=&
  \sum\limits_{i}\,
  \int
  dz\,
  f_{i}\left(z,\alpha_s(\mur),\mufs \right) \,
  \hat{\sigma}_{ei \to X}\left(z,Q^2,\alpha_s(\mur),\murs,\mufs \right)\,
  \, ,\qquad
\end{eqnarray}
for $l=e,\nu$ and in proton-proton collisions ($pp$) as,
\begin{eqnarray}
\label{eq:factorization-pp}
  \displaystyle
  \sigma_{pp \to X} &=&
  \sum\limits_{ij}\,
  \int
  dz_1\, dz_2\,
  f_{i}\left(z_1,\alpha_s(\mur),\mufs \right) \,
  f_{j}\left(z_2,\alpha_s(\mur),\mufs \right) \,
  \times
  \nonumber \\ &&
  \qquad\qquad\qquad\qquad
  \times \,
  \hat{\sigma}_{ij \to X}\left(z_1,z_2,Q^2,\alpha_s(\mur),\murs,\mufs \right)\,
  \, ,\qquad
\end{eqnarray}
where the PDFs in the nucleon $f_{i}$ ($i=q,{\bar q},g$) are the
objects of our primary interest. They describe the nucleon momentum
fraction $z$ (or $z_1$, $z_2$) carried by the parton and the sums in
eqs.~(\ref{eq:factorization-ep}) and (\ref{eq:factorization-pp}) run
over all light (anti-)quarks and the gluon. The parton cross
sections denoted $\hat{\sigma}$ are calculable in perturbation
theory in powers of the strong coupling constant $\alpha_s$ and
describe the hard interactions at short distances of order ${\cal
O}(1/Q)$. We have also displayed all implicit and explicit
dependence on the renormalization and factorization scales, $\mur$
and $\muf$. Throughout our analysis, however, we will identify them, 
$\mur = \muf = \mu$. All dependencies of $\sigma_{ep \to lX}$ and
$\sigma_{pp \to X}$ on the kinematics and, likewise the integration
boundaries of the convolutions, has been suppressed in
eqs.~(\ref{eq:factorization-ep}) and (\ref{eq:factorization-pp}), as
these are specific to the observable under consideration.

In standard DIS, the (semi-)inclusive cross section $\sigma_{ep \to lX}$ 
in eq.~(\ref{eq:factorization-ep}) depends on the Bjorken 
variable $x$, the inelasticity $y$ and on $Q^2$, the (space-like)
momentum-transfer between the scattered lepton and the nucleon.
Moreover, it admits a decomposition in terms of the well-known DIS
(unpolarized) structure functions $F_i$, $i=1,2,3$. 
QCD factorization applied to the DIS structure functions implies
\begin{eqnarray}
\label{eq:SFfactorization}
  \displaystyle
  a_k\, F_k(x,Q^2) &=&
  \sum\limits_{i}\,
  \int\limits_x^1\,
  \frac{dz}{z}\,
  f_{i}\left(z,\alpha_s(\mu),\mus \right) \,
  C_{k,i}\left(\frac{x}{z},Q^2,\alpha_s(\mu),\mus \right)\,
  \, ,\qquad
\end{eqnarray}
where $a_1=2$, $a_2=1/x$, $a_3=1$, and $C_{k,i}$ denote the Wilson
coefficients. 
$F_L = F_2 - F_T$ defines the longitudinal structure function in terms 
of the tranverse structure function $F_T = 2xF_1$, 
see also eq.~(\ref{eq:Rratio}) below for the relation including target masses.
Eq.~(\ref{eq:SFfactorization}) integrated over $x$ gives rise to the standard Mellin moments, 
\begin{eqnarray}
\label{eq:Mellin-def}
  \displaystyle
  F_k(N,Q^2) &=&
  \int\limits_0^1\,
  dx\, x^{N-1}\, a_k\, F_i(x,Q^2)
  \, .\qquad
\end{eqnarray}
These link the theoretical description of DIS to the
operator-product expansion (OPE) on the light-cone.
The OPE allows to express the DIS
structure functions as a product of (Mellin moments of) the Wilson
coefficients $C_{k,i}$ and operator matrix elements (OMEs) of
leading twist (twist-2). Moreover, it admits a well-defined
extension in powers of $1/Q^2$ (twist-4, twist-6 and so on), 
cf. Sec.~\ref{sec:power-cor}.

The scale dependence of the PDFs is contained in the well-known
evolution equations
\begin{equation}
\label{eq:evolution}
\frac{d}{d \ln \mu^2}\,
  \left( \begin{array}{c} f_{q_i}(x,\mu^2)  \\ f_{g}(x,\mu^2)  \end{array} \right)
\: =\: \sum_{j}\, \int\limits_x^1\, {dz \over z}\,
\left( \begin{array}{cc} P_{q_iq_j}(z) & P_{q_ig}(z) \\
  P_{gq_j}(z) & P_{gg}(z) \end{array} \right)\,
  \left( \begin{array}{c} f_{q_j}(x/z,\mu^2)  \\ f_{g}(x/z,\mu^2)  \end{array} \right)
\, ,
\end{equation}
at leading twist, 
which is a system of coupled integro-differential equations
corresponding to the different possible parton splittings. The
splitting functions $P_{ij}$ in eq.~(\ref{eq:evolution}) have been
determined at NNLO in~\cite{Moch:2004pa,Vogt:2004mw}, which implies
knowledge on the first three terms in the powers series in
$\alpha_s$ (suppressing parton indices),
\begin{eqnarray}
\label{eq:P-exp-alphas}
  P &=&
    \alpha_s \, \sum_{l=0}^\infty \alpha_s^l P^{(l)}
  \, .
\end{eqnarray}
The PDFs $f_i$ are subject to sum rule constraints due to conservation 
of the quark number and the momentum in the nucleon, which imply 
at each order in perturbation theory 
a vanishing first (second) Mellin moment for specific (combinations of) 
splitting functions $P_{ij}$ in eq.~(\ref{eq:P-exp-alphas}).
These sum rule constraints relate the PDF fit parameters used
in the parametrizations of the input distributions, see Sec.~\ref{sec:results}.
The accuracy of the numerical solution of the differential eq.~(\ref{eq:evolution}) 
up to NNLO was tested by comparison to
programs such as {\tt QCD-PEGASUS}~\cite{Vogt:2004ns} 
or {\tt HOPPET}~\cite{Salam:2008qg}.

For the massless DIS structure functions we will be using the
following input from perturbative QCD at leading twist,
\begin{eqnarray}
  \label{eq:F23L}
  F_k \,=\,
    \sum_{l=0}^\infty \alpha_s^l F_k^{(l)}
    \, , \qquad k=2,3,
    \qquad\qquad
  F_L \,=\,
    \alpha_s \, \sum_{l=0}^\infty \alpha_s^l F_L^{(l)}
\, ,
\end{eqnarray}
where, again, (NLO) NNLO accuracy is defined by the first (two)
three terms in the power series in $\alpha_s$ , cf.
eq.~(\ref{eq:P-exp-alphas}). The Wilson coefficients for $F_2$,
$F_3$ in eq.~(\ref{eq:F23L}) are known to NNLO from
~\cite{vanNeerven:1991nn,Zijlstra:1991qc,Zijlstra:1992qd,Zijlstra:1992kj,Moch:1999eb},
and, actually, even to next-to-next-to-next-to-leading order
(N$^3$LO) from~\cite{Vermaseren:2005qc,Moch:2008fj}, and for $F_L$
to NNLO from~\cite{Moch:2004xu,Vermaseren:2005qc}. Note that in the
latter case the perturbative expansion starts at order $\alpha_s$,
thus NNLO accuracy for $F_L$ actually requires three-loop
information, which is numerically not unimportant.

Likewise, for the partonic cross sections of the DY process 
in eq.~(\ref{eq:factorization-pp}), i.e., for hadronic $W$- and $Z$-boson production, 
we use
\begin{eqnarray}
  \label{eq:sigmahat}
  \hat{\sigma}_{ij \to W^\pm/Z} &=&
    \sum_{l=0}^\infty \alpha_s^l \hat{\sigma}_{ij}^{(l)}
  \, ,
\end{eqnarray}
with the NNLO results
of~\cite{Hamberg:1990np,Harlander:2002wh,Anastasiou:2003yy,Anastasiou:2003ds,Catani:2009sm}.

At the level of NNLO accuracy, QCD perturbation theory is expected to provide precise predictions 
as generally indicated by the numerical size of the radiative corrections at successive 
higher orders and their pattern of apparent convergence.
The residual theoretical uncertainty from the truncation of the perturbative expansion 
is conventionally estimated by studying the scale stability of the prediction,
i.e., by variation of the renormalization and factorization scales $\mur$ and $\muf$
in eqs.~(\ref{eq:factorization-ep}) and (\ref{eq:factorization-pp}).
As stated above, we set $\mur = \muf = \mu$ in our analysis, and moreover, 
identify the scale $\mu$ with the relevant kinematics of the process, e.g., $\mu = Q$ for DIS.
Currently no PDF fits with an independent variation of $\mur$ and $\muf$ are
available and we leave this issue for future studies.

One important aspect is the production of heavy quarks in DIS both
for the neutral-current (NC) and the charged-current (CC) exchange. In the
former case, pair-production of charm-quarks accounts for a 
considerable part of the inclusive DIS cross section measured at
HERA, especially at small Bjorken-$x$, while the latter case is
needed in the description of neutrino-nucleon DIS. 
At not too large values of $Q^2$, the NC reaction is dominated by
the photon-gluon fusion process $\gamma^{\,\ast}g \to c \bar{c}\,X$,
while the CC case proceeds through $W^{\,\ast}s \to c$, so that the
perturbative expansion of the respective heavy-quark structure
functions reads,
\begin{eqnarray}
  \label{eq:hqNC-CC}
  F_{k,\rm NC}^q(x,Q^2,m_q^2) \,=\,
    \alpha_s \, \sum_{l=0}^\infty \alpha_s^l F_k^{q,(l)}
  \, ,\qquad\qquad
  F_{k,\rm CC}^q(x,Q^2,m_q^2) \,=\,
    \sum_{l=0}^\infty \alpha_s^l F_k^{q,(l)}
\, ,
\end{eqnarray}
where $k=2,3,L$ and $m_q$ is the heavy-quark mass. The heavy-quark
Wilson coefficients are known exactly to NLO, both for
NC~\cite{Laenen:1992zk} and
CC~\cite{Gottschalk:1980rv,Gluck:1996ve}. The NNLO results for
$F_{k,\rm NC}^q$ are, at present, approximate only and based on the
logarithmically enhanced terms near
threshold~\cite{Laenen:1998kp,Alekhin:2008hc,Presti:2010pd}
(see~\cite{Corcella:2003ib} for threshold resummation in the CC
case).
As well known~\cite{Buza:1995ie}, 
the heavy-flavor corrections to $F_2$ are represented with an accuracy 
of $O(1\%)$ and better for $Q^2/m_q^2 \gsim 10$. 
Under this condition the Wilson coefficients are given by Mellin convolutions 
of massive OMEs~\cite{Buza:1995ie,Buza:1996wv,Bierenbaum:2007qe,Bierenbaum:2009zt} 
and the massless Wilson coefficients~\cite{vanNeerven:1991nn,Zijlstra:1991qc,Zijlstra:1992qd,Zijlstra:1992kj,Moch:1999eb,Moch:2004xu,Vermaseren:2005qc}.
Fixed Mellin moments of the heavy-quark OMEs have also been computed at three loops 
in~\cite{Bierenbaum:2009mv} 
and first results for general values of Mellin-$N$ have been calculated in~\cite{Ablinger:2010ty}.
Mellin space expressions for the 
NC and CC Wilson coefficients up to ${\cal O}(\alpha_s^2)$ are available in~\cite{Alekhin:2003ev,Blumlein:2011zu}.

In the current PDF analysis, the bulk of data from DIS experiments
can be described in a scheme with $n_f=3$ light flavors. 
At asymptotically large scales $Q \gg m_c, m_b$ the genuine 
contributions for heavy charm- and bottom-quarks in a FFNS with $n_f=3$ 
grow as $\alpha_s(Q) \ln(Q^2/m^2)$, as the quark masses screen the collinear divergence.
The standard PDF evolution equations in eq.~(\ref{eq:evolution})
resum these logarithms at the expense of matching the effective
theories, i.e. QCD with $n_f$ and $n_f+1$ light flavors. This
defines a variable-flavor number scheme (VFNS) and gives rise to the
so-called heavy-quark PDFs for charm- and bottom-quarks in QCD with
effectively $n_f=4$ and $n_f=5$ light flavors. The heavy-quark PDFs
are generated from the light flavor PDFs in a $n_f=3$-flavor FFNS as
convolutions with OMEs, see e.g.~\cite{Buza:1996wv,Alekhin:2009ni}. 
The VFNS requires the matching conditions both for the strong coupling $\alpha_s$ and the
PDFs (through the corresponding OMEs), which are known to
N$^3$LO~\cite{Larin:1994va,Chetyrkin:1997sg} for $\alpha_s$ and to
NNLO for the OMEs~\cite{Buza:1996wv,Bierenbaum:2009zt}. 
An extensive discussion of the VFNS implementation has been presented in our previous
analysis~\cite{Alekhin:2009ni}.

The heavy-quark masses in eq.~(\ref{eq:hqNC-CC}) are well-defined within 
a specific renormalization scheme, 
the most popular ones being the on-shell and the \MSbar-scheme. 
The former uses the so-called pole-mass $m_q$,
defined to coincide with the pole of the heavy-quark propagator at
each order in perturbative QCD, and known to have intrinsic
theoretical limitations. As a novelty of our analysis, we employ the
\MSbar-scheme for $m_q$, which enters both in the massive OMEs and
in the Wilson coefficients and introduces a running mass $m_q(\mu)$
depending on the scale $\mu$ of the hard scattering in complete
analogy to the running coupling $\alpha_s(\mu)$. As a benefit,
predictions for the heavy-quark structure functions in terms of the
\MSbar-mass display better convergence properties and greater
perturbative stability at higher orders~\cite{Alekhin:2010sv}, thus
reducing the inherent theoretical uncertainty.

The {\tt Fortran} code {\tt OPENQCDRAD} for the numerical computation of 
all hard scattering cross sections within the present PDF analysis 
is publicly available~\cite{openqcdrad:2011}.
It comprises in particular the theory predictions for the DIS structure functions including the
heavy-quark contributions as well as  for the hadronic $W$- and $Z$-boson production 
and it is capable of computing 
of the benchmark cross sections to NNLO accuracy in QCD in Sec.~\ref{sec:crs}.

We neglect all effects due to Quantum Electrodynamics (QED) on the
PDF evolution. For reasons of consistency, QED effects (including a
photon PDF, see e.g., the analysis in~ see~\cite{Martin:2004dh}) are
sometimes needed in computations of cross sections including
electroweak corrections at higher orders. Quite generally the
effects are small, though. The NNLO QCD corrections to the photon's
parton structure are known~\cite{Moch:2001im,Vogt:2005dw} and we
will address this issue in a future publication.

%%
%% ---------------------------------------------------------------------
%%
\subsection{Power corrections}
\label{sec:power-cor}

The leading twist approximation to the QCD improved parton model is
valid only at asymptotically large momentum transfers $Q^2$ and the
factorization underlying eqs.~(\ref{eq:factorization-ep}) and
(\ref{eq:factorization-pp}) is not sensitive to the finite hadron
size effects or, equivalently, to soft hadronic scales like the
nucleon mass $M_N$. At low momentum transfer comparable to the
nucleon mass such hadronic effects cannot
be ignored and the standard factorization ansatz acquires power
corrections in $1/Q^2$. 
In the case of PDF analyses the higher twist terms are especially important 
for the DIS data since they cover a kinematical range down to $Q^2 \sim M_N^2$. 
The power corrections for the kinematics of DY data used in our 
fit (cf. Sec.~\ref{sec:data}) are negligible 
due to the large momentum transfer $Q^2 \gg M_N^2$ in this case. Therefore
we do not consider power corrections for the DY process.

In DIS the power corrections arise from kinematic considerations 
once the hadron mass effects are taken into account, 
i.e., the so-called target mass correction (TMC). 
The TMC can be calculated in a straightforward way from the
leading twist PDFs within the OPE~\cite{Georgi:1976ve}.
In our analysis the TMC are taken into account in the form of the
Georgi-Politzer prescription~\cite{Georgi:1976ve}. 
For relevant observables, i.e., 
the structure function $F_2$ and the transverse one $F_T$ it reads
\begin{equation}
\label{eq:tmct}
F_T^{\rm TMC}(x,Q^2) \,=\,
    \frac{x^2}{\xi^2\gamma}\, F_T(\xi,Q^2) 
    + 2 \frac{x^3M_N^2}{Q^2\gamma^2}\, \int\limits_\xi^1 \frac{d\xi'}{{\xi'}^2}\, F_2(\xi',Q^2) 
\, ,
\end{equation}
and
\begin{equation}
\label{eq:tmc2}
F_2^{\rm TMC}(x,Q^2) \,=\,
    \frac{x^2}{\xi^2\gamma^3}\, F_2(\xi,Q^2) 
    + 6\frac{x^3M_N^2}{Q^2\gamma^4}\, \int\limits_\xi^1 \frac{d\xi'}{{\xi'}^2}\, F_2(\xi',Q^2)
\, ,
\end{equation}
respectively, which holds up to ${\cal O}(M_N^2/Q^2)$. 
Here $\xi=2x/(1+\gamma)$ and $\gamma=(1+4x^2M_N^2/Q^2)^{1/2}$ 
is the Nachtmann variable~\cite{Nachtmann:1973mr}.
The quantities on the right hand side of eqs.~(\ref{eq:tmct}) and (\ref{eq:tmc2})
are the leading twist structure functions introduced in eq.~(\ref{eq:F23L}) above.

Power corrections can also arise dynamically as
so-called higher twist terms from correlations of the partons inside the hadron. 
The twist-4 terms in the nucleon structure function $F_2$ turns out to be non-negligible at 
large $x$~\cite{Virchaux:1991jc,Blumlein:2006be}. 
Moreover, the higher twist terms in the
longitudinal structure function appear to be necessary for the 
description of the NMC data at moderate $x$~\cite{Alekhin:2011ey,Blumlein:2008kz}
and the SLAC data on the structure function  
$R=\sigma_L/\sigma_T$~\cite{Whitlow:1990gk}, where 
$\sigma_L$ and $\sigma_T$ are the absorption cross sections for the 
longitudinally and the transversely polarized virtual photons, respectively 
(see also eq.~\ref{eq:Rratio}).
The OPE, for lepton-nucleon DIS provides the 
framework for the systematic classification of the higher twist terms 
referring to local composite operators of twist-4 and higher~\cite{Shuryak:1981kj}.
Nonetheless the shapes of the higher twist terms are poorly known. 
Therefore they cannot be accounted for on the same solid theoretical footing as the 
leading twist contributions discussed in Sec.~\ref{sec:pQCD}.
Furthermore, both, the scaling violations and the Wilson coefficients for the
various higher twist contributions have not been computed to the same order in
perturbation theory as for the leading twist part.

Basically two strategies exist to address the issue of power corrections in the PDF analysis. 
The first one imposes kinematical cuts on the data. For DIS, these cuts
are performed at high hadronic invariant masses $W^2 =
Q^2(1/x-1)+M_N^2$ where the nucleon mass is
included in the kinematical considerations. In this way, one aims at
a data sample with reduced sensitivity to power corrections. Typical
values for cuts on $W^2$ are of the order of $12$~GeV$^2$. As a
drawback of this procedure one eliminates a rather large fraction of
data at low $Q^2$ with excellent statistical precision. A more
serious concern, however, is due to the generally poor theoretical
understanding of those non-perturbative QCD effects beyond leading
twist factorization. One simply cannot estimate from first
principles the region of $Q^2$ (or $W^2$), where power corrections
can be safely neglected. Therefore, the present analysis
(following~\cite{Alekhin:2000ch}) examines both, TMC
and higher twist contributions, in detail in order to control and
quantify their impact in the determination of the standard leading
twist PDFs. 

In practice  higher twist contributions are usually parameterized independently from the
leading twist one with some function of $x$, which is typically
polynomial in $x$. In our analysis the power corrections are
non-negligible for the case of the DIS data and are defined
within an entirely phenomenologically motivated ansatz, as follows
\begin{eqnarray}
\label{eq:htwist}
  \displaystyle
  F_i^{\rm ht}(x,Q^2) &=&
    F_i^{\rm TMC}(x,Q^2)
  +
    \frac{H_i^{\tau=4}(x)}{Q^2}
  +
    \frac{H_i^{\tau=6}(x)}{Q^4}
  + \dots
  \, ,\qquad
\end{eqnarray}
where $F_i^{\rm TMC}$ are given by eqs.~(\ref{eq:tmct}) and (\ref{eq:tmc2}).
The coefficients $H_i$ are parameterized by a cubic spline with the 
spline nodes selected at x=0, 0.1, 0.3, 0.5, 0.7, 0.9, and 1. 
This choice provides sufficient flexibility of the coefficients $H_i$ 
with respect to the data analysed and, at the same time, keeps a reasonable number of nodes.  
The values of $H_i(1)$ are fixed at zero due to kinematic constraints.
The values of $H_i(0)$ are also put to zero in view of the fact 
that no clear signs of any power-like terms can be found in the low-$x$ HERA data.
The rest of the spline-node values of $H_i$ were fitted to the data 
simultaneously with the PDF parameters and the value of $\alpha_s$. 
We neglect the $Q^2$-dependence of the higher twist
operators due to the QCD evolution. Therefore the coefficients $H_i$
do not depend on $Q^2$. This treatment could be further refined by
considering the individual (quasi-)partonic OMEs along with their
renormalization, i.e., their $Q^2$-dependence, which is known for
the twist-4 operators to first order in
$\alpha_s$~\cite{Braun:2009vc}. 
Another complication is the emergence of more Bjorken-like variables $x_i$, 
the number of which grows with increasing twist.
Experimental information on the other hand is only available for 
the variable $x=\sum_i x_i$ and $Q^2$.
We leave these aspects for future studies. 

With the kinematical cuts imposed on the DIS data in our analysis 
(cf. Sec.~\ref{sec:data}) the twist-6 terms are 
irrelevant~\cite{Alekhin:2007fh} therefore the 
coefficients $H_i^{\tau=6}$ in eq.~(\ref{eq:htwist}) are washed out. 
The target dependence of the higher twist parametrization in eq.~(\ref{eq:htwist}) 
has been studied in~\cite{Alekhin:2003qq,Alekhin:2008mb,Alekhin:2008ua}.
The isospin asymmetry in $H_T$ is poorly constrained by the data used in our fit. 
It is comparable with zero within the uncertainties~\cite{Alekhin:2003qq} and 
therefore it was put to zero in our analysis. 
The isospin asymmetry in $H_2$ is also numerically small, however, due 
to lower uncertainties it cannot be put to zero without deterioration 
of the fit quality. 
In summary, we fit three twist-4 coefficients, 
for the proton $H_2^{\rm p}$, for the neutron $H_2^{\rm n}$ and for the nucleon $H_T^{\rm N}$, 
in addition to the leading twist terms. 
The impact of the power corrections on the DIS neutrino-nucleon di-muon production 
data used in our fit is marginal~\cite{Alekhin:2008mb} 
as well as on the inclusive charge-current data~\cite{herapdf:2009wt}.
Therefore they are not considered 
for the case of charged-current structure functions.

%%
%% ---------------------------------------------------------------------
%%
\subsection{Corrections for nuclear effects}
\label{sec:nucl-cor}

\begin{figure}[t!]
\centerline{
  \includegraphics[width=15.5cm]{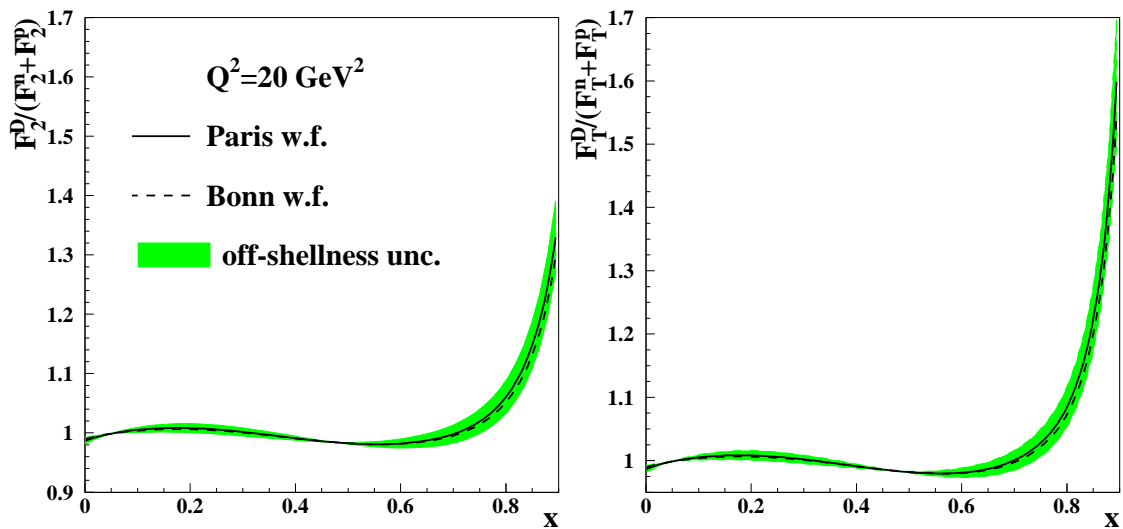}}
  \vspace*{-5mm}
  \caption{\small
    \label{fig:deut}
      The ratio of the deuterium structure function $F_2$ (left) and 
      $F_T$ (right) with account of the Fermi-motion and off-shellness
      effects of~\cite{Kulagin:2004ie} calculated 
      for the Paris potential of~\cite{Lacombe:1980dr} (solid)
      and the Bonn potential of~\cite{Machleidt:1987hj} (dashes)
      at the momentum transfer of $20~{\rm GeV}^2$
      to the sum of those for free proton and neutron versus $x$.
      The shaded area around the solid line 
      gives the uncertainty due to a variation of the off-shell effects by 50\%.
      The calculations are performed 
      at NNLO QCD accuracy using the PDFs and the twist-4 terms 
      obtained in~\cite{Alekhin:2009ni}. 
}
\end{figure}

For our analysis we select primarily the data obtained off proton targets.
However in some cases the necessary constraints on PDFs come only from 
nuclear target data. For example the neutral-current
DIS off deuterium targets allows the separation 
of the large-$x$ $u$- and $d$-quark distributions, which contribute to the 
DIS structure functions in the form of linear combination weighted with 
the quark charges. However the analysis of the deuterium data requires modeling
of nuclear effects. They include the Fermi-motion and 
off-shellness of the nucleons, excess of pions in the nuclear 
matter, Glauber shadowing, etc., cf.~\cite{Kulagin:2004ie,Arrington:2011qt} for reviews. 
Among them only the Fermi-motion effects can be calculated 
with an uncertainty better or comparable to the uncertainty in the 
existing experimental data. The Fermi-motion correction is given as  
a convolution of the free 
nucleon structure functions with the deuterium wave function, which 
in turn is constrained by the low-energy electron-nuclei scattering data. 
The parameterization of the off-shell effect used in our fit was obtained from the 
analysis of the world data on DIS off heavy nuclear targets and 
extrapolated to the deuterium target. 
In this way we assume that the nuclear model suggested in~\cite{Kulagin:2004ie} 
can be applied to the case of light nuclei, like 
deuterium. This assumption has been recently confirmed for 
the case of the $^3{\rm He}$ and $^9{\rm Be}$ targets~\cite{Kulagin:2010gd}. 
However, in order to 
take a conservative estimate of the deuteron correction uncertainty
due to the off-shellness effect we vary its magnitude by 50\%.
The uncertainty obtained in this way is comparable with one given in~\cite{Arrington:2011qt}.
Other nuclear effects, like shadowing and pion excesses in nuclei
considered in~\cite{Kulagin:2004ie} were found numerically negligible 
for the case of deuterium. Thus our model of the nuclear effects for 
deuterium is based on the combination of the Fermi-motion and 
the off-shellness effects only. The nuclear corrections 
depend both on the deuterium wave function and the 
free nucleon structure functions, while the latter include the 
target-mass corrections and the twist-4 terms, cf. eq.~(\ref{eq:htwist}). 
Due to the structure function dependence the value of the 
correction is sensitive to the fitted parameters and ideally it should be 
re-calculated iteratively in the fit. 
However, this approach turns out to be rather time-consuming. 
Therefore we calculate the deuteron correction once at the 
beginning with the PDFs and twist-4 terms which were obtained in~\cite{Alekhin:2009ni}.
Since the deuteron model employed in our fit is the same as 
one of~\cite{Alekhin:2009ni}
this approach introduces only a marginal bias into the fit.
The nuclear correction for the 
representative kinematics of the deuterium data is given 
in Fig.~\ref{fig:deut} for the cases of the deuterium wave function obtained with the Paris 
potential of~\cite{Lacombe:1980dr} used in our analysis
and the Bonn potential of~\cite{Machleidt:1987hj}. The difference 
between these two cases is marginal as compared to the errors in the data. 
The data on di-muon production in the (anti-)neutrino-nucleon scattering
which are used in our analysis in order to constrain the strange sea distribution, 
were obtained on iron targets. In contrary to the 
neutral-current case a particular shape of the nuclear correction at
large $x$ is unimportant since data do not populate the region of 
$x \gtrsim 0.3$.  
At small $x$ the neutrino-nucleon DIS nuclear corrections 
are enhanced due to the parity non-conserving 
part of the charged current~\cite{Kulagin:2007ju}, however their
impact on the strange sea distribution is still smaller than 
its error~\cite{Alekhin:2008mb}. 
Therefore in the modeling of the di-muon production data 
we employ the nuclear corrections of~\cite{Kulagin:2007ju}
without consideration of their uncertainties. 
The DY data used in our fit span the moderate-$x$ region. 
For such kinematics the nuclear effects are quite smaller than the 
errors in data~\cite{Alde:1990im} and they are not considered
in our analysis.

%%
%% ---------------------------------------------------------------------------
%%
\renewcommand{\theequation}{\thesection.\arabic{equation}}
\setcounter{equation}{0}
\renewcommand{\thefigure}{\thesection.\arabic{figure}}
\setcounter{figure}{0}
\renewcommand{\thetable}{\thesection.\arabic{table}}
\setcounter{table}{0}
\section{Data}
\label{sec:data}

The nucleon PDFs are usually extracted from a combination of hard-scattering data, 
which provides complementary constraints on the different PDF species.  
A particular choice of which processes to be used in an analysis is commonly driven by 
the theoretical accuracy of the data modeling and/or the experimental uncertainties in the data. 
In our fit we employ the data on inclusive DIS, the DY process, and 
di-muon production in neutrino-nucleon DIS. 
In combination they allow for a good separation of the quark flavors in wide range of $x$ and 
provide good constraints on the gluon distribution at small 
values of $x$, which are mostly important for the collider phenomenology.

%%
%% ---------------------------------------------------------------------
%%
\subsection{Inclusive DIS}
\label{sec:incldis}

Studies of inclusive DIS date back to the early days of QCD and since that time 
a wealth of the accurate data has been collected. 
The first fixed-target DIS experiments at SLAC were followed by data from CERN and Fermilab and then 
at the electron-proton collider HERA at DESY. 
The most accurate data of these 
experiments obtained on the proton and deuterium targets are included into 
our analysis~\cite{Bodek:1979rx,Atwood:1976ys,Mestayer:1982ba,Gomez:1993ri,
Dasu:1993vk,Benvenuti:1989rh,Benvenuti:1989fm,Arneodo:1996qe,Aaron:2010ry}.

In all cases we employ the data on the inclusive cross section, 
which is related to the DIS structure functions as follows 
\begin{eqnarray}
\label{eq:sigma}
{\lefteqn{
     \frac{d^2\, \sigma(x,Q^2)}{dx dQ^2} \,=\,}}
\nonumber\\ && 
    \frac{4 \pi \alpha^2}{x Q^4} \, \Biggl\{ \Biggl( 1 - y - x y
    \frac{M_N^2}{s} \Biggr)  F_2^{\rm ht}(x,Q^2) 
    + \frac{y^2}{2} \left(1-\frac{2 m_l^2}{Q^2}\right) F_{T}^{\rm ht}(x,Q^2) 
    \pm y\Biggl(1 - \frac{y}{2}\Biggr) xF_{3}^{\rm ht}(x,Q^2) 
    \Biggr\} \,
   \, ,
\end{eqnarray}
with the mass of the incident charged lepton $m_l$
and $\pm F_{3}^{\rm ht}$ corresponding to different polarizations 
for the case of the charged current.
The nucleon structure functions 
$F_{2}^{\rm ht}$, $F_{T}^{\rm ht}$ and $F_{3}^{\rm ht}$ are calculated 
with account of the nuclear correction described 
in Sec.~\ref{sec:nucl-cor}, if relevant. 
Note, that higher twist contributions to $F_{3}^{\rm ht}$ are set to zero,
cf. Sec.~\ref{sec:power-cor}.
In this way we provide a consistent treatment of the data, 
contrary to the common procedure in global PDF fits 
which are based on the data for the structure function $F_2$. 
The structure functions $F_{2}$ and $F_{T}$ in eq.~(\ref{eq:sigma}) 
also enter in the ratio of the longitudinally to transversely 
polarized virtual photon absorption cross sections (see e.g.~\cite{Arbuzov:1995id}), 
\begin{eqnarray}
  \label{eq:Rratio}
  R(x,Q^2) &=& \frac{F_L(x,Q^2)}{F_T(x,Q^2)}
 =\frac{F_2}{F_T}\left(1+\frac{4M_N^2x^2}{Q^2}\right)-1
  \, .
\end{eqnarray}
In order to avoid contributions from nucleon resonances and 
the twist-6 terms we do not include into the analysis any inclusive DIS data with 
\begin{equation}
\label{eq:discuts}
Q^2<2.5~{\rm GeV}^2
\, ,\qquad\qquad
W<1.8~{\rm GeV}
\, . 
\end{equation}

The kinematics spanned by each DIS data set used in our fit and 
their systematic uncertainties are described in the following subsections. 
The normalization uncertainty is a particular case of the systematics. 
However it is considered separately since very often 
the absolute normalization of the DIS experiment is not independently determined. 
Instead, in such cases it is usually tuned to a selected set of other 
DIS experiments, which in turn provide the absolute normalization. 
The wealth of the DIS data used in our fit allows us to extend the 
basis for this normalization tuning.
Therefore, for the experiments lacking an absolute normalization 
we consider general normalization factors 
which are fitted simultaneously with other parameters of our data model.
Within this approach we introduce free normalization parameters for 
the separate early SLAC experiments 
of~\cite{Bodek:1979rx,Atwood:1976ys,Mestayer:1982ba,Gomez:1993ri} 
and for the NMC data of~\cite{Arneodo:1996qe} at each beam energy.
The errors in the normalization factors obtained in our fit 
are included into the general covariance matrix calculation. 
In this way we account for the impact of the absolute normalization 
uncertainty in the data on the PDFs, the higher twist terms and on the 
value of $\alpha_s$.

Our procedure for the treatment of the DIS data normalization in PDF fits 
differs substantially from other approaches. 
For instance, in the MSTW PDF fit~\cite{Martin:2009iq}
free normalization parameters are introduced for all data sets, 
including even those where the absolute normalization has been determined experimentally. 
Other PDF fits also commonly employ the NMC data averaged over the beam energies 
and combined data from the SLAC experiments, rather than the respective individual data sets.

\subsubsection{HERA}

In our analysis we use the HERA data 
on the inclusive neutral-current and the 
charged-current cross sections~\cite{herapdf:2009wt}.
This sample was obtained by a combination of the run~I data of 
the H1 and ZEUS experiments, and includes in particular the 
data of~\cite{Adloff:2000qk,Chekanov:2001qu}
used earlier in the ABKM09 fit~\cite{Alekhin:2009ni}.
The HERA data span the region of $Q^2$ up to $30000~{\rm GeV}^2$.
However, we impose an additional cut of $Q^2<1000~{\rm GeV}^2$ on the neutral-current sample.
This allows to neglect the $Z$-boson exchange contribution, which is of the
order $\sim 1\%$ at $Q^2=1000 {\rm GeV}^2$. 
At the same time, the high-$Q^2$ part of those data displays only a poor
sensitivity to our PDF fit, since the accuracy of those HERA data is ${\cal O}(10\%)$.
Therefore the chosen cut does not distort the fit in any way.
The normalization uncertainty in the HERA data of~\cite{herapdf:2009wt}
is 0.5\%, much better than one in the HERA data of~\cite{Adloff:2000qk,Chekanov:2001qu}. 
In particular due to the improvement in normalization 
the new data of~\cite{herapdf:2009wt} somewhat overshoot 
the previous H1 data of~\cite{Adloff:2000qk}.  
The total number of correlated systematic uncertainties in the 
HERA data of~\cite{herapdf:2009wt} is 114, including the uncertainties
due to the combination procedure and the general normalization. 
Many of them are improved as compared to the separate experiment samples
as a results of cross-calibration in the process of combination.   

A complementary set of the inclusive HERA data was obtained by the H1
collaboration in the run with a reduced collision energy~\cite{Aaron:2010ry}. 
These data are particularly sensitive to the structure function $F_L$ and 
thereby to the small-$x$ shape of the gluon distribution. 
The normalization uncertainty in the low-energy 
data of~\cite{Aaron:2010ry} is 3\%. 
The point-to-point correlated systematic uncertainties come from 8  
independent sources and there is also a number of uncorrelated systematic 
uncertainties in the data. 

\subsubsection{BCDMS}

The BCDMS data of~\cite{Benvenuti:1989rh,Benvenuti:1989fm}
used in our fit were collected at the CERN muon beam at 
energies of 100, 120, 200, and 280 GeV for the incident muons. 
Due to the use of both proton and deuterium targets 
in the same experiment these data facilitate flavor separation of PDFs at large $x$. 
The BCDMS absolute normalization was monitored for the beam energy of 200 GeV.
The general normalization uncertainty in the data due to this monitor is 
as big as 3\%. The absolute normalization 
of the data obtained at the beam energies of 100, 120, and 280 GeV was calibrated 
with respect to the case of the beam energy setting of 200 GeV. The additional 
normalization uncertainty due to this calibration ranges from 
1\% to 1.5\% depending on the beam energy. 
Other systematic uncertainties in the BCDMS data stem from 5 sources
with the most important contributions due to incident 
and scattered muon energy calibration and the spectrometer resolution. 
Every source generates a point-to-point correlated uncertainty in the data, 
while the sources itself are uncorrelated with each other. 

\subsubsection{NMC}

The NMC experiment was performed like BCDMS at the CERN muon beam at 
incident muon energies of 90, 120, 200, and 280 GeV. 
However, the NMC data span lower values of $x$ and $Q^2$ as compared to BCDMS and overlap 
with the HERA data at the edge of the respective kinematics.
We use in our fit the NMC cross section data 
of~\cite{Arneodo:1996qe} for the proton and deuterium targets. 
Due to better coverage of the small-$x$ region those data are also sensitive to 
the isospin asymmetry in the sea distribution. 
The absolute normalization for the NMC data of~\cite{Arneodo:1996qe} was determined from tuning 
for each particular energy setting separately 
to the BCDMS and SLAC data, which overlap partially with NMC.
This tuning in~\cite{Arneodo:1996qe} 
was based on an empirical data model motivated basically by leading-order QCD calculations. 

In our analysis, therefore, we fit the NMC normalization factors for each incident beam energy 
and target simultaneously with the other parameters. 
In this manner, we ensure consistency with our data model, 
which in particular includes QCD corrections up to the NNLO, 
see~\cite{Alekhin:2011ey} for a detailed study of the impact of the NNLO QCD corrections  
on the interpretation of the NMC data.
The normalization factors obtained in the NNLO variant of our fit are given in Tab.~\ref{tab:NMC}.
In general they are within the uncertainty of 2\% quoted for the NMC data in~\cite{Arneodo:1995cq}.
However, the normalization factors for the proton target are somewhat larger 
than for the case of deuterium. This is explained by impact 
of the HERA data of~\cite{herapdf:2009wt}, which slightly overshoot 
the NMC data in the region of their overlap.  
The systematic uncertainties in the NMC data are due to 
the incident and scattered muon energy calibration, the reconstruction efficiency, 
acceptance, and the electroweak radiative corrections. 
Some of the systematic uncertainties are correlated for all data, some of them between the proton 
and deuterium data, and some between beam energies (cf.~\cite{Arneodo:1996qe} for details).  
In summary this gives 12 independent sources of systematic uncertainties for the NMC data 
used in our fit. 

\begin{table}[h]
\begin{center}
\begin{tabular}{|c|c|c|}
\hline
Beam energy (GeV) & proton & deuterium \\\hline 
90  & 1.012(12) & 0.990(12) \\\hline 
120  & 1.026(11) & 1.005(11) \\\hline
200  & 1.034(12) & 1.014(11) \\\hline
280  & 1.026(11) & 1.007(11) \\\hline
\end{tabular}
\caption{
\small
The NMC normalization factors obtained in our NNLO fit 
for different incident beam energies and targets. 
}
\label{tab:NMC}
\end{center}
\end{table}

\subsubsection{SLAC}

The SLAC experiments used in our fit and the number of data points
for each experiment after the cut of eq.~(\ref{eq:discuts})
are listed in Tab.~\ref{tab:SLAC}. 
The last and most elaborated in this series is experiment 
E-140~\cite{Dasu:1993vk}. 
In particular, it took advantages of the improved electroweak radiative 
corrections and the accurate determination of the data absolute normalization, which 
is as big as 1.8\% for the deuterium sample. Other point-to-point 
correlated systematic uncertainties are due to background contamination, 
the spectrometer acceptance, and the electroweak
radiative corrections~\cite{Whitlow:1990dr}. The rest of 
systematic error sources for the experiment E-140 are 
uncorrelated.

\begin{table}[h]
\begin{center}
\begin{tabular}{|c|c|c|c|c|c|}
\hline
Experiment & Target & NDP & NSE & Normalization & Normalization \\
 & &  & & (our fit) & (Ref.~\cite{Whitlow:1990gk})    \\\hline
E-49a~\cite{Bodek:1979rx} & proton &59  &3 & $1.022(11)$ & 1.012  \\
 & deuterium & 59 &3 & $0.999(10)$ & 1.001   \\\hline

E-49b~\cite{Bodek:1979rx} & proton & 154& 3 & $1.028(10)$ & 0.981  \\
 & deuterium & 145 &3 & $1.008(10)$ & 0.981   \\\hline

E-87~\cite{Bodek:1979rx} & proton & 109 &3 & $1.032(10)$ & 0.982  \\
 & deuterium & 109 &3 & $1.017(10)$ & 0.986   \\\hline

E-89a~\cite{Atwood:1976ys} & proton & 77 &4 & $1.$ & 0.989  \\
 & deuterium &71 &5 & $1.$ & 0.985   \\\hline

E-89b~\cite{Mestayer:1982ba} & proton & 90 &3 & $1.016(10)$ & 0.953  \\
 & deuterium & 72 &3 & $0.996(10)$ & 0.949   \\\hline

E-139~\cite{Gomez:1993ri} & deuterium & 17 &3  & $1.014(10)$ & 1.008   \\\hline

E-140~\cite{Dasu:1993vk} & deuterium & 26 &5 & $1.$ & 1.   \\\hline

\end{tabular}
\caption{
\small
The list of SLAC experiments used in our fit 
(first column: the experiment number; 
third column: the number of data points (NDP) used in the fit; 
fourth column: the number of correlated systematic errors (NSE) in the data; 
fifth column: the normalization factor applied to the data in our fit; 
sixth column: the normalization factor applied to the data in the re-analysis of~\cite{Whitlow:1990gk}). 
Note, that the normalization factors of the fifth column apply to the data, which were 
re-normalized in~\cite{Whitlow:1990gk} by the factors given in the sixth column. 
}
\label{tab:SLAC}
\end{center}
\end{table}

The earlier SLAC data used in our fit
are collected with various experimental setups and data 
processing chains. In particular, the 
electroweak radiative corrections applied to the data differ in details, 
various methods are used to determine absolute normalization of the data,
etc. To overcome this diversity the early SLAC data 
of~\cite{Bodek:1979rx,Atwood:1976ys,Mestayer:1982ba,Gomez:1993ri} 
were reanalysed within a uniform approach and the leveled set of the SLAC
data was obtained in~\cite{Whitlow:1990gk}.
As a part of this leveling the absolute normalization 
factors for the data of~\cite{Bodek:1979rx,Atwood:1976ys,Mestayer:1982ba,Gomez:1993ri}
were calibrated with the help of the E-140 data of~\cite{Dasu:1993vk}. Due to the lack  
of the E-140 proton data this calibration is straightforward only for 
the deuterium case. The proton data normalization tuning 
was performed in two steps. First, the normalization of 
the deuterium data of experiment E-49b was determined with the help of 
the E-140 deuterium data. Then, the proton data normalization 
for all other experiments was tuned to the E-49b proton data, 
assuming equal normalization for the proton and deuterium samples of 
the E-49b experiment. The data of experiment E-89a are kinematically 
separated from other SLAC experiments considered. Therefore, 
their normalization tuning was based on the elastic scattering samples 
obtained in the experiments E-89a, E-89b, and E-140 
(cf.~\cite{Whitlow:1990gk} for the details). 
In view of the fact, that we do not include elastic data 
in the fit we keep the normalization of the 
SLAC experiment E-89a at the value obtained in~\cite{Whitlow:1990gk}.
At the same time in order to take into account the uncertainties in the 
E-89a data normalization we add to those data the general 
normalization uncertainty of 2.8\% and an additional 
normalization uncertainty of 0.5\% for the case of deuterium, which are
quoted in~\cite{Whitlow:1990gk}.
The normalization factors for the early SLAC experiments of~\cite{Bodek:1979rx,Mestayer:1982ba,Gomez:1993ri}
are considered as free parameters of the fit. 
The SLAC normalization factors for the NNLO variant of our fit 
are given in Tab.~\ref{tab:SLAC} in comparison with
the ones of~\cite{Whitlow:1990gk}, 
which were obtained with an empirical QCD-motivated model of the data. 
The deuterium normalization factors obtained in our fit are in a good agreement 
with the ones of~\cite{Whitlow:1990gk}. For the proton target case 
our normalization factors are somewhat bigger, in particular due to 
wider set of data is used for the normalization tuning in our case.

%%
%% ---------------------------------------------------------------------
%%
\subsection{Drell-Yan process}
\label{sec:drell-yan}

The data for the DY process provide a complementary constraint on the PDFs.  
In particular they allow to separate the sea and the 
valence quark distributions in combination with the DIS data. 
We use for this purpose the data obtained by the 
fixed-target Fermilab experiments 
E-605~\cite{Moreno:1990sf} and E-866~\cite{Towell:2001nh}.

The experiment E-605 collected proton-copper collisions data 
at the center-of-mass energy of 38.8 GeV for di-muon invariant masses 
in the range of $7\div17~{\rm GeV}$. At this kinematics the DY data are sensitive to the PDFs down to $x\sim0.03$. The 
normalization uncertainty in the E-605 data is 
15\%. However other systematic uncertainties in the data are not fully 
documented in~\cite{Moreno:1990sf}. 
The point-to-point correlated systematic is estimated as $+10\%$
for low di-muon masses and $-10\%$ for higher masses. 
Due to lacking details in~\cite{Moreno:1990sf} we assume a linear dependence
of this systematic error on the di-muon mass.  
Additional uncorrelated systematic uncertainties in the E-605 data due to the Monte Carlo 
acceptance calculation are combined with the statistical ones in quadrature. 

The data of the E-866 experiment on the ratio of the 
proton-proton and proton-deuterium collision cross 
sections~\cite{Towell:2001nh} 
are particularly sensitive to the isospin asymmetry of the 
sea quark distributions. The absolute normalization uncertainty cancels
in this ratio. Other E-866 systematic uncertainties 
stem from 5 independent sources with the biggest contributions 
due to the deuterium composition and the event 
detection and reconstruction. The unpublished data on the absolute 
DY cross sections for the proton and deuterium targets 
are also available~\cite{Webb:2003ps}. However these
data are in poor agreement with the 
DIS data (cf.~\cite{Alekhin:2006zm} for a detailed comparison). 
Therefore, we do not employ the data of~\cite{Webb:2003ps}
in our fit. Note, that in the MSTW PDF fit~\cite{Martin:2009iq}
the E-866 data on the absolute cross sections are  
shifted upwards by 8.7\% in order to bring them into agreement with 
the other data sets.

%%
%% ---------------------------------------------------------------------
%%
\subsection{Di-muon production in {\boldmath $\nu N$} DIS}
\label{sec:di-muon}

The production of di-muons in neutrino-nucleon collisions 
provides unique information about the strange sea distribution in the nucleon. 
One of the muons produced in this reaction may be resulting from the decay of
a charmed hadron.
Thus, the production of the $c$-quarks in neutrino-nucleon collisions is
directly related to initial-state strange quarks. 
Therfore, by relying upon a $c$-quark fragmentation model one can determine the 
(anti-)strange-sea distribution from the data on di-muon production
in an (anti-)neutrino beam. 
The details of the fragmentation model are quite important in this context
due to kinematic cuts imposed to suppress a background of muons coming 
from the light mesons. Herewith the absolute normalization of the model 
is defined by the semi-leptonic branching ratio $B_\mu$ of the charmed hadrons. 
The value of $B_\mu$ is poorly known due to the uncertainty 
in the hadronic charm production rate for the neutrino-nucleon interactions.
On the other hand, the value of $B_\mu$ is also constrained by the 
di-muon data themselves~\cite{Bazarko:1994tt,Alekhin:2008mb}.
Therefore, for consistency, we fit the value of $B_\mu$ simultaneously with the PDF parameters 
imposing available independent constraints on $B_\mu$ coming from emulsion experiments
(cf.~\cite{Alekhin:2008mb} for the details).

We use in the fit the di-muon data provided by two Fermilab experiments, 
CCFR and NuTeV, and corrected for the cut of 5 GeV imposed on the 
muon decay energy in order to suppress the light-meson 
background~\cite{Goncharov:2001qe,Mason:2006qa}. 
The data of the NuTeV experiment were normalized through the use of 
the inclusive single muon event rates. Therefore, the normalization 
error in the data is marginal and it is not considered in our fit. 
Besides, 8 independent sources contribute to the point-to-point
correlated systematic uncertainties. 
The neighboring NuTeV data points 
are also correlated due to  smearing of the kinematic variables.
These correlations are not documented in~\cite{Goncharov:2001qe}. Instead 
the errors in the data are inflated in such a way that 
the fit of a model to the data with inflated errors is 
equivalent to the regular fit with account of the data correlations  
(cf.~\cite{Goncharov:2001qe} for details).
The average data error inflation factor is about 1.4. 
Therefore, the normal value of $\chi^2$ for the inflated-error fit is about 
one half of the NDP. 
The CCFR data of~\cite{Goncharov:2001qe} were processed 
similarly to the NuTeV ones. In particular, the errors in the data 
were also inflated by factor of about 1.4 in order to take into account 
the data point correlations. However, only 
the combined systematic errors in the CCFR data are available. 
In view of the lack of any detailed information about 
the systematic error correlations we employ in our fit the combined 
systematic errors quoted in~\cite{Goncharov:2001qe} 
assuming them to be fully point-to-point correlated. 

%%
%% ---------------------------------------------------------------------------
%%
\renewcommand{\theequation}{\thesection.\arabic{equation}}
\setcounter{equation}{0}
\renewcommand{\thefigure}{\thesection.\arabic{figure}}
\setcounter{figure}{0}
\renewcommand{\thetable}{\thesection.\arabic{table}}
\setcounter{table}{0}
\section{Results}
\label{sec:results}

We are now in a position to present the results of our analysis ABM11 to NLO and NNLO in QCD for $n_f=3$ in a FFNS. 
The PDF sets for $n_f=4$ and $5$ are then generated by matching as described above 
and we will comment on the changes in the PDFs obtained compared to the ABKM09 set~\cite{Alekhin:2009ni}.
In addition to the fit results and the covariance matrix for the 
correlations of the fit parameters, we also present the pulls for separate
experiments, which reflect the compatibility of these data sets with respect
to the experimental systematics.
The discussion of the value of the strong coupling obtained in ABM11 is supplemented 
by a compilation of $\alpha_s(M_Z)$ determinations in NLO and NNLO analyses extending~\cite{Alekhin:2011gj}.
For a valence distribution we also compute the lowest Mellin moment of our PDFs and compare with the latest
available data from lattice simulations.

%%
%% ---------------------------------------------------------------------
%%
\subsection{PDF parameters}
\label{sec:pdffit}

In the new analysis the shape of the PDFs has been updated and the number of
fit parameters has been slightly enlarged compared to ABKM09~\cite{Alekhin:2009ni}.
In detail, we are using the following parametrizations 
at the starting scale $\mu^2 = Q_0^2 = 9.0~{\rm GeV}^2$ in the scheme with $n_f=3$ flavors, 
\begin{eqnarray}
\label{eq:pdf1}
  x q_{v}(x,Q_0^2)
  &=&
  \displaystyle
  \frac{ 2 \delta_{qu}+\delta_{qd}} {N^{v}_q}x^{a_{q}}(1-x)^{b_q}x^{P_{qv}(x)}
  \, , 
  \qquad 
  \\
\label{eq:pdf2}
  x u_{s}(x,Q_0^2) 
  \,=\,
  x\bar{u}_{s}(x,Q_0^2) 
  &=& 
  A_{us} x^{a_{us}}(1-x)^{b_{us}} x^{a_{us}\, P_{us}(x) }
  \, ,
  \qquad 
\\
\label{eq:pdf3}
  x\Delta(x,Q_0^2)
  \,=\,
  xd_{s}(x,Q_0^2)-xu_{s}(x,Q_0^2) 
  &=&
  A_\Delta x^{a_{\Delta}}(1-x)^{b_{\Delta}} x^{P_{\Delta}(x)}
  \, ,
  \qquad
\\
\label{eq:pdf4}
  xs(x,Q_0^2) = x\bar{s}(x,Q_0^2) 
  &=& 
  A_{s} x^{a_{s}}(1-x)^{b_{s}}
  \, ,
\\
\label{eq:pdf5}
  xg(x,Q_0^2)
  &=&
  A_{g}x^{a_{g}}(1-x)^{b_{g}} x^{a_{g}\, P_{g}(x) }
  \, ,
  \qquad
\end{eqnarray}
where $q = u,d$ and $\delta_{qq^\prime}$ denotes the Kronecker function in eq.~(\ref{eq:pdf1})  
and the strange-quark distribution is taken to be charge-symmetric, cf.~\cite{Alekhin:2008mb}. 
The polynomials $P(x)$ in eqs.~(\ref{eq:pdf1})--(\ref{eq:pdf5}) are given by
\begin{eqnarray}
\label{eq:Ppdf1}
  P_{qv}(x) &=& \gamma_{1,q}\, x + \gamma_{2,q}\, x^2 + \gamma_{3,q}\, x^3
  \, , 
  \\
\label{eq:Ppdf2}
  P_{us}(x) &=& \gamma_{3,us}\, \ln x\, (1 + \gamma_{1,us}\, x + \gamma_{2,us}\, x^2 ) 
  \, ,  
  \qquad
\\
\label{eq:Ppdf3}
  P_{\Delta}(x) &=& \gamma_{1,{\Delta}}\, x
  \, ,  
  \qquad
\\
\label{eq:Ppdf5}
  P_{g}(x) &=& \gamma_{1,g}\, x
  \, .
  \qquad
\end{eqnarray}
The new functional form with the additional parameters $\gamma_{3,u}$ and $\gamma_{3,us}$ provides sufficient flexibility 
in the small-$x$ $u$-quark distribution with respect to the analysed data and we have checked that no additional terms 
are required to improve the quality of the fit.
All 24 PDF parameters are given in Tab.~\ref{tab:fitvalues} 
together with their $1\sigma$ uncertainties computed from  
the propagation of the statistical and systematic errors in the data, cf. App.~\ref{sec:appA}.
Note that the normalization parameters for the valence quarks, $N^{v}_q$, and gluons, $A_{g}$,
are related to the other PDF parameters due to 
conservation of fermion number and of momentum, respectively.

\begin{table}[h]
\renewcommand{\arraystretch}{1.5}
\begin{center} 
\footnotesize
\begin{tabular}{|c|c|c|c|c|c|c|} 
\hline 
\multicolumn{1}{|c|}{ } & 
\multicolumn{1}{c|}{$a$} & 
\multicolumn{1}{c|}{$b$} & 
\multicolumn{1}{c|}{$\gamma_1$} & 
\multicolumn{1}{c|}{$\gamma_2$} & 
\multicolumn{1}{c|}{$\gamma_3$} & 
\multicolumn{1}{c|}{$A$} 
\\ \hline 
$u_v$
 &  0.712 $\pm$ 0.081
 &  3.637  $\pm$ 0.138
 &  0.593  $\pm$ 0.774
 &  -3.607  $\pm$ 0.762
 &  3.718  $\pm$ 1.148
  &
\\
$d_v$
 &  0.741  $\pm$ 0.157
 &  5.123  $\pm$ 0.394
 &  1.122  $\pm$ 1.232
 &  -2.984  $\pm$ 1.077
  &
  &
\\
$u_s$
 & -0.363 $\pm$ 0.035
 &  7.861  $\pm$ 0.433
 &  4.339  $\pm$ 1.790
  &
 & 0.0280 $\pm$ 0.0036
 & 0.0808 $\pm$ 0.0122
\\
$\Delta$
 &   0.70  $\pm$ 0.28
 &  11.75  $\pm$ 1.97
 &  -2.57  $\pm$ 3.12
  &
  &
 &  0.316  $\pm$ 0.385
\\
$s$ 
 & -0.240 $\pm$ 0.055
 &  7.98  $\pm$ 0.65
  &
  &
  &
 & 0.085 $\pm$ 0.017
\\
$g$ 
 & -0.170 $\pm$ 0.012
 &  10.71  $\pm$ 1.43
 &  4.00  $\pm$ 4.21
  &
  &
  &
\\
\hline 
\end{tabular} 
\caption{ \small
The parameters of the PDFs in eqs.~(\ref{eq:pdf1})--(\ref{eq:pdf5}) 
and their $1\sigma$ errors obtained in the scheme with $n_f=3$ flavors.}
\label{tab:fitvalues} 
\end{center} 
\end{table} 

As in our previous analysis ABKM09~\cite{Alekhin:2009ni},
the small-$x$ exponent $a_{\Delta}$ for the difference between the up- and the
down-quark sea is fixed to $a_{\Delta}=0.7$ in eq.~(\ref{eq:pdf3}) as an ansatz, 
because of lacking neutron-target data in this region of small values of $x$.
This is in agreement with the values obtained for the small-$x$ exponents 
of the valence quark distributions.
The uncertainty on $a_{\Delta}$ is then determined with help of 
an additional pseudo-measurement of $a_{\Delta}=0.7 \pm 0.3$ added to the data
set (and with the error on $a_{\Delta}$ released) 
in order to quantify the impact on the other parameters of the PDF fit.
This provides us with the result given in Tab.~\ref{tab:fitvalues}.
The value of the charmed-hadron
semi-leptonic branching ratio $B_\mu$ obtained from our NNLO fit is $0.0917\pm 0.0034$.
This is in agreement with the earlier determination~\cite{Alekhin:2008mb}
within the errors.

\begin{figure}[ht!]
\centerline{
  \includegraphics[width=15.5cm]{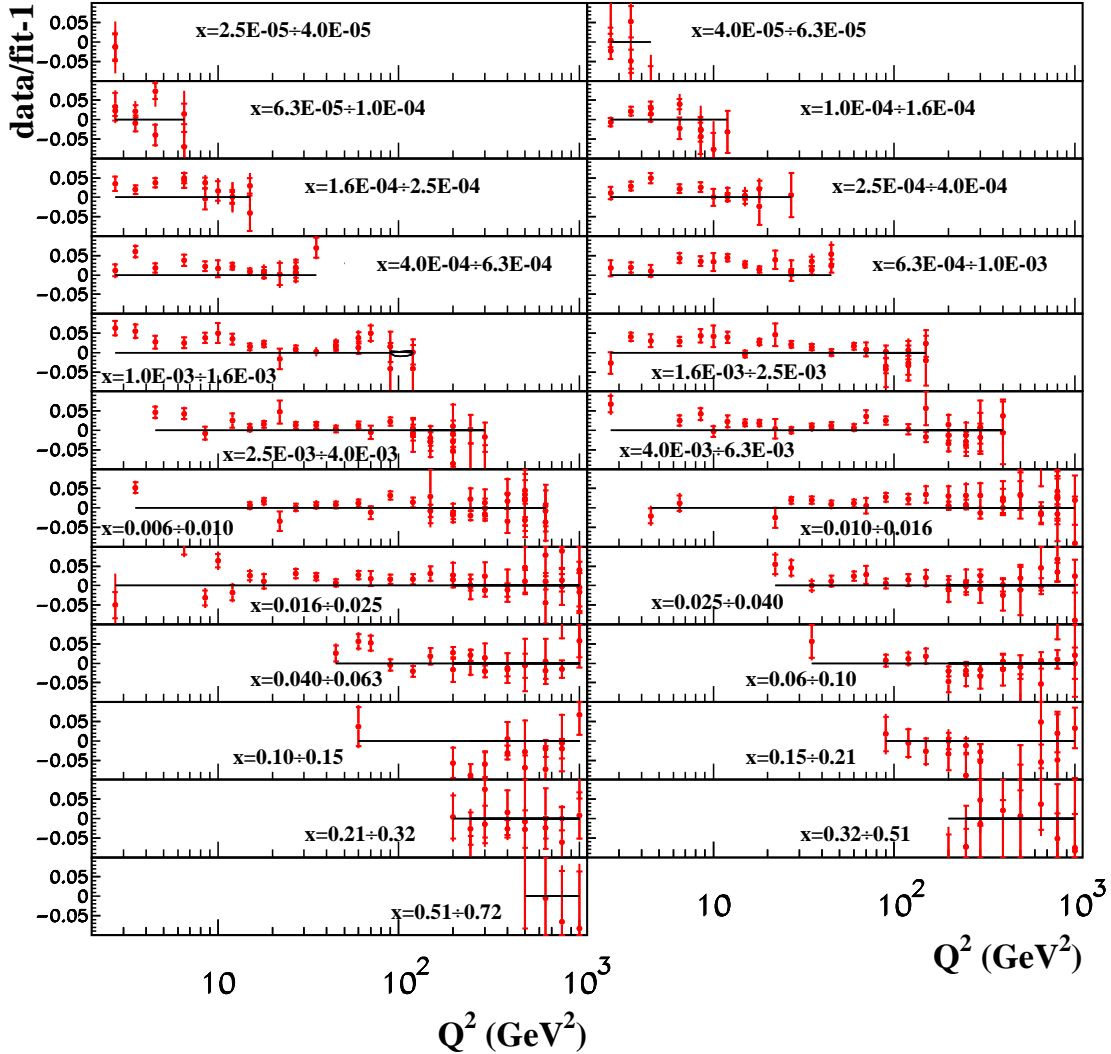}}
  \vspace*{-5mm}
  \caption{\small
    \label{fig:heranc}
      The pulls versus momentum transfer $Q^2$ for 
      the HERA neutral-current inclusive DIS cross section 
      data of~\cite{herapdf:2009wt}
      binned in $x$ with respect to our NNLO fit. 
      The data points with different inelasticity $y$ still may overlap in the plot. 
      The inner bars show statistical errors in data and the outer bars the 
      statistical and systematic errors combined in quadrature. 
}
\end{figure}
\begin{figure}[ht!]
\centerline{
  \includegraphics[width=15.5cm]{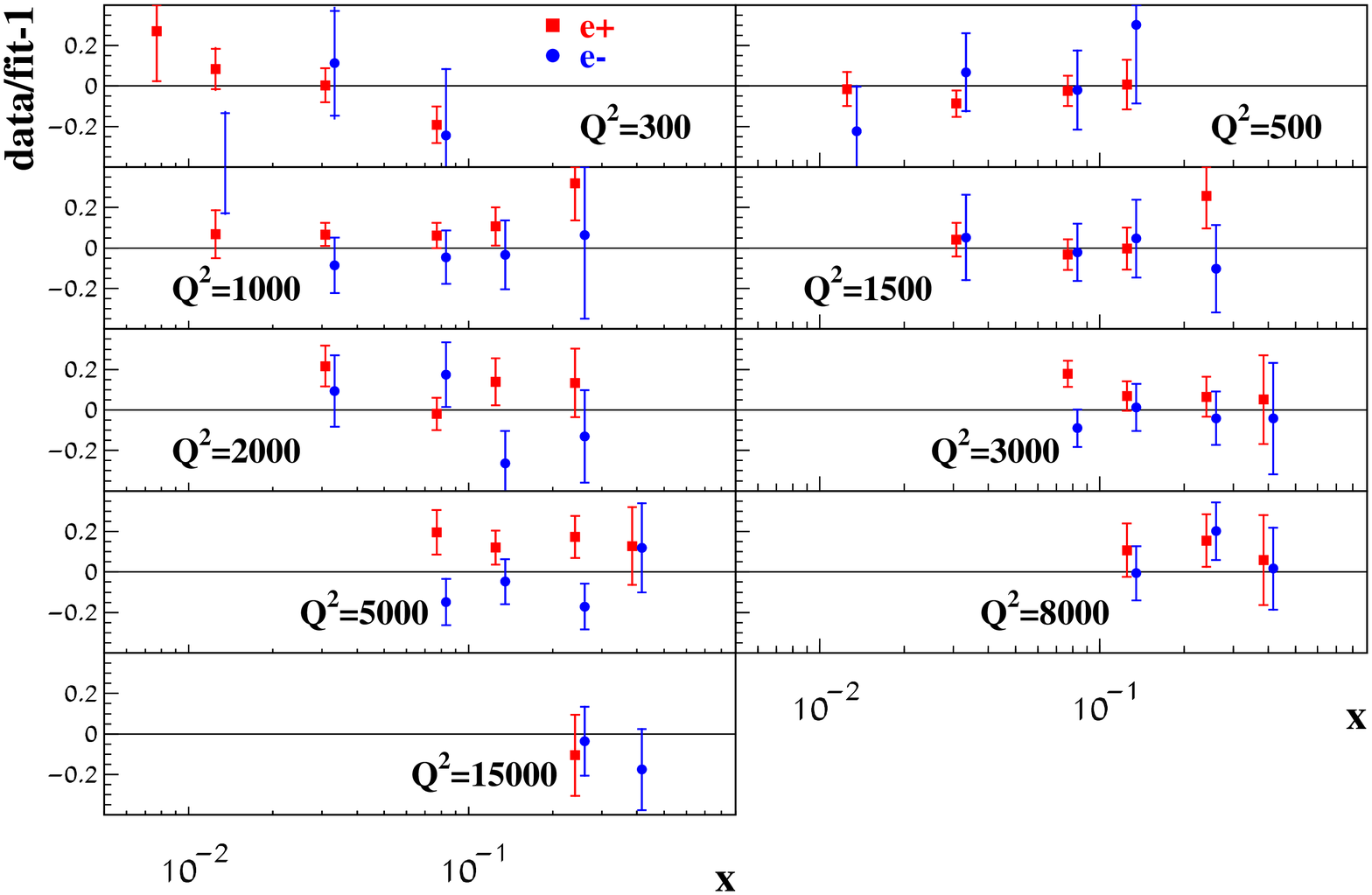}}
  \vspace*{-5mm}
  \caption{\small
    \label{fig:heracc}
      The same as Fig.~\ref{fig:heranc} for the pulls of
      the HERA charged-current inclusive DIS cross section 
      data of~\cite{herapdf:2009wt}
      binned in the momentum transfer $Q^2$ in units of ${\rm GeV}^2$
      versus $x$ (squares: positron beam; circles: electron beam).  
}
%\end{figure}  
%
  \vspace*{5mm}
%
%\begin{figure}[h!]
\centerline{
  \includegraphics[width=15.5cm]{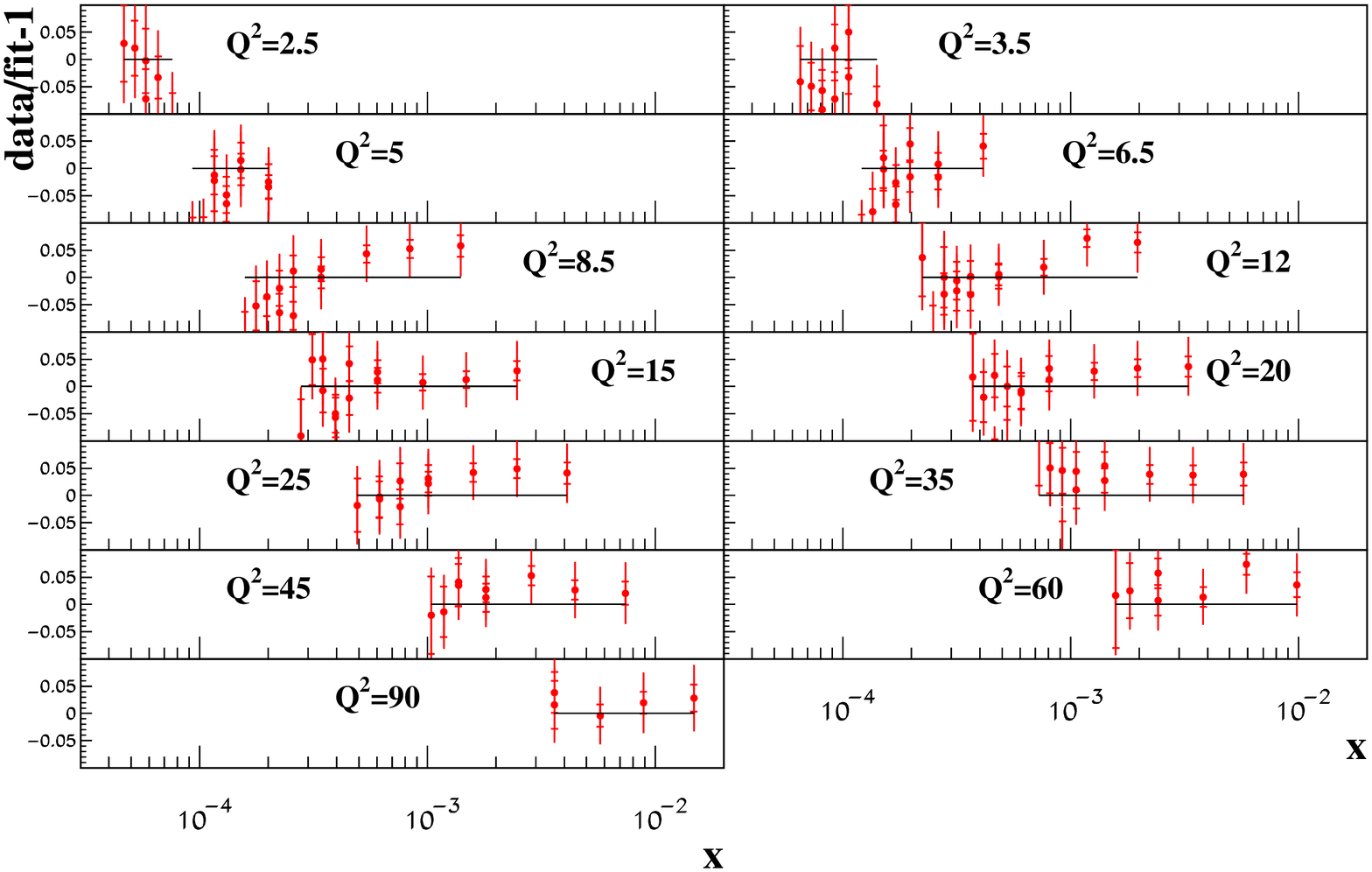}}
  \vspace*{-5mm}
  \caption{\small
    \label{fig:heralow}
      The same as Fig.~\ref{fig:heranc} for the pulls of
      the H1 neutral-current inclusive DIS cross section 
      data of~\cite{Aaron:2010ry}
      binned in the momentum transfer $Q^2$  in units of ${\rm GeV}^2$
      versus $x$.
}
\end{figure}

The three other parameters of our fit to be discussed in detail in Secs.~\ref{sec:alphas} and~\ref{sec:hq-mass},
are the strong coupling constant $\alpha_s$ in the \MSbar-scheme 
and the heavy-quark masses $m_c$ and $m_b$, which we take in the \MSbar-scheme as well.
The latter represents a novel feature of our analysis as all previous PDF determinations 
have always used the pole mass definition for $m_c$ and $m_b$.
As an advantage, we can constrain the central values of both, $m_c(m_c)$ and $m_b(m_b)$ directly to their 
particle data group (PDG) 
\cleardoublepage
\newpage
\noindent
results~\cite{Nakamura:2010pdg} 
without having to rely on a perturbative scheme transformation between a running \MSbar- and a pole mass. 
It is well known that at low scales such as $\mu \simeq m_c$ 
this scheme transformation is poorly convergent in perturbation theory.
Thus, in the present analysis, we add the following pseudo-data as input
\begin{eqnarray}
  \label{eq:mcmbinp}
  m_c(m_c) \,=\, 1.27 \pm 0.08\,\, {\rm GeV}
  \, ,
  \qquad\qquad
  m_b(m_b) \,=\, 4.19 \pm 0.13\,\, {\rm GeV}
  \, ,
\end{eqnarray}
and, subsequently, release the uncertainty of the quark masses to test 
its sensitivity to the other PDF parameters.
The value for $\alpha_s(M_Z)$ on the other hand is determined entirely from
data in the 
fit, cf. Sec.~\ref{sec:alphas}.
The 24 PDF parameters of Tab.~\ref{tab:fitvalues}, $\alpha_s$, $m_c$ and $m_b$
provide us in our analysis in total with 27 correlated parameters.
Their covariance matrix is presented in Tabs.~\ref{tab:pdfco1}--\ref{tab:pdfco3}.

\begin{figure}[h!]
\centerline{
  \includegraphics[width=15.5cm]{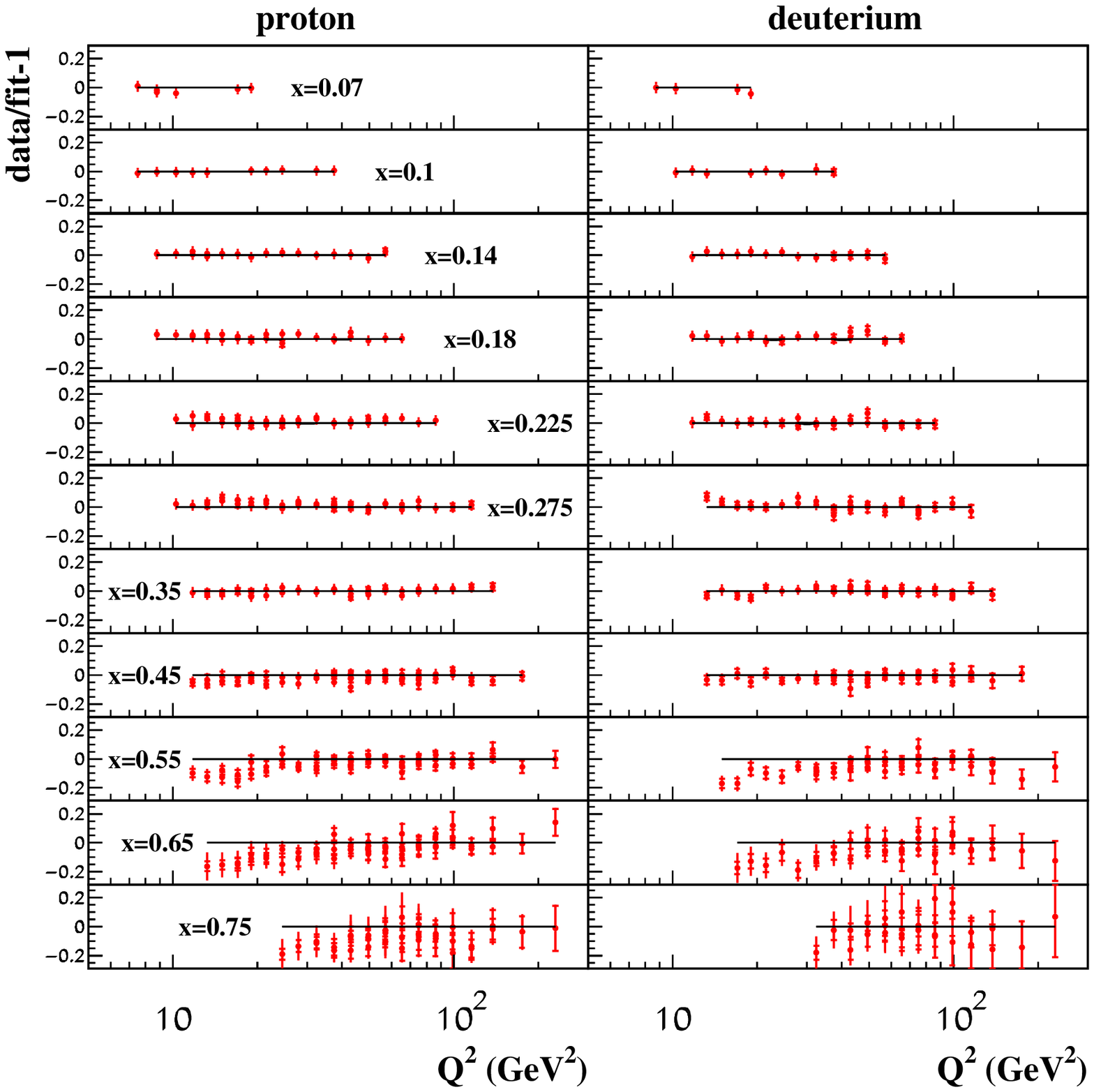}}
  \vspace*{-5mm}
  \caption{\small
    \label{fig:bcdms}
      The same as Fig.~\ref{fig:heranc} for the pulls of
      the BCDMS inclusive DIS cross section 
      data of~\cite{Benvenuti:1989rh,Benvenuti:1989fm}
      for the proton target (left) and for the deuterium target (right).
}
\end{figure}
\begin{figure}[h!]
\centerline{
  \includegraphics[width=15.5cm]{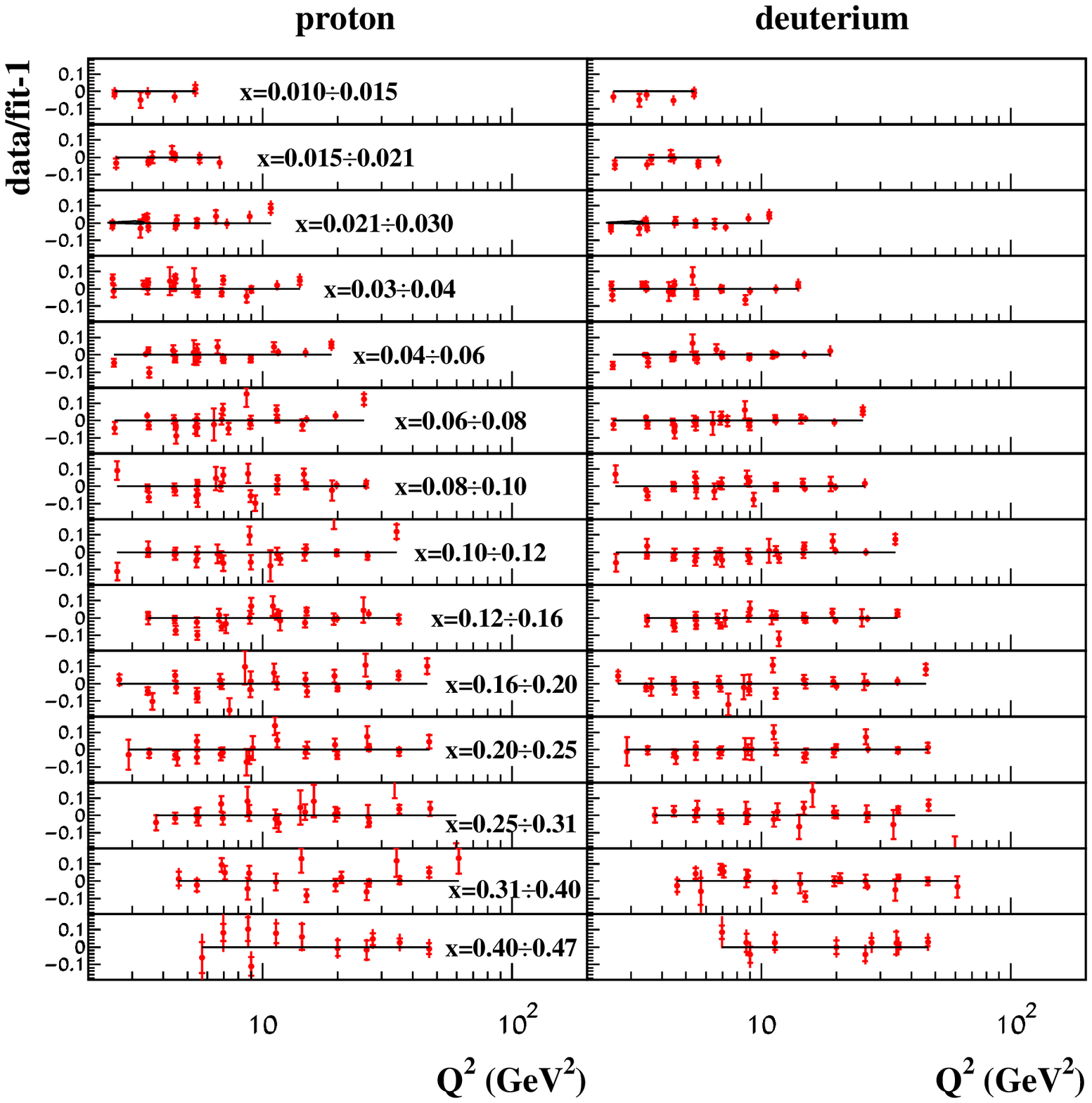}}
  \vspace*{-5mm}
  \caption{\small
    \label{fig:nmc}
      The same as Fig.~\ref{fig:heranc} for the pulls of
      the NMC inclusive DIS cross section 
      data of~\cite{Arneodo:1996qe}
      for the proton target (left) and for the deuterium target (right).
}
\end{figure}
\begin{figure}[h!]
\centerline{
  \includegraphics[width=15.5cm]{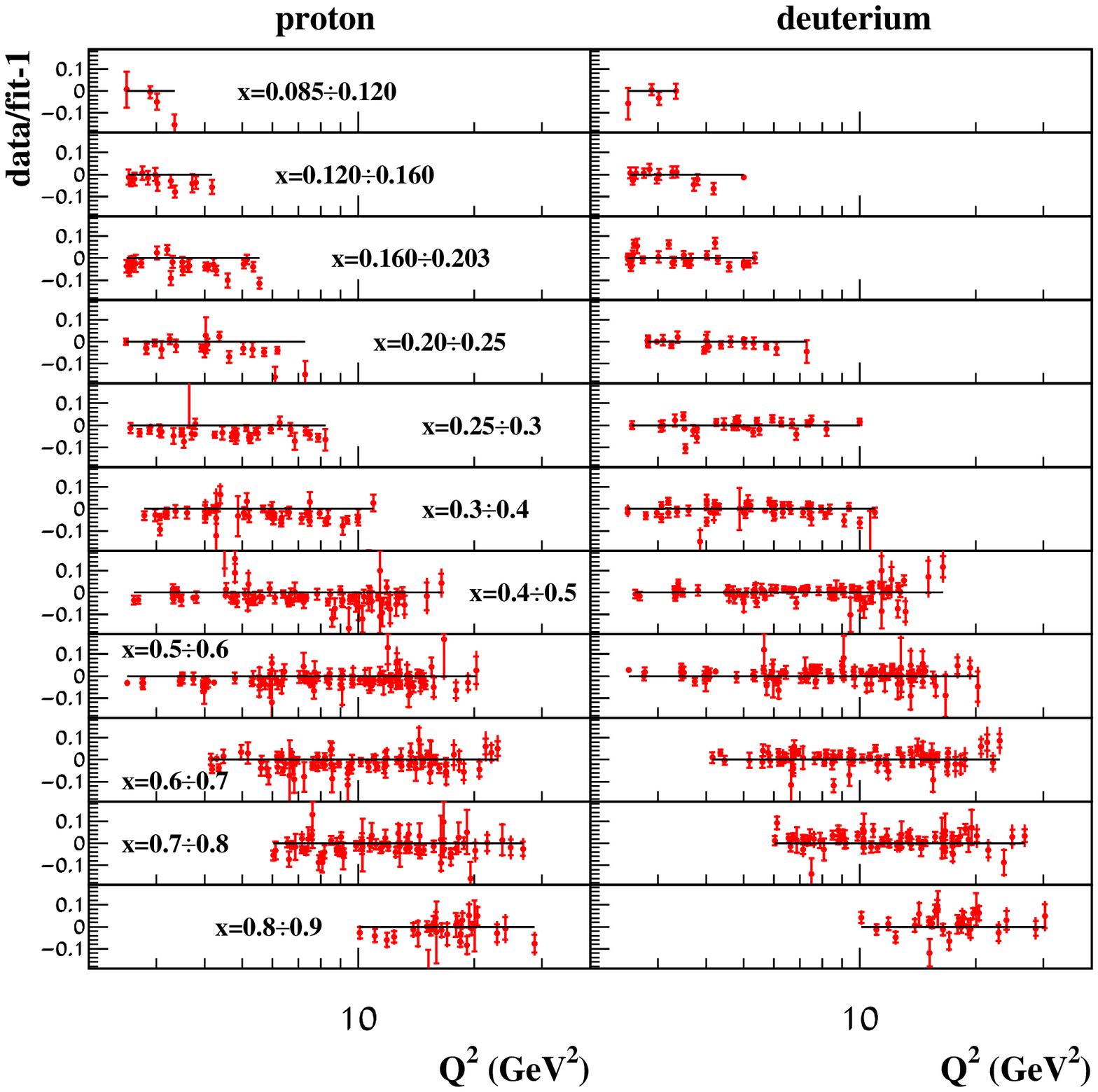}}
  \vspace*{-5mm}
  \caption{\small
    \label{fig:slac}
      The same as Fig.~\ref{fig:heranc} for the pulls of
      the SLAC inclusive DIS cross section 
      data of~\cite{Atwood:1976ys,Bodek:1979rx,Dasu:1993vk,Gomez:1993ri,Mestayer:1982ba}
      for the proton target (left) and for the deuterium target (right).
}
\end{figure}
\begin{figure}[h!]
\centerline{
  \includegraphics[width=15.5cm]{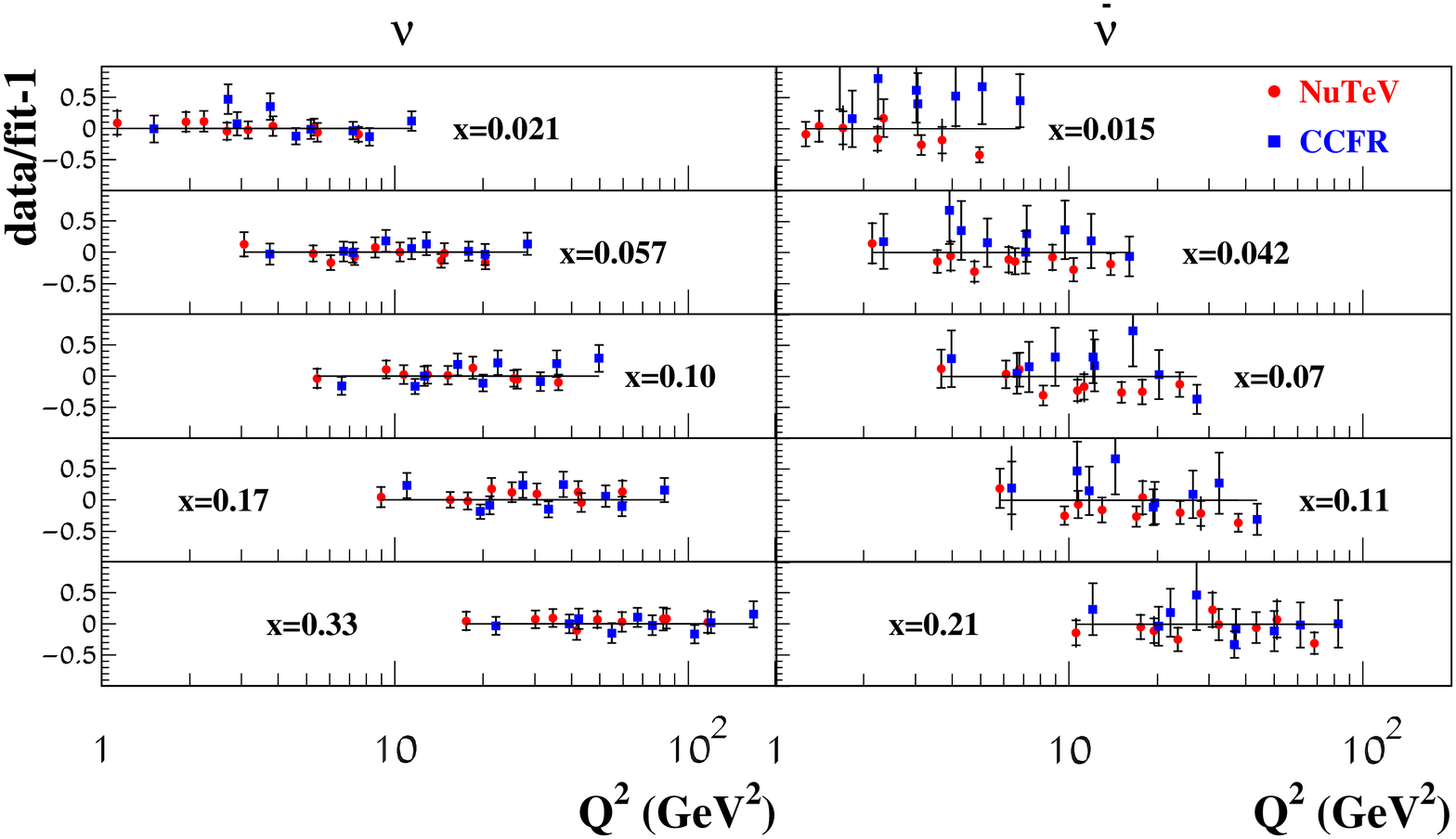}}
  \vspace*{-5mm}
  \caption{\small
    \label{fig:dimuon}
      The same as Fig.~\ref{fig:heranc} for the pulls of
      neutrino (left) and anti-neutrino (right) induced di-muon 
      production cross section 
      data of~\cite{Goncharov:2001qe} (circles: NuTeV experiment, 
      squares: CCFR experiment).
}
%\end{figure}
%
  \vspace*{5mm}
%
%\begin{figure}[ht!]
\centerline{
  \includegraphics[width=8.25cm]{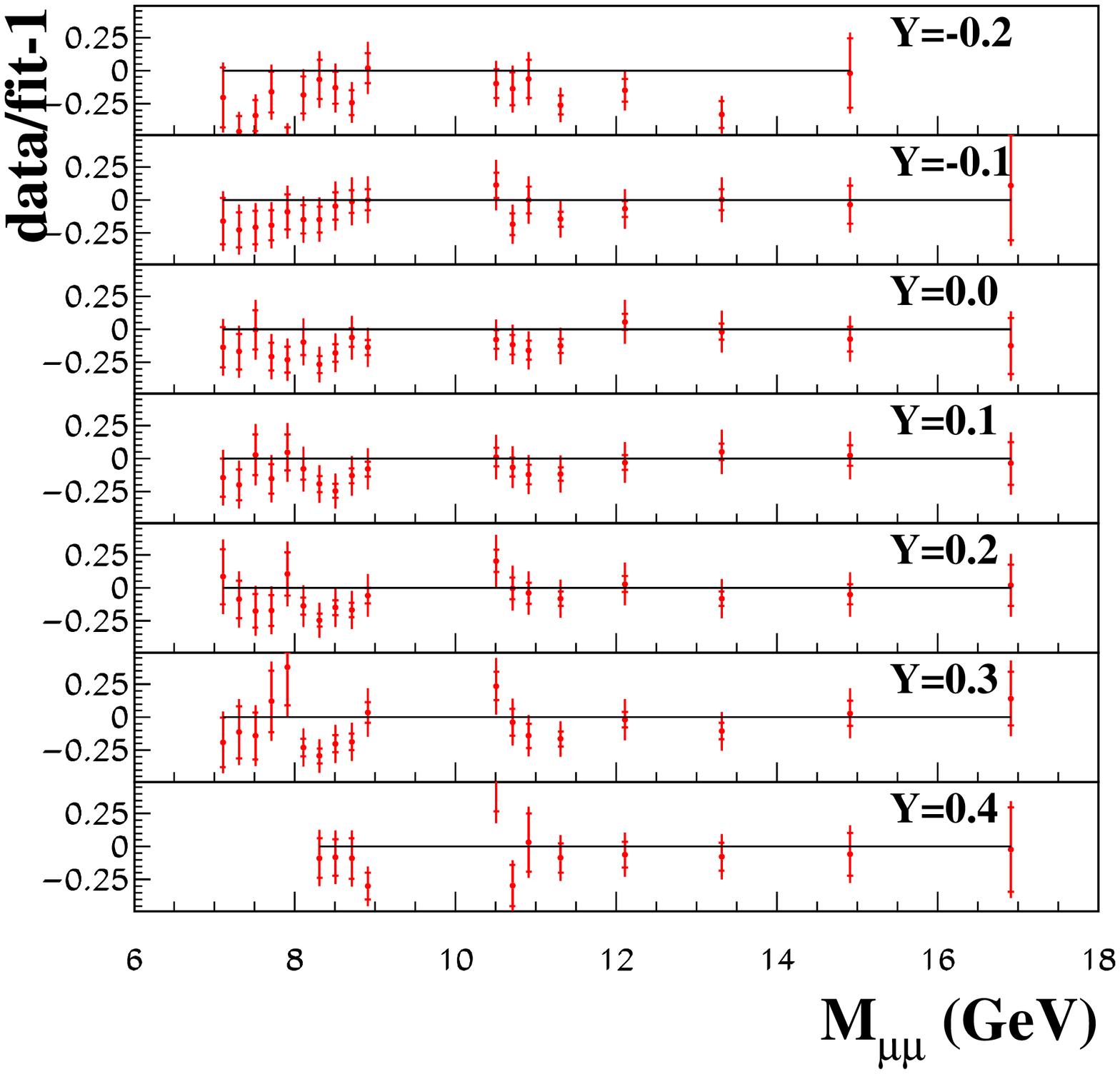}
  \includegraphics[width=8.25cm]{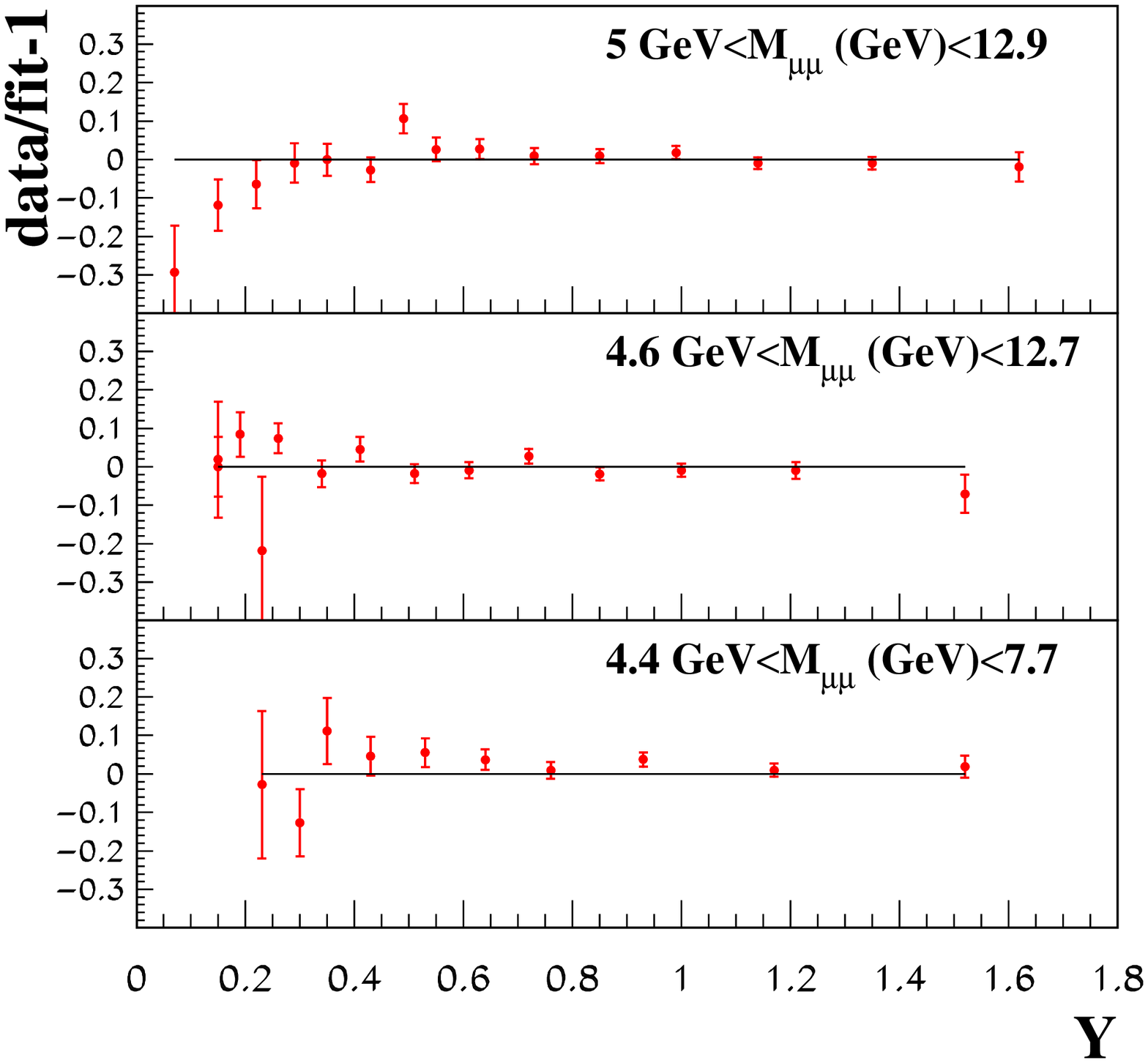}}
  \vspace*{-5mm}
  \caption{\small
    \label{fig:dy}
      The same as Fig.~\ref{fig:heranc} for the pulls of
      the DY process cross section 
      data of~\cite{Moreno:1990sf}
      binned in the muon pair rapidity $Y$ 
      versus the invariant mass $M_{\mu \mu}$ of the muon pair (left)
      and the ones of~\cite{Towell:2001nh} binned in 
      $M_{\mu \mu}$ versus $Y$ (right).
}
\end{figure}

It is instructive to study the pulls of the individual data sets included in the fit.
This provides a mean of assessing the quality of the fit in detail 
and allows for an investigation of specific kinematical regions.
In Figs.~\ref{fig:heranc}--\ref{fig:heralow} we display 
the detailed dependence of the pulls on the momentum transfer $Q^2$ and $x$ 
for the HERA NC and CC inclusive DIS cross section data of~\cite{herapdf:2009wt} 
as well as the low $Q^2$ data of~\cite{Aaron:2010ry} with respect to our NNLO fit.
We find overall a very good description of the data,
even at the edges of the kinematical region of HERA, i.e., at smallest values
of $x$ and largest values of $Q^2$.
The respective $\chi^2$ values for the fit at NLO and NNLO are given in Tab.~\ref{tab:chi2tab}.

\begin{table}[h]
\begin{center}
\begin{tabular}{|l|l|c|c|c|}
\hline
& Experiment & NDP & $\chi^2({\rm NNLO})$ & $\chi^2({\rm NLO})$ \\
\hline
DIS inclusive &H1\&ZEUS~\cite{herapdf:2009wt} &486 & 537 & 531  \\ 
&H1~\cite{Aaron:2010ry} &130 & 137 & 132  \\ 
&BCDMS~\cite{Benvenuti:1989rh,Benvenuti:1989fm} &605 & 705 & 695 \\ 
&NMC~\cite{Arneodo:1996qe} &490 & 665 & 661 \\ 
&SLAC-E-49a~\cite{Bodek:1979rx} &118 & 63 & 63  \\ 
&SLAC-E-49b~\cite{Bodek:1979rx} &299 & 357 & 357 \\ 
&SLAC-E-87~\cite{Bodek:1979rx} &218 & 210 & 219  \\ 
&SLAC-E-89a~\cite{Atwood:1976ys} &148 & 219 & 215 \\ 
&SLAC-E-89b~\cite{Mestayer:1982ba} &162 & 133 & 132  \\ 
&SLAC-E-139~\cite{Gomez:1993ri} &17 & 11 & 11 \\ 
&SLAC-E-140~\cite{Dasu:1993vk} &26 & 28 & 29  \\ 
\hline
Drell-Yan  
&FNAL-E-605~\cite{Moreno:1990sf} &119 & 167 & 167 \\ 
&FNAL-E-866~\cite{Towell:2001nh} &39 & 52 & 55  \\ 
\hline
DIS di-muon  
&NuTeV~\cite{Goncharov:2001qe} &89 & 46 & 49 \\ 
&CCFR~\cite{Goncharov:2001qe} &89 & 61 & 62 \\ 
\hline
Total  &    & 3036 & 3391 & 3378 \\ 
\hline 
\end{tabular}
\caption{\small 
  The value of $\chi^2$ obtained in the NNLO and NLO fits for different data sets. 
}
\label{tab:chi2tab}
\end{center}
\end{table}

Next, in Figs.~\ref{fig:bcdms}--\ref{fig:slac} we show the respective pulls of the 
BCDMS~\cite{Benvenuti:1989rh,Benvenuti:1989fm},
NMC~\cite{Arneodo:1996qe} and 
SLAC~\cite{Atwood:1976ys,Bodek:1979rx,Dasu:1993vk,Gomez:1993ri,Mestayer:1982ba}
inclusive DIS cross section data 
as a function of $x$ and binned in the momentum transfer $Q^2$.
Again, our fit provides a very good description (see Tab.~\ref{tab:chi2tab} for the $\chi^2$ values),
especially at low $W^2$ thanks the phenomenological ansatz for the structure
functions with the higher twist terms of Tab.~\ref{tab:htsvalues}.

In Fig.~\ref{fig:dimuon} we plot the data for the 
(anti-)neutrino induced di-muon production cross section of ~\cite{Goncharov:2001qe}
which constrains the strange PDF. We give both, the pulls for the NuTeV and for
the CCFR experiment. 
Finally, in Fig.~\ref{fig:dy} we display the DY cross section 
data of~\cite{Moreno:1990sf,Towell:2001nh} which 
depends on the muon pair rapidity $Y$ and the invariant mass $M_{\mu \mu}$ of
the muon pair and which assists in the flavor separation of the PDF fit.
It is obvious from Figs.~\ref{fig:dimuon}, \ref{fig:dy} and Tab.~\ref{tab:chi2tab} 
that we achieve again a very good description in all cases.

\begin{figure}[th!]
  \centerline{
    \includegraphics[width=15.5cm]{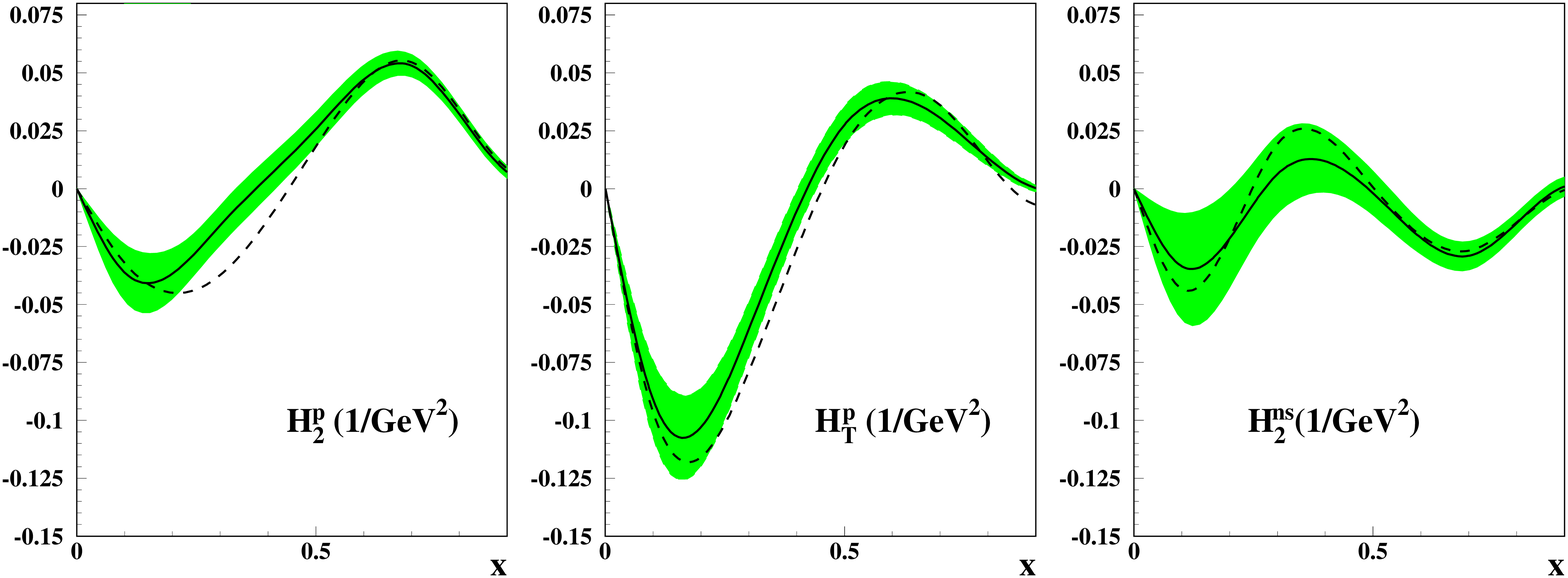}}
  \caption{\small
    \label{fig:hts}
    The central values (solid line) and the 1$\sigma$ bands (shaded area) 
    for the coefficients of the twist-4 terms of eq.~(\ref{eq:htwist})
    in the inclusive DIS structure functions obtained from our NNLO fit
    (left panel: $F_2$ of the proton, central panel: $F_T$ of the proton,   
    right panel: non-singlet $F_2$). The central values of the 
    twist-4 coefficients obtained from our NLO fit are 
    shown for comparison (dashes). 
  }
\end{figure}

\begin{table}[h!]
\renewcommand{\arraystretch}{1.5}
\begin{center} 
%\small
\begin{tabular}{|c|c|c|c|} 
\hline 
\multicolumn{1}{|c|}{ } & 
\multicolumn{1}{c|}{$H_2^{\rm p}(x)/{\rm GeV}^2$} & 
\multicolumn{1}{c|}{$H_2^{\rm ns}(x)/{\rm GeV}^2$} & 
\multicolumn{1}{c|}{$H_T^{\rm p}(x)/{\rm GeV}^2$} 
\\ \hline 
$x=0.1$
 &  -0.036  $\pm$ 0.012
 &  -0.034  $\pm$ 0.023
 &  -0.091  $\pm$ 0.017
\\ 
$x=0.3$
 &  -0.016  $\pm$ 0.008
 &  0.006  $\pm$ 0.017
 &  -0.061  $\pm$ 0.012
\\
$x=0.5$
 &  0.026  $\pm$ 0.007
 & -0.0020  $\pm$ 0.0094
 &  0.0276 $\pm$ 0.0081
\\
$x=0.7$
 &  0.053  $\pm$ 0.005
 &  -0.029  $\pm$ 0.006
 &  0.031  $\pm$ 0.006
\\
$x=0.9$
 &  0.0071  $\pm$ 0.0026
 & 0.0009  $\pm$ 0.0041
 & 0.0002  $\pm$ 0.0015
\\
\hline 
\end{tabular} 
\caption{ \small
The parameters of the twist-4 contribution to the DIS structure functions 
in eq.~(\ref{eq:htwist}) for the fit to NNLO accuracy in QCD.}
\label{tab:htsvalues} 
\end{center} 
\end{table}

\begin{figure}[th!]
  \centerline{
    \includegraphics[width=15.5cm]{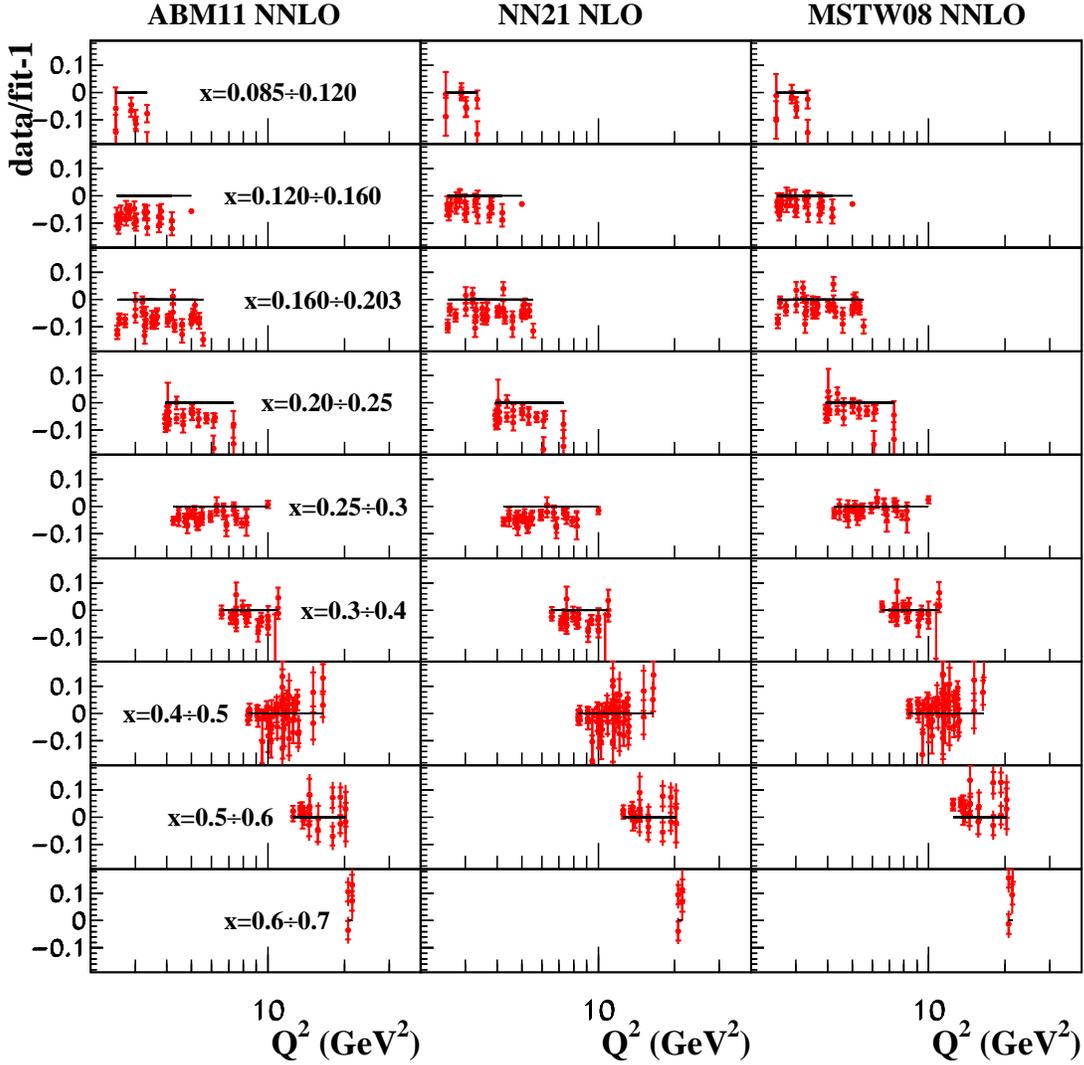}}
  \caption{\small
    \label{fig:slacw}
    The same as Fig.~\ref{fig:slac} 
    without the HT terms taken into account and for various 
    3-flavor PDFs (left panel: present analysis, 
    right panels: MSTW~\cite{Martin:2009iq}).
    The NLO calculations based on the 3-flavor NLO NN21 PDFs~\cite{Ball:2011mu} 
    are given for comparison (central panels). 
    Only the  data surviving after the 
    cut of eq.~(\ref{eq:softcut}) are shown;  
    the proton and deuterium data points are superimposed.
  }
\end{figure}

\begin{figure}[th!]
\centerline{
  \includegraphics[width=15.5cm]{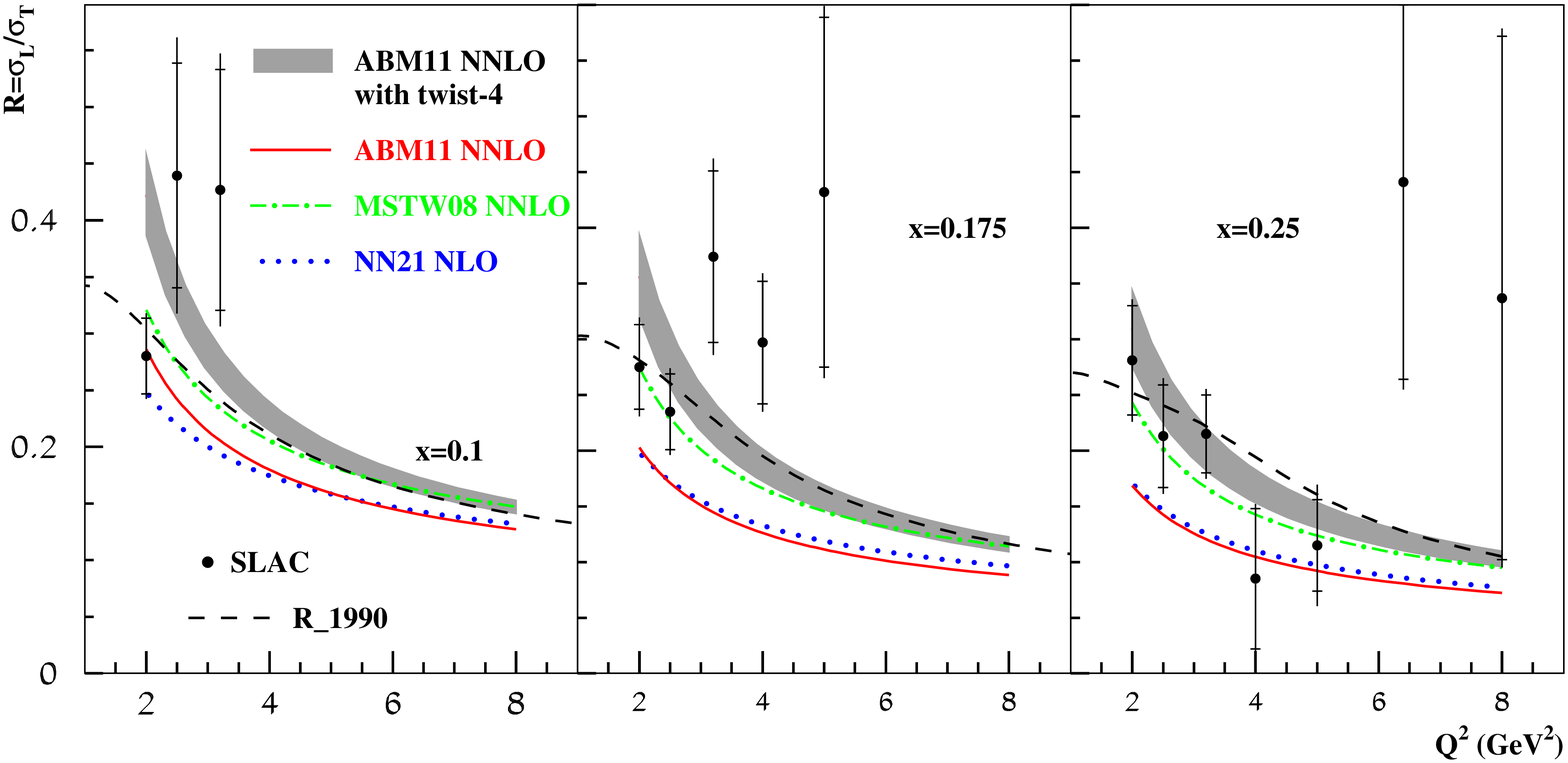}}
  \caption{\small
    \label{fig:rslac}
     The shaded area gives 1$\sigma$ band of the NNLO predictions for 
     the ratio $R=\sigma_L/\sigma_T$, cf. eq.~(\ref{eq:Rratio}), 
     based on the ABM11 PDFs and 
     the twist-4 terms obtained from our fit  
     at different values of $x$ versus the momentum transfer $Q^2$. 
     The central values of the 
     NNLO predictions for $R$ based on the MSTW 
     PDFs~\cite{Martin:2009iq} (dashed dots), the NNLO predictions 
     based on the ABM11 PDFs (solid line),
     and the NLO predictions based on the NN21 PDFs~\cite{Ball:2011mu}
     (dots), all taken in the 3-flavor scheme and
     without twist-4 terms, are given for comparison.  
     The data points show values of $R$ extracted from the 
     SLAC proton and deuterium data~\cite{Whitlow:1990gk} with 
     the empirical parameterization of those data $R_{1990}$ obtained 
     in~\cite{Whitlow:1990gk} superimposed (dashes).
}
\end{figure}

The last missing piece of information on the PDF fit concerns the shape of the higher twist terms 
for the inclusive DIS structure functions introduced in eq.~(\ref{eq:htwist}).
As outlined in Sec.~\ref{sec:power-cor} we fit three twist-4 coefficients for
a complete description of both, proton and nucleon targets.
In detail, the proton $H_2^{\rm p}$, 
the non-singlet $H_2^{\rm ns}=H_2^{\rm p}-H_2^{\rm n}$ and the proton $H_T^{\rm p}$  
contribute 15 parameters in total and we assume $H_T^{\rm ns}=0$.
The respective coefficients are listed in Tab.~\ref{tab:htsvalues}
and shown in Fig.~\ref{fig:hts}, all in units of ${\rm GeV}^{-2}$,
They are in agreement with the earlier results 
of~\cite{Virchaux:1991jc,Blumlein:2006be} up to the parameterization 
of the HT contribution (compare eq.~(\ref{eq:htwist}) with eq.~(35) of~\cite{Blumlein:2006be}). 
The magnitude of the HT terms reduces from the NLO to the NNLO case, 
see Fig.~\ref{fig:hts}, however the change is comparable with the coefficient uncertainties and 
the NNLO twist-4 coefficients do not still vanish, 
in line with the results of~\cite{Blumlein:2006be}. 
The non-singlet twist-4 term in $F_T$ is comparable to zero within uncertainties 
therefore it was fixed at zero in our analysis as discussed in Sec.~\ref{sec:power-cor}.
The non-singlet twist-4 term in $F_2$ is negative at $x \lesssim 0.5$.  
This is also in line with earlier results~\cite{Virchaux:1991jc}, 
again taking into account the difference in the HT parameterizations. 
The HT terms are mostly important at small hadronic invariant mass $W$. 
This was confirmed in a comparison of the low-$W$ JLAB data~\cite{Malace:2009kw}
with predictions based on the ABKM09 PDFs with account of the twist-4 terms, 
which were extracted in the analysis of~\cite{Alekhin:2009ni} similarly to the present one. 
However even with the cut of $W^2>12.5~{\rm GeV}^2$ as commonly imposed in global PDF fits 
the HT terms are numerically important for the region of $x\lesssim 0.3$, 
which is not affected by this cut. 

From Fig.~\ref{fig:slacw} it is evident, that calculations, 
which are based on our NNLO PDFs, but do not include the HT terms, 
systematically overshoot the SLAC data at $x\lesssim 0.3$ 
due to the HT terms being negative in this region.
The value of $\chi^2/{\rm NDP}$ for the SLAC data at 
\begin{equation}
\label{eq:softcut}
W^2 > 12.5~{\rm GeV}^2\, ,\qquad\qquad 
Q^2 > 2.5~{\rm GeV}^2\, ,
\end{equation}
is 699/246 in this case. This is much worse than the value of $\chi^2/{\rm NDP}=292/246$
obtained in our fit for the same subset of data.
We have also performed similar comparisons 
taking the published 3-flavor NNLO MSTW~\cite{Martin:2009iq} 
and NLO NN21~\cite{Ball:2011mu} PDFs as an input of our fitting code. 
The NNLO MSTW and NLO NN21 predictions obtained in this way 
without accounting for the HT terms and with the cut of eq.~(\ref{eq:softcut}) imposed 
lead to poor a description of the SLAC data, cf. Fig.~\ref{fig:slacw}. 
For the case of NN21 PDFs the agreement with the data is particularly bad, 
with an off-set reaching up to $\sim 10\%$ at $x\sim 0.15$ and 
a value of $\chi^2/{\rm NDP}=518/246$. 
The MSTW value of $\chi^2/{\rm NDP}=514/246$ is also far from ideal in this case, 
obviously due to the missing HT terms.
Also, MSTW does not take into account the target mass corrections.
Note, that for the comparison performed without the twist-4 terms the 
ABM11 value of $\chi^2$ is worse than ones of MSTW and NN21 since 
those PDFs are obtained disregarding the HT terms. 
This also shows that parts of the twist-4 terms obtained in our fit
are effectively absorbed into the MSTW and NN21 PDFs. 

The relative contribution of the higher twist terms to the ratio $R=\sigma_L/\sigma_T$ 
in eq.~(\ref{eq:Rratio}) is particularly important reaching up to one half at moderate 
$x$~\cite{SanchezGuillen:1990iq}. 
The value of $R$ calculated including the NNLO QCD corrections and 
the twist-4 terms of Tab.~\ref{tab:htsvalues} 
is in reasonable agreement with the SLAC data 
on $R$~\cite{Whitlow:1990gk} and the parameterization of those data 
$R_{1990}$, see cf. Fig.~\ref{fig:rslac}. 
The latter is based on the empirical combination of the QCD-like terms with the twist-4 and twist-6 
terms, pretending to describe the data down to scales $Q^2\sim 1~{\rm GeV}^2$. 
Due to the twist-6 term, which provides saturation of $R_{1990}$ at small $Q$,
the shape of $R_{1990}$ is somewhat different from our calculation, while 
both agree with the data at $Q^2>2~{\rm GeV}^2$ within the errors. 
The leading twist NNLO contribution to $R$ undershoots the full calculation
by a factor of $1.5-2$, depending on $x$, cf. Fig.~\ref{fig:rslac}.
The leading-twist NLO calculations based on the 3-flavor NN21 PDFs are in a good 
agreement with our NNLO leading-twist term and go by factor of $1.5-2$ 
lower than the data as well. 
The leading-twist NNLO calculations for the 3-flavor MSTW PDFs 
are larger than the NLO NN21 ones and are in better agreement with the data. 
Note, that this is related to the fact that the data on $R$ of~\cite{Whitlow:1990gk} are included into the MSTW fit 
allowing for a better description of the SLAC cross section data 
as compared to the NN21 case, see Fig.~\ref{fig:slacw} and the related discussion above.
At the same time this leads to an effective absorption of the twist-4 terms into the fitted PDFs.  

\begin{figure}[th!]
\centerline{
  \includegraphics[width=15.5cm]{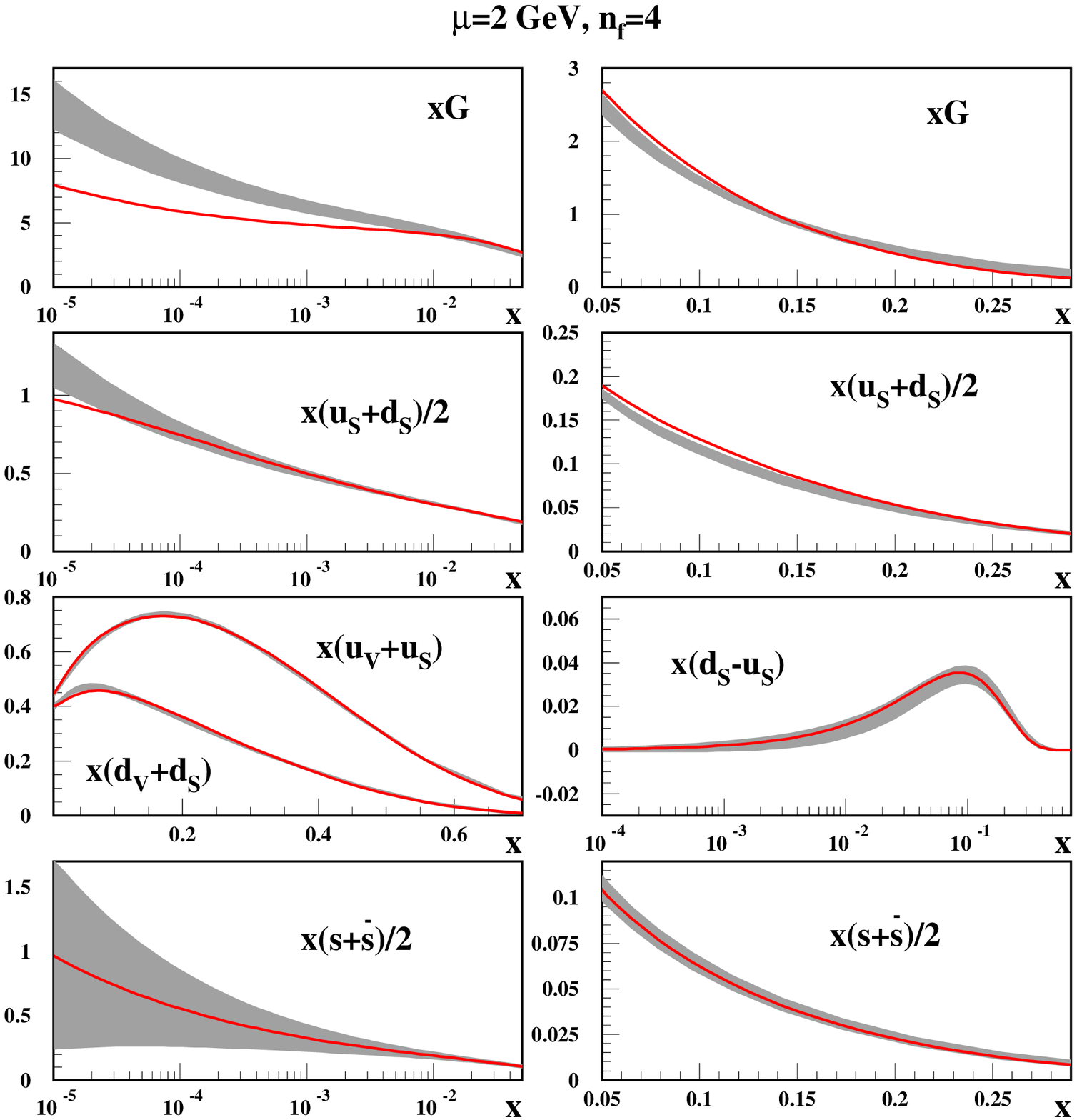}}
  \caption{\small
    \label{fig:pdf1}
     The 1$\sigma$ band for the 4-flavor NNLO 
     ABKM09 PDFs~\cite{Alekhin:2009ni} at the scale of
     $\mu=2~{\rm GeV}$ versus $x$ (shaded area) compared with
     the central values for ones of this analysis (solid lines).   
}
\end{figure}

\begin{figure}[th!]
\centerline{
  \includegraphics[width=15.5cm]{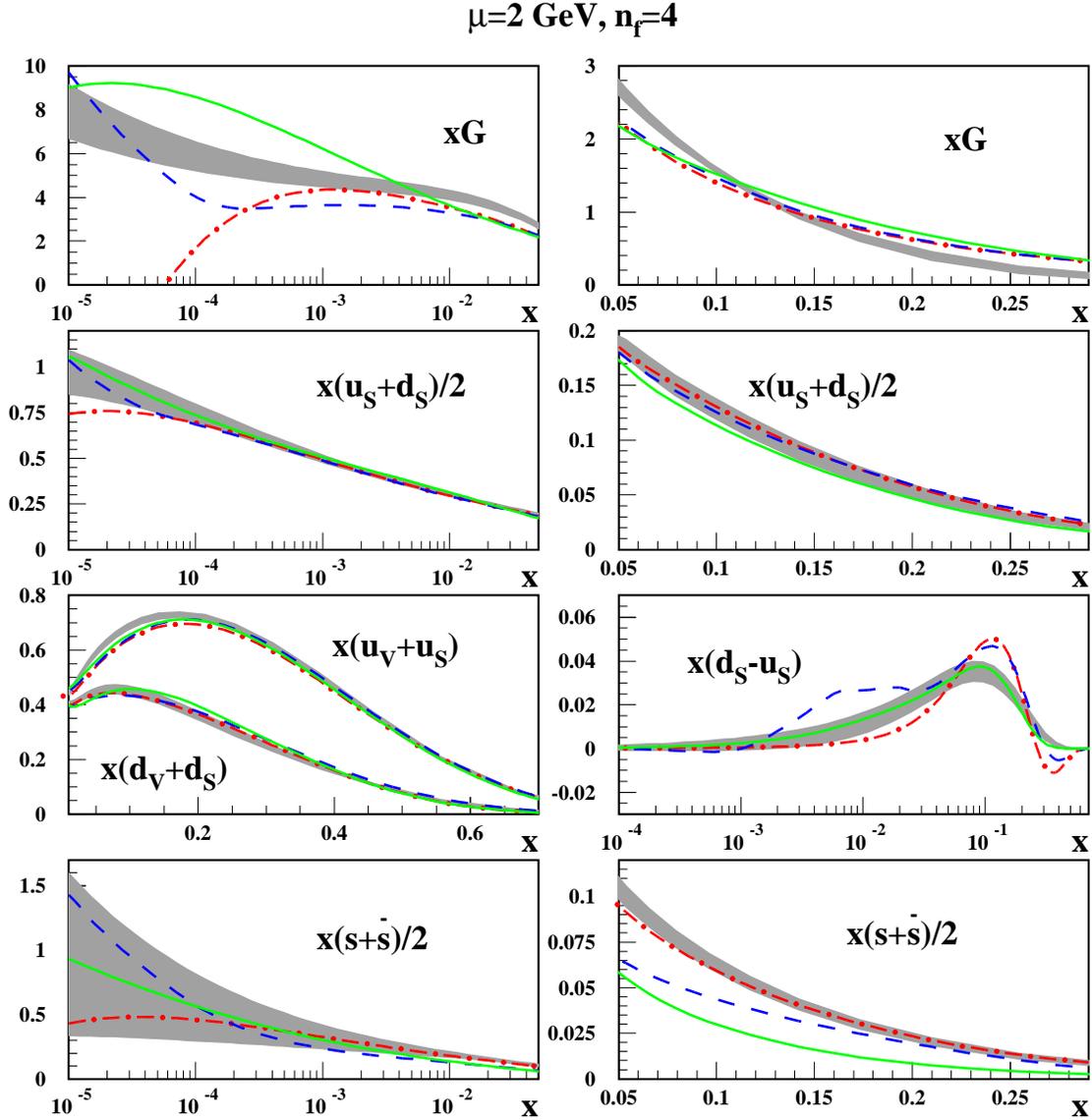}}
  \caption{\small
    \label{fig:pdf3}
     The 1$\sigma$ band for the 4-flavor NNLO ABM11 PDFs at the scale 
     of $\mu=2~{\rm GeV}$ versus $x$ obtained  in this analysis (shaded area) 
     compared with the ones obtained by other groups 
     (solid lines: JR09~\cite{JimenezDelgado:2008hf},
     dashed dots: MSTW~\cite{Martin:2009iq}, 
     dashes: NN21~\cite{Ball:2011uy}).
}
\end{figure}

\begin{figure}[th!]
\centerline{
  \includegraphics[width=9.0cm]{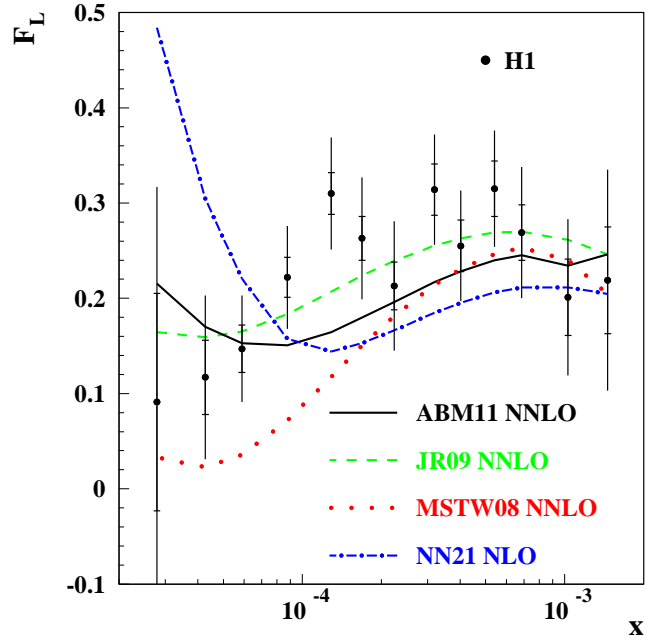}}
  \caption{\small
    \label{fig:flh1}
     The data on $F_L$ versus $x$ obtained by the H1
     collaboration~\cite{Aaron:2010ry} 
     confronted with the 3-flavor scheme NNLO predictions based on the 
     different PDFs (solid line: this analysis, dashes: 
     JR09~\cite{JimenezDelgado:2008hf}, dots: MSTW~\cite{Martin:2009iq}). 
     The NLO predictions based on the 3-flavor NN21 
     PDFs~\cite{Ball:2011mu} are given for comparison (dashed dots). 
     The value of $Q^2$ for the data points and the curves in the plot 
     rises with $x$ in the range of $1.5 \div 45~{\rm GeV}^2$.
  }
\end{figure}

\begin{figure}[th!]
\centerline{
  \includegraphics[width=15.5cm]{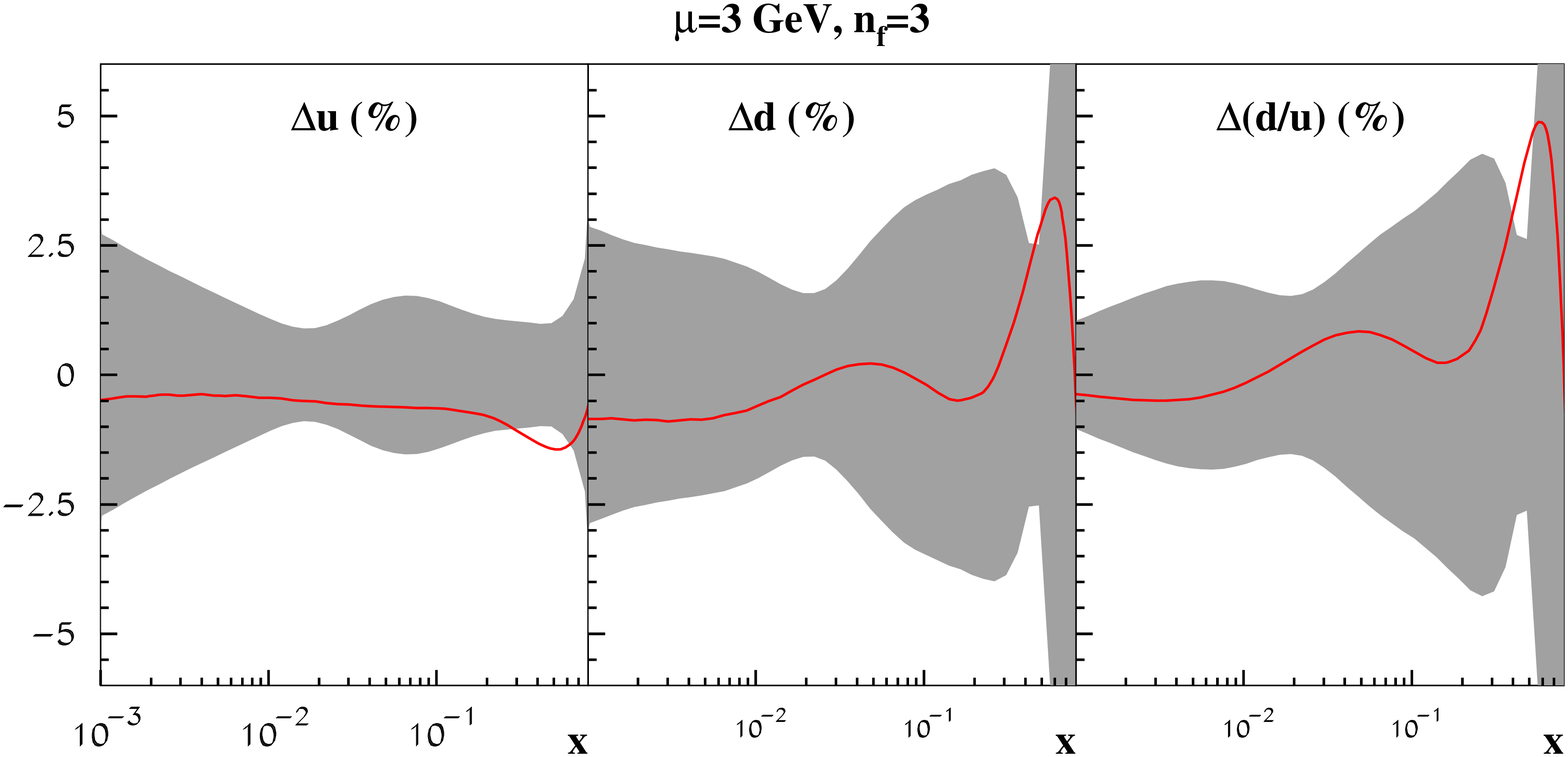}}
  \caption{\small
    \label{fig:norm}
     The 1$\sigma$ band for the 3-flavor NNLO 
     $u$-quark distribution (left panel), $d$-quark distribution 
     (central panel), and the $d/u$ ratio 
     (right panel) at the scale of $\mu=2~{\rm GeV}$ 
     versus $x$ in comparison to 
     the central value of the fit variant
     with the data normalization changed from the settings of 
     Sec.~\ref{sec:data} to the ones of~\cite{Martin:2009iq}
     (solid curves).
  }
\end{figure}

The NNLO ABM11 PDFs obtained in the present analysis 
are compared in Fig.~\ref{fig:pdf1} with our earlier ABKM09 PDFs. 
The biggest change between these two sets is observed for the small-$x$
gluon and sea distributions. 
Firstly, this change happens since the HERA NC inclusive data~\cite{herapdf:2009wt} 
used in the present analysis lie by several percent higher than 
the HERA data of~\cite{Chekanov:2001qu,Adloff:2000qk} used in the 
ABKM09 analysis, due to improvements in the monitor calibration.  
Secondly, the small-$x$ PDFs are particularly sensitive to the 
treatment of the heavy-quark electro-production and, therefore, they 
change due to the  NNLO corrections and the running-mass scheme implemented in the present 
PDF fit. Other ABM11 PDFs are in agreement with the ABKM09 ones within the 
uncertainties.   

The NNLO PDFs obtained by other groups are compared with the NNLO ABM11 PDFs in Fig.~\ref{fig:pdf3}. 
The agreement between the various PDFs is not ideal, 
a fact that may be explained by the differences in the data sets used 
to constrain the PDFs, by the factorization scheme employed, 
by the treatment of the data error correlation and so on. 
A detailed clarification of these issues is beyond the scope of the present paper. 
Therefore we discuss only the most significant differences, e.g., 
the gluon distributions at small $x\lesssim0.001$, 
which are quite different for all PDF sets considered in Fig.~\ref{fig:pdf3}.

To that end, we compare in Fig.~\ref{fig:flh1} the small-$x$ data on $F_L$ 
obtained by the H1 collaboration~\cite{Aaron:2010ry} with the predictions based on these PDFs.
The $F_L$ data are quite sensitive to the small-$x$ gluon PDFs.
Moreover, in order to provide a consistent comparison all predictions are 
taken in the running-mass 3-flavor scheme with the heavy-quark masses 
of eq.~(\ref{eq:mcmbinp}) and with the 3-flavor PDFs.
The H1 data on $F_L$ are in a good agreement with the NNLO ABM11 predictions. 
Although these data were not included into the fit
of~\cite{JimenezDelgado:2008hf} they are also in a good agreement with the NNLO JR09 predictions. 
The NNLO MSTW and NLO NN21 predictions on the other hand miss the H1 data. 
Thus, the latter can be used to consolidate the small-$x$ behavior of the
gluon PDFs provided by different groups. 
Likewise, the SLAC DIS cross section data of~\cite{Atwood:1976ys,Bodek:1979rx,Dasu:1993vk,Gomez:1993ri,Mestayer:1982ba} 
can also be of help in consolidating the results of the different PDF fits. 
As one can see in Fig.~\ref{fig:slacw} the MSTW and NN21  
predictions systematically overshoot the SLAC data at $x\sim0.2$. 
As we discussed above, this happens due to the omission of the higher twist terms. 
Once the latter are neglected in the fit, the power corrections are partially absorbed in the leading twist PDFs. 
Therefore, this discrepancy is evidently also related to the difference of those PDFs with 
the ABM11 ones at moderate $x$, cf. Fig.~\ref{fig:pdf3}. 
On the other hand, the ABM11 large-$x$ gluon distribution goes lower 
than the NN21 and MSTW ones, because we do not include the Tevatron inclusive jet data 
into the fit, cf.~\cite{Alekhin:2011cf} and Sec.~\ref{sec:jets}.

Another striking difference in Fig.~\ref{fig:pdf3} is related to the 
strange sea distribution, which is commonly constrained by the data on the di-muon 
production in the $\nu N$ DIS in all PDF fits considered. 
Nonetheless, for the NN21 and JR09 sets it goes significantly lower at
$x\gtrsim 0.02$ than for the MSTW and ABM11 ones. 
The difference between the NN21 and ABM11 strange sea distributions 
appears to be due to eq.~(34) of~\cite{Ball:2011mu} for the di-muon production cross section, 
which contains an additional factor of $(1+m_c^2/Q^2)$ as compared, e.g., 
to eq.~(3) of~\cite{Gluck:1996ve} employed in our analysis, cf. also~\cite{Blumlein:2011zu}. 
At small $Q^2$ this factor reaches a numerical value of 2 and 
the strange sea is suppressed correspondingly in the fit to the data. 
We have convinced ourselves that with this factor taken into account the NN21 PDFs deliver 
a satisfactory description of the CCFR and NuTeV di-muon data.
On the other hand, the discrepancy between the results of JR09 and ABM11 in Fig.~\ref{fig:pdf3} can 
directly be traced back to the ansatz $s(x,Q_0^2) = \bar{s}(x,Q_0^2) = 0$ of JR09,
i.e., the assumption of vanishing strangeness at the low starting scale of 
$Q_0^2 <1~{\rm GeV}^2$ in the dynamical valence-like PDF model of JR09.
This is different from ours, cf. eq.~(\ref{eq:pdf4}). 
Moreover, JR09 has not used the data on di-muon production in neutrino-nucleon
collisions in their fit, see Sec.~\ref{sec:di-muon}.

Finally, the difference in the large-$x$ non-strange quark distributions 
appears partly due to the general normalization of the data, which is often
a matter of choice in the PDF fits.
In order to quantify impact of the choice of the data normalization on the PDFs (cf. Sec.~\ref{sec:data}),
we have performed a variant of our NNLO fit with the same normalization factor settings 
as employed in the MSTW fit~\cite{Martin:2009iq}. 
The relative difference between the $u$- and $d$-quark distributions 
obtained in this variant of the fit and our nominal one is displayed in Fig.~\ref{fig:norm}. 
Clearly, the impact of the data normalization choice is most pronounced at large $x$, 
where the trend is different for the cases of $d$- and $u$-quarks. 
Therefore the effect is amplified in the ratio $d/u$ 
which is important for the interpretation of the charged-lepton asymmetry data 
from hadron colliders, cf. Sec.~\ref{sec:wzprod}.
As shown in Fig.~\ref{fig:norm}, the relative difference in the ratio $d/u$ 
reaches up to 5\% at $x\sim 0.5$.

%%
%% ---------------------------------------------------------------------
%%
\subsection{Strong coupling constant}
\label{sec:alphas}
\begin{figure}[t!]
\centerline{
  \includegraphics[width=8.5cm]{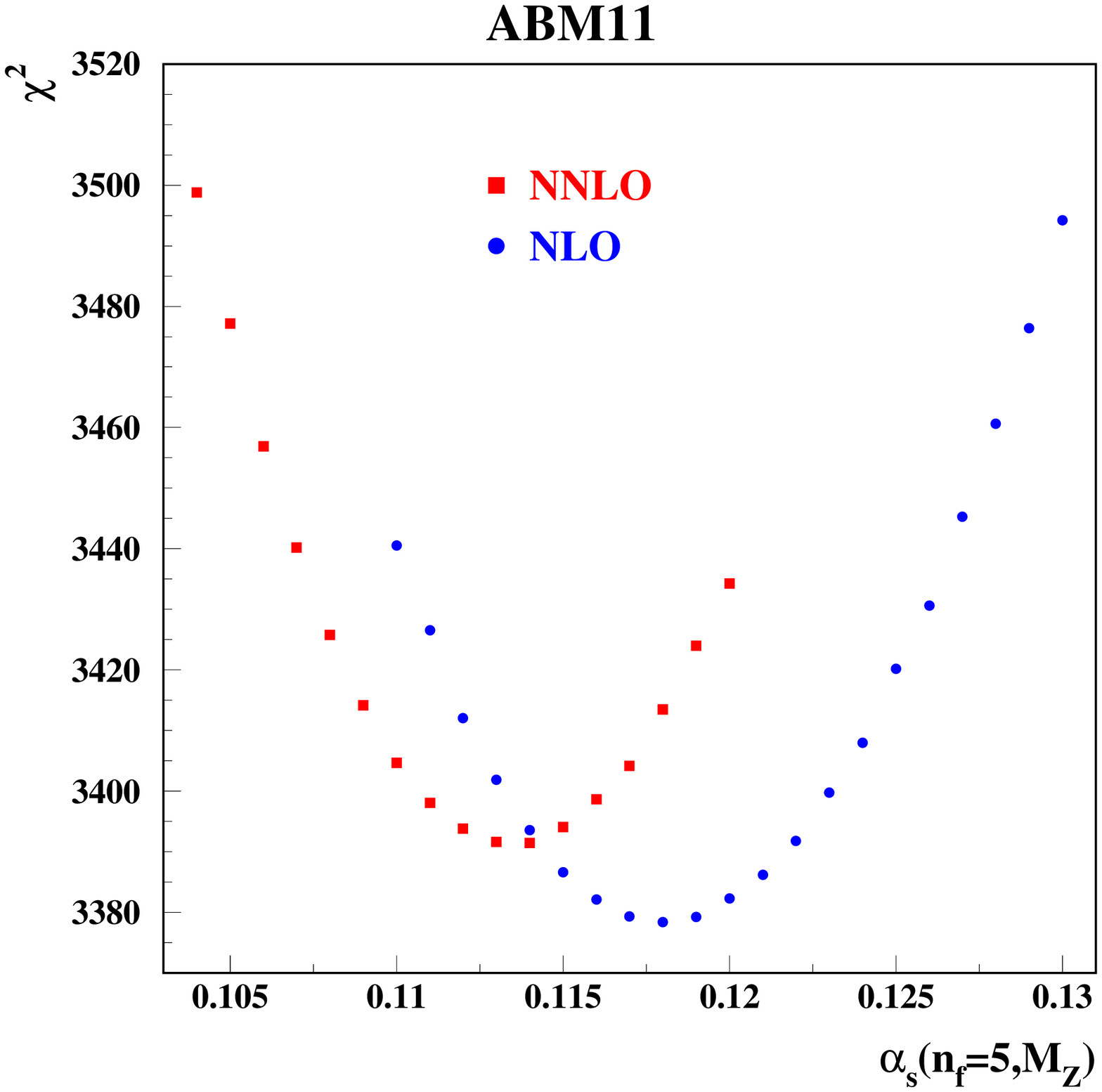}}
  \vspace*{-5mm}
  \caption{\small
    \label{fig:aschi2}
      The $\chi^2$-profile eq.~(\ref{eq:EQCHI}) 
      as a function of $\alpha_s(M_Z)$ in the present analysis 
      at NLO (circles) and NNLO (squares).
  }
\end{figure}

\begin{figure}[th!]
\centerline{
  \includegraphics[width=15.5cm]{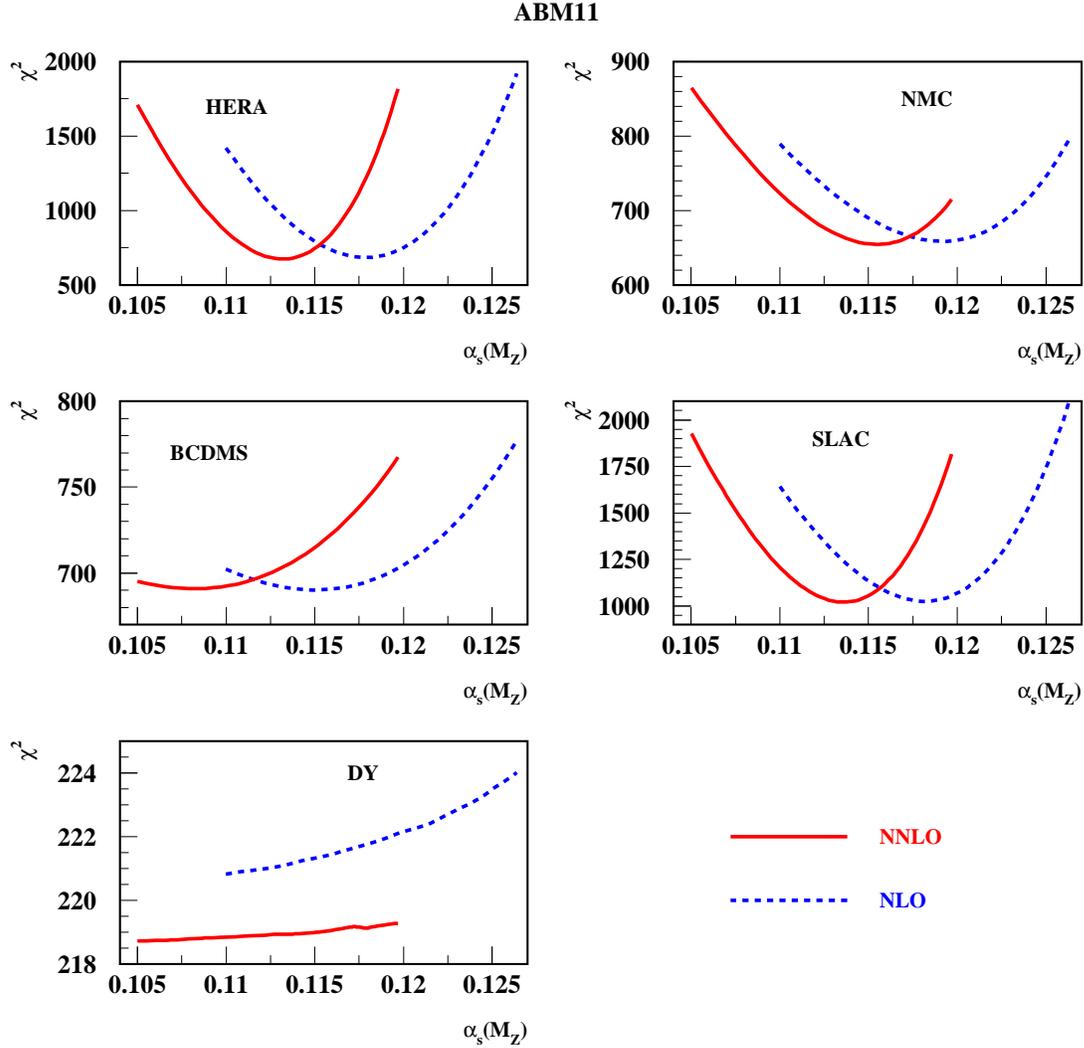}}
  \caption{\small
    \label{fig:scanfixed}
      The $\chi^2$-profile versus the value of $\alpha_s(M_Z)$, eq.~(\ref{eq:EQCHI}),
      for the data sets of Tab.~\ref{tab:alphas1}, all calculated with the 
      PDF and HT parameters fixed at the values obtained from the fits with 
      $\alpha_s(M_Z)$ released (solid lines: NNLO fit, dashes: NLO one).
  }
\end{figure}

\begin{table}[h!]\centering
\begin{tabular}{|l|c|c|c|}
\hline
\multicolumn{1}{|c|}{Experiment} &
\multicolumn{3}{c|}{$\alpha_s({M_Z})$} \\
\cline{2-4}
\multicolumn{1}{|c|}{ } &
\multicolumn{1}{c|}{NLO$_{exp}$} &
\multicolumn{1}{c|}{NLO} &
\multicolumn{1}{c|}{NNLO} 
\\ 
\hline 
BCDMS      &$0.1111 \pm 0.0018$ &$0.1150  \pm 0.0012  $ &$0.1084 \pm 0.0013$ \\
NMC        &$0.117\phantom{0}~^{+~0.011}_{-~0.016}\phantom{0}\phantom{0}$ 
                                &$0.1182 \pm 0.0007$&$0.1152 \pm 0.0007$\\
SLAC       &                    &$0.1173 \pm 0.0003$&$0.1128 \pm 0.0003$\\
HERA comb. &                    &$0.1174 \pm 0.0003$&$0.1126 \pm 0.0002$\\            
DY         &                    &$0.108\phantom{0} \pm 0.010\phantom{0}$&$0.101\phantom{0} \pm 0.025\phantom{0}$\\
\hline 
ABM11      &                    &$0.1180 \pm 0.0012$ & $0.1134 \pm 0.0011$ \\
\hline
\end{tabular} 
\caption{\small
  \label{tab:alphas1} 
  Comparison of the values of $\alpha_s({M_Z})$ obtained 
  by BCDMS~\cite{Benvenuti:1989fm} and NMC~\cite{Arneodo:1993kz} at NLO 
  with the individual results of the fit in the present analysis 
  at NLO and NNLO 
  for the HERA data~\cite{herapdf:2009wt,Aaron:2010ry},
  the NMC data~\cite{Arneodo:1996qe}, 
  the BCDMS data~\cite{Benvenuti:1989rh,Benvenuti:1989fm}, 
  the SLAC data~\cite{Whitlow:1990gk,Bodek:1979rx,Atwood:1976ys,Mestayer:1982ba,Gomez:1993ri,Dasu:1993vk}, 
  and the DY data~\cite{Moreno:1990sf,Towell:2001nh}.}
\end{table}
\begin{table}[h!]\centering
\begin{tabular}{|l|c|c|c|}
\hline
\multicolumn{1}{|c|}{Experiment} &
\multicolumn{3}{c|}{$\alpha_s(M_Z)$} \\
\cline{2-4}
\multicolumn{1}{|c|}{ } &
\multicolumn{1}{c|}{NLO$_{exp}$} &
\multicolumn{1}{c|}{NLO} &
\multicolumn{1}{c|}{NNLO$^*$} \\
\hline 
D0  1 jet             & $0.1161~^{+~0.0041}_{-~0.0048}$      
                      & $0.1190 \pm 0.0011$ & $0.1149 \pm 0.0012$ \\
D0  2 jet             &      & $0.1174 \pm 0.0009$ & $0.1145 \pm 0.0009$ \\   
CDF 1 jet (cone)      &      & $0.1181 \pm 0.0009$ & $0.1134 \pm 0.0009$ \\   
CDF 1 jet ($k_\perp$)  &      & $0.1181 \pm 0.0010$ & $0.1143 \pm 0.0009$ \\   
\hline
ABM11                 &      & $0.1180 \pm 0.0012$ & $0.1134 \pm 0.0011$ \\   
\hline 
\end{tabular} 
\renewcommand{\arraystretch}{1}  
\caption{ \small
\label{tab:alphas2}
Comparison of the values of $\alpha_s({M_Z})$ obtained by D0 in~\cite{Abazov:2009nc} 
with the ones based on including individual 
data sets of Tevatron jet data~\cite{Abulencia:2007ez,Aaltonen:2008eq,Abazov:2008hua,Abazov:2010fr} 
into the analysis at NLO. 
The NNLO$^*$ fit refers to the NNLO analysis of the DIS and DY data together with 
the NLO and soft gluon resummation corrections (next-to-leading logarithmic accuracy) 
for the 1 jet inclusive data, cf. \cite{Kidonakis:2000gi,Alekhin:2011cf}.}
\end{table}
\begin{figure}[th!]
\centerline{
  \includegraphics[width=15.5cm]{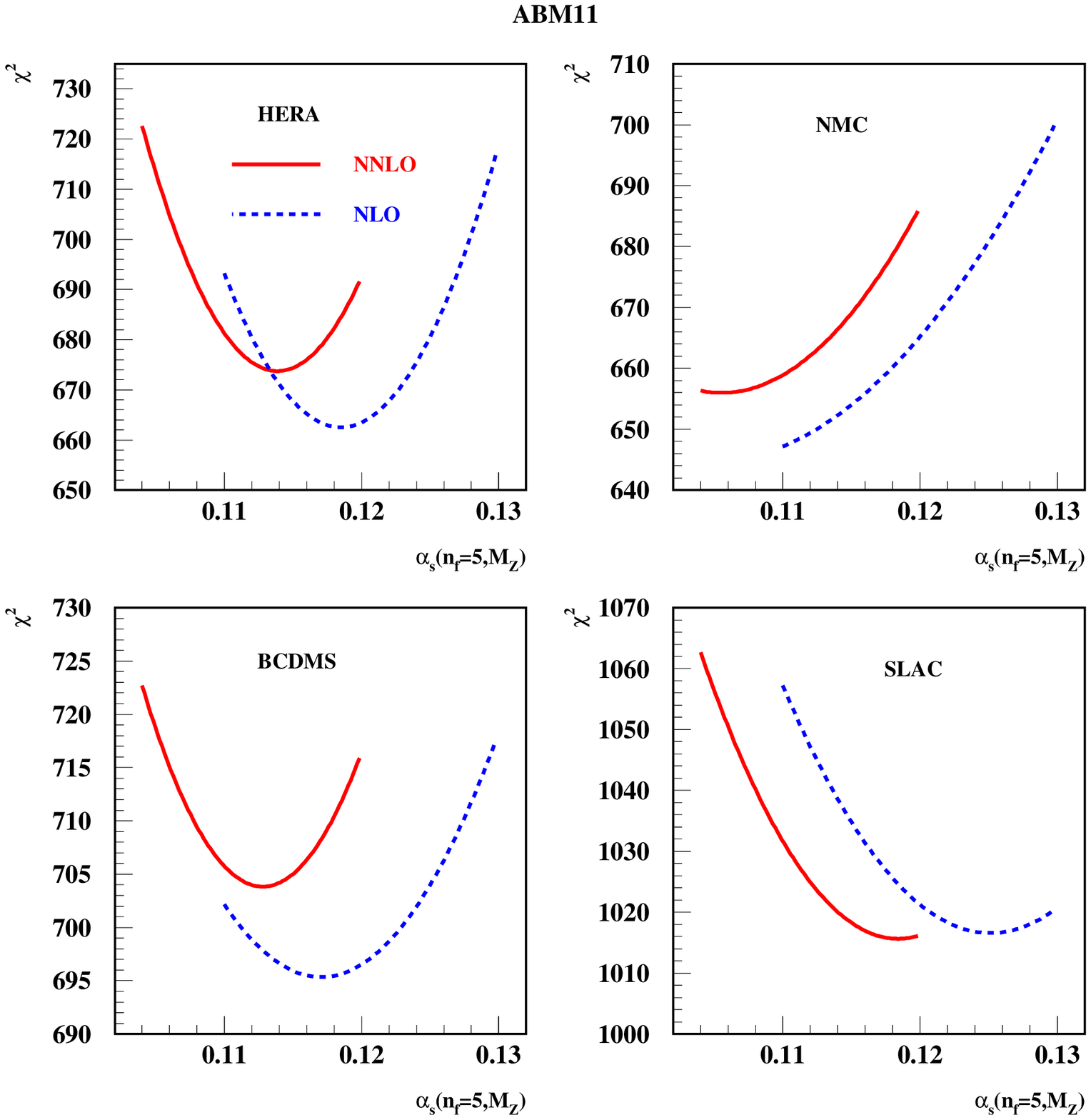}}
  \caption{\small
    \label{fig:scanas}
    The $\chi^2$-profile versus the value of $\alpha_s(M_Z)$, eq.~(\ref{eq:EQCHI}),
    for the data sets of Tab.~\ref{tab:alphas1} 
    all obtained in variants of the present analysis with the value of $\alpha_s$ fixed 
    and all other parameters fitted
    (solid lines: NNLO fit, dashes: NLO fit).
}
\end{figure}

For a precision determination of $\alpha_s(M_Z)$ the fit of systematically
compatible data sets is a necessary prerequisite.
Here the individual data sets determine the average in such a way
that their individual effect is closely compatible with the central value within the errors.
Enhancing the precision from NLO to NNLO, and in some cases to even higher orders,
the central values and the values obtained for the individual data sets both stabilize. 
Moreover, the values of $\alpha_s(M_Z)$ obtained by individual experiments, capable to measure 
$\alpha_s$ from their data alone, have to be consistently reproduced. 
One observes a decreasing sequence of differences $|\Delta \alpha_s(M_Z)|$ between the sequential orders, 
cf. e.g.,~\cite{Blumlein:2006be,BBprep}.

In the present analysis based on the measured scattering cross sections and with account of higher
twist contributions, c.f. also \cite{Alekhin:2011ey} we obtain 
from the data sets described in Sec.~\ref{sec:data},
\begin{eqnarray}
\label{eq:abm11as-nlo}
\alpha_s(M_Z)  &=& 0.1180 \pm 0.0012\, \qquad\qquad \rm{at~NLO}\, ,
\\
\label{eq:abm11as-nnlo}
\alpha_s(M_Z) &=& 0.1134 \pm 0.0011\, \qquad\qquad \rm{at~NNLO}\, .
\end{eqnarray}
The value at NNLO is shifted by $\Delta \alpha_s(M_Z) = 0.0046$ downward if compared to the NLO value. 
This range of uncertainty is well compatible with the scale uncertainty
observed in a variation of the factorization and renormalization scales at NLO of $O(0.0050)$, 
cf.~\cite{Blumlein:1996gv}. 
The present data allow a measurement of $\alpha_s(M_Z)$ with an accuracy of $O(1\%)$. 
Therefore NNLO analyses are mandatory, since the NLO results exhibit much too large theory errors.
The response to the fitted $\alpha_s$-dependence is measured
using the $\chi^2$ functional of eq.~(\ref{eq:EQCHI}).
In Fig.~\ref{fig:aschi2} the dependence of $\chi^2$ of $\alpha_s(M_Z)$ is illustrated at NLO and NNLO.

In Tab.~\ref{tab:alphas1} we compare the values for $\alpha_s(M_Z)$ obtained for the individual data sets
at NLO and NNLO in the present fit and with results obtained by some of the experiments.  
Here the $\alpha_s$-value for BCDMS was re-evaluated using the value of 
$\Lambda_{\rm QCD}^{NLO} = 224$ MeV and $\alpha_s(10~{\rm GeV}) = 0.160$ \cite{Benvenuti:1989fm}. 
These values correspond to a NLO fit with $n_f = 4$ in the \MSbar-scheme. 
We evolved this value 
back to the charm threshold keeping $n_f = 4$ and determined then $\alpha_s(M_Z)$ evolving forward
passing the bottom threshold.
In \cite{Virchaux:1991jc} higher twist contributions
and $\Lambda_{QCD}^{n_f = 4}$ were fitted together for the BCDMS $\mu p$ and $\mu d$ data 
resulting in the somewhat larger value $\Lambda_{QCD}^{n_f = 4} = 263 \pm 42$~MeV
and the NLO value $\alpha_s(M_Z) = 0.113 \pm 0.003~{\rm (exp)}$, 
which was also obtained~\cite{Blumlein:2006be,BBprep}. Both values are compatible within errors.

In Fig.~\ref{fig:scanfixed} we plot the $\chi^2$-profile using eq.~(\ref{eq:EQCHI}) at NLO and NNLO.
To that end, we compare the fit result with the $\alpha_s$-behavior of the individual 
data set, fixing all other parameters. 
The minimum and variation ($\Delta \chi^2 \equiv 1$) then determine the values given in 
Tab.~\ref{tab:alphas1}, see also Fig.~\ref{fig:scanfixed}. 
For BCDMS and NMC we find complete consistency to the values given by the 
experiments and the present analysis. 
The downward shift $\Delta \alpha_s(M_Z)$ which is consistently observed when going from NLO to NNLO 
amounts to values between 0.0030 and 0.0055, with a lower sensitivity for the DY 
data, which yield rather low values with large errors. 
The fitted central values are well covered by the individual data sets. 
Fig.~\ref{fig:scanfixed} shows the response with respect to $\alpha_s(M_Z)$ 
of the individual data sets fixing the non-perturbative shape parameters in the global fit.
One may refit these parameters in changing $\alpha_s(M_Z)$, cf. Fig.~\ref{fig:scanas}.
However, here the change in the other PDF parameters remains undocumented, in particular 
if the corresponding covariance matrices are not publicly available, 
as is the case for some of the global fits. We have performed this analysis only to compare to the MSTW 
and NNPDF analyses below but still prefer the results of Fig.~\ref{fig:scanfixed}. 
Both results are given in Tab.~\ref{tab:alphas6} for comparison. 
Comparing Figs.~\ref{fig:scanfixed} and~\ref{fig:scanas} 
one finds that the shape parameters in case of the BCDMS 
and HERA data remain widely stable and larger shifts are introduced for the NMC and SLAC data.
The stability of the results, on the other 
hand, allows to conclude that fully compatible sets of precision data were used.

We have also performed NLO fits, including the Tevatron jet 
data~\cite{Abulencia:2007ez,Aaltonen:2008eq,Abazov:2008hua,Abazov:2010fr}.
Furthermore, we have formally extended the analysis fitting the DIS and DY data at NNLO 
while treating the Tevatron jet data at NLO and  
supplementing threshold corrections based on soft gluon resummation~\cite{Kidonakis:2000gi} 
for the single jet inclusive data. 
This latter approximation we denoted by NNLO$^*$, cf.~\cite{Alekhin:2011cf}.
A NLO measurement of $\alpha_s(M_Z)$ was also performed by CDF \cite{Affolder:2001hn},
with larger errors than in \cite{Abazov:2009nc}, 
$\alpha_s(M_Z) = 0.1178~^{+0.0081}_{-0.0095} ({\rm exp})~^{+0.0071}_{-0.0047}
({\rm scale})~\pm 0.0059~({\rm PDF})$.
At NLO the different sets of Tevatron jet data do not modify the value obtained in our standard
analysis. A consistent NNLO is not yet possible since the corresponding scattering cross sections
still have to be calculated. 
Again a systematic downward shift of $\Delta \alpha_s(M_Z) = 0.0029 - 0.0047$
is obtained upon going from NLO to NNLO$^*$.
The corresponding central values are 1$\sigma$ compatible with our NNLO
central value in eq.~(\ref{eq:abm11as-nnlo}).
We would like to mention that already our former ABKM09 results~\cite{Alekhin:2009ni} 
give a very good description of the CMS jet data~\cite{Rabbertz:1368241}
and also the Tevatron 3-jet data~\cite{Bandurin:2011sh}. 
We note that in a recent NLO analysis of the 5-jet cross section at LEP 
a value of $\alpha_s(M_Z) = 0.1156~^{+0.0041}_{-0.0034}$ was obtained \cite{Frederix:2010ne}.

\begin{figure}[th!]
\centerline{
  \includegraphics[width=8.0cm]{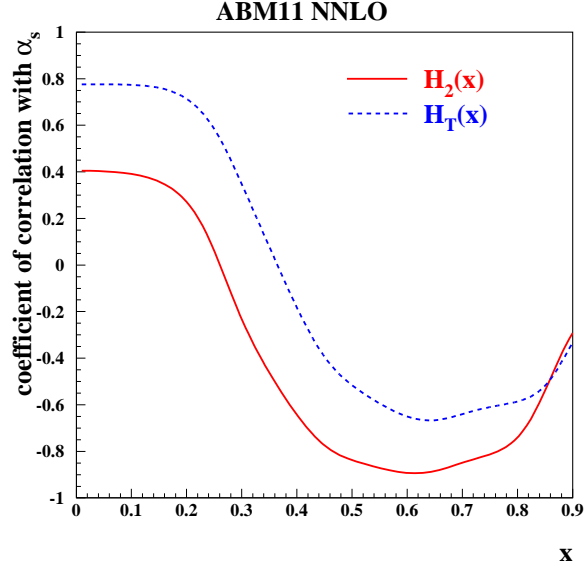}}
  \caption{\small
    \label{fig:corrht}
    The correlation coefficient of $\alpha_s(M_Z)$ with 
    the nucleon twist-4 coefficients 
    $H_2$ (solid line) and $H_T$ (dashes) versus $x$ as obtained in our NNLO fit. 
  }
\end{figure}
\begin{figure}[th!]
\centerline{
  \includegraphics[width=15.5cm]{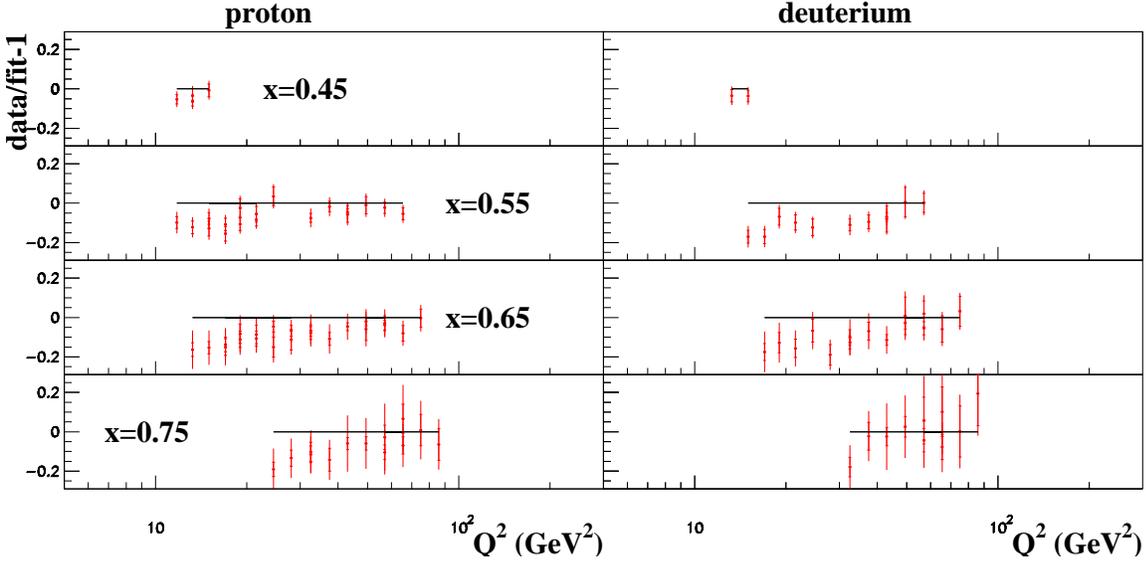}}
  \caption{\small
    \label{fig:kk}
    The same as in Fig.~\ref{fig:bcdms} for the data points rejected in 
    the analysis of~\cite{Shaikhatdenov:2009xd}.
  }
\end{figure}

The higher twist terms play an important role in the determination of 
$\alpha_s$ from the DIS data~\cite{Virchaux:1991jc}. 
In our analysis they contribute 
up to 10\% of the cross section at the low margin of $Q^2$ and $W$ given by eq.~(\ref{eq:discuts}).
As a result, the value of $\alpha_s$ is strongly correlated with the twist-4 coefficients, which are extracted from the 
fit simultaneously with $\alpha_s$, cf. Fig.~\ref{fig:corrht}.
The correlation is more pronounced for $H_2$ at large $x$ and for $H_T$ at small $x$. 
The latter affects the determination of $\alpha_s$ even in the 
case of a more stringent cut on $W$ since the low-$x$ part of the 
data is not sensitive to this cut. 
For the variant of our NNLO fit with the cut of eq.~(\ref{eq:softcut}) 
imposed and the higher twist terms set to zero, we obtain the value of $\alpha_s(M_Z)=0.1191(6)$, 
much bigger than our nominal result of eq.~(\ref{eq:abm11as-nnlo}). 
For comparison, the same fit with the higher twist terms fixed at the values of 
Tab.~\ref{tab:htsvalues} gives $\alpha_s(M_Z)=0.1131(5)$ comparable with 
eq.~(\ref{eq:abm11as-nnlo}).
Note that in both cases the error in $\alpha_s$ is 
smaller than one of eq.~(\ref{eq:abm11as-nnlo}) despite the reduced data set used in the fit. 
This says, that the uncertainty in $\alpha_s$ is essentially controlled by the
higher twist term variation.  
To get rid of the impact of the higher twist terms on $\alpha_s$ 
an even more stringent cut on $Q^2$ is necessary, in addition to the cut on $W$. 
With the NNLO variant of our fit and using 
\begin{equation}
  \label{eq:ssoftcut}
  W^2 > 12.5~{\rm GeV}^2\, ,\qquad\qquad 
  Q^2 > 10~{\rm GeV}^2\, ,
\end{equation}
the value of $\alpha_s(M_Z)=0.1134(8)$ is obtained, if  
the higher twist terms are set to zero and $\alpha_s(M_Z)=0.1135(8)$, 
if the higher twist terms are fixed at the values of Tab.~\ref{tab:htsvalues}. 
From this comparisons we conclude that $\alpha_s$ is pushed to larger values  
due to the neglect of the higher twist terms in the case of a cut as in
eq.~(\ref{eq:softcut}), which is commonly imposed in the global PDF fits. 
Likewise, it is less sensitive to the details of the fit ansatz. 
As we have found earlier~\cite{Alekhin:2011ey}, the value of $\alpha_s$ extracted from our 
fit is quite sensitive to the treatment of the NMC data~\cite{Arneodo:1995cq}. 
Normally, we use the NMC data on the cross section in the fit, cf. Sec.~\ref{sec:incldis}. 
If, however, we employ instead the NMC data on $F_2$ extracted by 
the NMC collaboration with their own assumptions about the value of $R$, 
the value of $\alpha_s(M_Z)$ increases by $+0.0035$ for the case of 
our earlier ABKM09 NNLO fit~\cite{Alekhin:2011ey}. 
In comparison, for the variant of the ABKM09 with the cut 
of eq.~(\ref{eq:softcut}) imposed and the high-twist terms set to zero, 
the value of $\alpha_s(M_Z)$ is shifted by $+0.0003$ only. 
This is in agreement with the size of the $\alpha_s$ variation  
obtained in the variant of the MSTW08 fit with an improved treatment 
of the NMC data~\cite{Thorne:2011kq}.
Note, however, that the details of the DIS data treatment employed 
in the improved analysis of~\cite{Thorne:2011kq}, are still different from ours. 
In places, this has an impact on the $\alpha_s$ value. 
E.g., if we combine the errors in the NMC and HERA data in quadrature, as it is done 
in~\cite{Thorne:2011kq,Martin:2009iq}, the NNLO  ABKM09 value of $\alpha_s(M_Z)$ 
is shifted upwards by $+0.0029$ and its sensitivity to the NMC data treatment 
is reduced to $+0.0011$. 
This effect may in particular explain the relatively big value of 
$\alpha_s(M_Z)=0.1164$ observed in~\cite{Thorne:2011kq} 
with the cuts similar to ones of eq.~(\ref{eq:ssoftcut}). 

The impact of the NMC data treatment on the fit has also been studied 
for NN21 in~\cite{Ball:2011we}, where little effect on the PDFs has been found.
However the value of $R$ obtained in that analysis is still much 
lower than the one obtained in our fit and closer to the 
value of $R$ used by NMC to extract the value of $F_2$ from the 
cross section (compare Fig.~2 in~\cite{Ball:2011we} and Fig.~1
in~\cite{Alekhin:2011ey}).
Note in this context that the value of $\chi^2/NDP\approx 1.7$
obtained in~\cite{Ball:2011we} for the NMC data is much bigger than in 
our case, cf.Tab.~\ref{tab:chi2tab}.
Furthermore, in the NN21 study the value of $\alpha_s$ is fixed. 
Therefore, its correlation with the PDFs is not considered. 
Finally, the expressions for cross sections used in the NN21 analysis 
do not include the power corrections in $Q^2$, which are numerically important 
at small $Q^2$ (compare eqs.~(2),~(3) in ~\cite{Ball:2011we} 
with eq.~(1) in~\cite{Arneodo:1995cq}, and
eqs.~(\ref{eq:sigma}),~(\ref{eq:Rratio}) of the present article).
These differences make a detailed comparison of the results of~\cite{Alekhin:2011ey}
and~\cite{Ball:2011we} difficult. 

The value of $\alpha_s$ obtained in our fit is substantially constrained by the BCDMS data 
of~\cite{Benvenuti:1989rh,Benvenuti:1989fm}, which almost 
entirely survive after the cut of eq.~(\ref{eq:discuts}). 
Meanwhile the authors of~\cite{Shaikhatdenov:2009xd} suggested to cut 
in addition the most inaccurate BCDMS data with low inelasticity $y$. 
The value of $\alpha_s(M_Z)$ reported in~\cite{Shaikhatdenov:2009xd} with such 
cut is by $\sim 0.009$ larger than the one obtained from the analysis of the 
whole set of the BCDMS data at NNLO.  
The pulls of the low-$y$ data rejected in the analysis of~\cite{Shaikhatdenov:2009xd} 
with respect to our NNLO fit are given in Fig.~\ref{fig:kk}. 
The fit is in reasonable agreement with data within the errors. 
Furthermore, rejecting these data points from the NNLO fit, 
we obtain a value of $\alpha_s(M_Z)=0.1139(12)$, 
which is somewhat bigger than the one in eq.~(\ref{eq:abm11as-nnlo}).
The statistical significance of the shift, however, is marginal. 
The discrepant findings of~\cite{Shaikhatdenov:2009xd} concerning the impact of the low-$y$ BCDMS 
data may appear due to the fact that the systematic uncertainties in the data are 
not taken into account in~\cite{Shaikhatdenov:2009xd}.  
In our case the systematic errors are included into the
value of $\chi^2$, cf. Sec.~\ref{sec:appA}. 
Therefore the low-$y$ data points with an enhanced systematic uncertainty 
have reduced weight and do not affect the fit. 

\begin{table}[h]\centering
\renewcommand{\arraystretch}{1}
\begin{tabular}{|l|c|c|c|c|}
\hline
\multicolumn{1}{|c|}{Experiment} &
\multicolumn{4}{c|}{$\alpha_s({M_Z})$} \\
\cline{2-5}
\multicolumn{1}{|c|}{ } &
\multicolumn{1}{c|}{NLO$_{exp}$} &
\multicolumn{1}{c|}{NLO} &
\multicolumn{1}{c|}{NNLO} &
\multicolumn{1}{c|}{N$^3$LO$^*$} 
\\ 
\hline 
BCDMS &$0.1111 \pm 0.0018$  &$0.1138 \pm 0.0007$ & $0.1126 \pm 0.0007$ & $0.1128 \pm 0.0006$ \\ 
NMC   &$0.117\phantom{0}~^{+~0.011}_{-~0.016}\phantom{0}\phantom{0}$
                            &$0.1166 \pm 0.0039$ & $0.1153 \pm 0.0039$ & $0.1153 \pm 0.0035$ \\ 
SLAC  &                     &$0.1147 \pm 0.0029$ & $0.1158 \pm 0.0033$ & $0.1152 \pm 0.0027$ \\ 
\hline 
BBG &                       &$0.1148 \pm 0.0019$ & $0.1134 \pm 0.0020$ & $0.1141 \pm 0.0021$\\ 
BB &                        &$0.1147 \pm 0.0021$ & $0.1132 \pm 0.0022$ & $0.1137 \pm 0.0022$ \\ 
\hline 
\end{tabular} 
\renewcommand{\arraystretch}{1}  
\caption{
\small
\label{tab:alphas3}
Comparison of the values of $\alpha_s({M_Z})$ obtained by 
BCDMS~\cite{Benvenuti:1989fm} and NMC~\cite{Arneodo:1993kz} at NLO 
with the results of the flavor non-singlet fits BBG~\cite{Blumlein:2006be} and
BB~\cite{BBprep}
of the DIS flavor non-singlet world data, 
at NLO, NNLO, and N$^3$LO$^*$ with the response 
of the individual data sets, combined for the experiments 
BCDMS~\cite{Benvenuti:1989rg,Benvenuti:1989fm,Benvenuti:1989gs}, 
NMC~\cite{Arneodo:1996qe}, and 
SLAC~\cite{Whitlow:1991uw}.}
\end{table}

We turn now to comparisons with other NNLO analyses which will be performed studying the contribution
of different data sets to $\alpha_s(M_Z)$. We first compare to the flavor non-singlet 
analyses~\cite{Blumlein:2006be,BBprep}. 
In~\cite{BBprep} the valence analysis is performed by accounting for the remnant sea-quark and gluon 
contributions to $F_2(x,Q^2)$ in the region $x > 0.35$ through the PDFs taken from~\cite{Alekhin:2009ni}.
The results are summarized in Tab.~\ref{tab:alphas3}.
The values of $\alpha_s(M_Z)$ at NLO turn out to be lower than those obtained in singlet analyses, 
cf. Tabs.~\ref{tab:alphas1} and~\ref{tab:alphas7}. 
However they are consistent within the scale variation errors.
At NNLO both analyses lead to the same values. 
Also note the anti-correlation of the size of higher twist contributions 
in the large-$x$ region with the inclusion of higher orders at leading twist, 
cf.~\cite{Alekhin:2003qq,Blumlein:2006be,Blumlein:2008kz,BBprep}.  
The next order, denoted by N$^3$LO$^*$, yields information on the remaining theoretical uncertainty. 
At N$^3$LO$^*$, the non-singlet three-loop 
Wilson coefficients are used~\cite{Moch:2004xu,Vermaseren:2005qc}
and the four-loop non-singlet anomalous dimension 
is estimated with a Pad\'e-approximation and accounting for a 100~\% error.
In fact the latter extrapolation agrees within $20\%$ with the second moment of the non-singlet
four-loop anomalous dimension \cite{Baikov:2006ai,Velizhanin:2011es}.
For the three experiments, which give the bulk information on $\alpha_s(M_Z)$,
the shift due to the N$^3$LO$^*$ contributions amount to 
$|\Delta \alpha_s(M_Z)| = 0.0002 - 0.0006$ and globally to 0.0007. 
At NNLO, the $\alpha_s(M_Z)$ values of the individual data sets vary by 0.0032, 
consistent within the 1$\sigma$ errors.

Next, in Tab.~\ref{tab:alphas4}, we compare with the fit results of NN21~\cite{Lionetti:2011pw,Ball:2011us} 
for individual data sets for DIS and other hadronic hard scattering data. 
The labels for those data sets in Tab.~\ref{tab:alphas4} follow the original
notation of NN21 in~\cite{Lionetti:2011pw,Ball:2011us} 
(and, likewise in Tab.~\ref{tab:alphas5} for MSTW~\cite{Martin:2009bu}).
The references corresponding to the data sets are given additionally.
At NLO the $\alpha_s(M_Z)$ values range from 0.1135 (E866, DY) to 0.1252 (NuTeV), with a corresponding 
range at NNLO of 0.1111 (D0 jet) to 0.1225 (CDF jet). 
The value of $\alpha_s(M_Z) = 0.1204 \pm 0.0015$ 
for the BCDMS data at NLO differs significantly from that given by the experiment 
$0.1111 \pm 0.0018$ \cite{Benvenuti:1989fm}. 
Comparing the change of the $\alpha_s(M_Z)$ values between the NLO and NNLO analyses one finds 
downward shifts between 0.0075 (NuTeV) and 0.003 (CDF R2KT)
and upward shifts between $0.0055$ (CDF Zrap) and $0.0061$ (ZEUS H2), 
see Tab.~\ref{tab:alphas4}. 
The values for $\alpha_s(M_Z)$ obtained 
for the SLAC data are found to be larger than 0.124 both at NLO and NNLO,
cf. Fig.~2 in~\cite{Ball:2011us}. 
The $\chi^2$ values for the scans in $\alpha_s(M_Z)$ at NNLO 
turn out to be worse than in NLO in the global analysis 
for the data sets of NMC, BCDMS, HERA~I, CHORUS, ZEUS F2C, DY E866, CDF Zrap, and D0 Zrap,
again see Fig.~2 in~\cite{Ball:2011us}. 
Comparing the DIS only fit to the global analysis it is found that 
the $\chi^2$ values improve significantly, except for NMCp, and to a lesser
extent for SLAC at higher values of $\alpha_s(M_Z)$, cf. Fig.~5 in~\cite{Ball:2011us}. 

\begin{table}[h!]\centering
\renewcommand{\arraystretch}{1}
\begin{tabular}{|l|c|c|c|}
\hline
\multicolumn{1}{|c|}{Experiment} &
\multicolumn{3}{c|}{$\alpha_s({M_Z})$} \\
\cline{2-4}
\multicolumn{1}{|c|}{ } &
\multicolumn{1}{c|}{NLO$_{exp}$} &
\multicolumn{1}{c|}{NLO} &
\multicolumn{1}{c|}{NNLO} 
\\ 
\hline 
BCDMS \cite{Benvenuti:1989rh,Benvenuti:1989fm}  
   &$0.1111 \pm 0.0018$&$0.1204 \pm 0.0015$ & $0.1158 \pm 0.0015$ \\
NMC$_p$ \cite{Arneodo:1996qe}
   &                   &$0.1192 \pm 0.0018$ & $0.1150 \pm 0.0020$ \\
NMC$_{pd}$ \cite{Arneodo:1996kd}
   &$0.117\phantom{0}~^{+~0.011}_{-~0.016}\phantom{0}\phantom{0}$
                       &                    & $0.1146 \pm 0.0107$ \\
SLAC \cite{Whitlow:1991uw} 
   &                    &$> 0.124$          & $> 0.124$ \\
HERA I \cite{herapdf:2009wt}
   &                   &$0.1223 \pm 0.0018$ & $0.1199 \pm 0.0019$ \\
ZEUS H2 \cite{Chekanov:2008aa,Chekanov:2009gm}
   &                   &$0.1170 \pm 0.0027$ & $0.1231 \pm 0.0030$ \\
ZEUS F2C \cite{Breitweg:1999ad,Chekanov:2003rb,Chekanov:2008yd,Chekanov:2009kj}
   &                   &$0.1144 \pm 0.0060$ &                     \\
NuTeV \cite{Goncharov:2001qe,Mason:2006qa} 
   &                   &$0.1252 \pm 0.0068$ & $0.1177 \pm 0.0039$ \\
\hline
E605 \cite{Moreno:1990sf} 
   &                   &$0.1168 \pm 0.0100$ &                     \\ 
E866 \cite{Webb:2003ps,Webb:2003bj,Towell:2001nh}
   &                   &$0.1135 \pm 0.0029$ &                     \\
CDF Wasy \cite{Aaltonen:2009ta}
   &                   &$0.1181 \pm 0.0060$ &                     \\
CDF Zrap \cite{Aaltonen:2010zza}
   &                   &$0.1150 \pm 0.0034$ & $0.1205 \pm 0.0081$ \\
D0 Zrap  \cite{Abazov:2007jy}
   &                   &$0.1227 \pm 0.0067$ &                     \\
\hline
CDF R2KT \cite{Abulencia:2007ez}
   &                   &$0.1228 \pm 0.0021$ & $0.1225 \pm 0.0021$ \\
D0 R2CON \cite{Abazov:2008hua}
   & $0.1161~^{+~0.0041}_{-~0.0048}\phantom{0}$ 
                       &$0.1141 \pm 0.0031$ & $0.1111 \pm 0.0029$ \\[0.5ex]
\hline 
NN21 
   &                   &$0.1191 \pm 0.0006$ & $0.1173 \pm 0.0007$ \\
\hline
\end{tabular} 
\renewcommand{\arraystretch}{1}  
\caption{\small
\label{tab:alphas4}
Comparison of the values of $\alpha_s({M_Z})$ obtained by 
BCDMS~\cite{Benvenuti:1989fm}, NMC~\cite{Arneodo:1993kz}, and D0~\cite{Abazov:2009nc} 
at NLO with the results of NN21~\cite{Lionetti:2011pw,Ball:2011us} 
for the fits to DIS and other hard scattering data at NLO and NNLO 
and the corresponding response of the different data sets analysed.}
\end{table}
\begin{table}[h!]\centering
\renewcommand{\arraystretch}{1}
\begin{tabular}{|l|c|c|c|}
\hline
\multicolumn{1}{|c|}{Experiment} &
\multicolumn{3}{c|}{$\alpha_s({M_Z})$} \\
\cline{2-4}
\multicolumn{1}{|c|}{ } &
\multicolumn{1}{c|}{NLO$_{exp}$} &
\multicolumn{1}{c|}{NLO} &
\multicolumn{1}{c|}{NNLO} 
\\ 
\hline 
BCDMS $\mu p, F_2$ \cite{Benvenuti:1989rh}  
   &$0.1111 \pm 0.0018$&$-$                &$0.1085 \pm 0.0095$  \\
BCDMS $\mu d, F_2$ \cite{Benvenuti:1989fm}  
   &                   &$0.1135 \pm 0.0155$&$0.1117 \pm 0.0093$  \\
NMC   $\mu p, F_2$ \cite{Arneodo:1996qe} 
   &$0.117\phantom{0}~^{+~0.011}_{-~0.016}\phantom{0}\phantom{0}$
                       &$0.1275 \pm 0.0105$&$0.1217 \pm 0.0077$  \\
NMC $\mu d, F_2$ \cite{Arneodo:1996qe} 
   &                   &$0.1265 \pm 0.0115$&$0.1215 \pm  0.0070$ \\
NMC $\mu n/\mu p$ \cite{Arneodo:1996kd}
   &                   &$0.1280 \phantom{\pm0.0000}$
                                           &$0.1160 \phantom{\pm0.0000}$  \\
E665  $\mu p,  F_2$ \cite{Adams:1996gu} 
   &                   &$0.1203 \phantom{\pm0.0000}$
                                           &$-$                  \\
E665  $\mu d,  F_2$ \cite{Adams:1996gu} 
   &                   &$-$                &$-$                  \\
SLAC  $ep,  F_2$ \cite{Whitlow:1990gk,Whitlow:1990dr} 
   &                   &$0.1180 \pm 0.0060$&$0.1140 \pm 0.0060$  \\
SLAC  $ed,  F_2$ \cite{Whitlow:1990gk,Whitlow:1990dr} 
   &                   &$0.1270 \pm 0.0090$&$0.1220 \pm 0.0060$  \\
NMC,BCDMS,SLAC, $F_L$ \cite{Benvenuti:1989rh,Arneodo:1996qe,Whitlow:1991uw} 
   &                   &$0.1285 \pm 0.0115$&$0.1200 \pm 0.0060$  \\
\hline
E886/NuSea $pp$, DY \cite{Webb:2003bj} 
   &                   &$-$                &$0.1132 \pm 0.0088$  \\
E886/NuSea $pd/pp$, DY \cite{Towell:2001nh}
   &                   &$0.1173 \pm 0.107\phantom{0}$ 
                                           &$0.1140 \pm 0.0110$   \\
\hline
NuTeV  $\nu N, F_2$ \cite{Tzanov:2005kr}
   &                   &$0.1207 \pm 0.0067$&$0.1170 \pm 0.0060$  \\
CHORUS $\nu N, F_2$  \cite{Onengut:2005kv}
   &                   &$0.1230 \pm 0.0110$&$0.1150 \pm 0.0090$  \\
NuTeV  $\nu N, xF_3$ \cite{Tzanov:2005kr}
   &                   &$0.1270 \pm 0.0090$&$0.1225 \pm 0.0075$  \\
CHORUS $\nu N, xF_3$ \cite{Onengut:2005kv} 
   &                   &$0.1215 \pm 0.0105$&$0.1185 \pm 0.0075$  \\
CCFR  \cite{Goncharov:2001qe,Mason:2006qa} 
   &                   &$0.1190  \phantom{\pm0.0000}$
                                           &$-$                  \\
NuTeV  $\nu N \rightarrow \mu \mu X$ \cite{Goncharov:2001qe,Mason:2006qa}
   &                   &$0.1150 \pm 0.0170$&$-$                  \\
H1 $ep$ 97-00, $\sigma_r^{\rm NC}$ \cite{Lobodzinska:2003yd,Adloff:2003uh,Adloff:2000qk,Adloff:2000qj}
   &                   &$0.1250 \pm 0.0070$&$0.1205 \pm 0.0055$  \\
ZEUS $ep$ 95-00, $\sigma_r^{\rm NC}$ \cite{Breitweg:1998dz,Chekanov:2001qu,Chekanov:2002ej,Chekanov:2003yv}
   &                   &$0.1235 \pm 0.0065$&$0.1210 \pm 0.0060$  \\
H1 $ep$ 99-00, $\sigma_r^{\rm CC}$  \cite{Adloff:2003uh} 
   &                   &$0.1285 \pm 0.0225$&$0.1270 \pm 0.0200$  \\
ZEUS $ep$ 99-00, $\sigma_r^{\rm CC}$ \cite{Chekanov:2003vw} 
   &                   &$0.1125 \pm 0.0195$&$0.1165 \pm 0.0095$  \\   
H1/ZEUS $ep, F_2^{\rm charm}$ \cite{Breitweg:1999ad,Chekanov:2003rb,Adloff:1996xq,Adloff:2001zj,Aktas:2005iw,Aktas:2004az}
   &                   &$-$                &$0.1165 \pm 0.0095$  \\
\hline
H1 $ep$ 99-00 incl. jets \cite{Aaron:2009vs,Aktas:2007pb} 
   &$0.1168~^{+~0.0049}_{-~0.0035}\phantom{0}$ 
                       &$0.1127 \pm 0.0093$&                     \\[0.5ex]
ZEUS $ep$ 96-00 incl. jets  \cite{ZEUSas:2010,Chekanov:2002be,Chekanov:2006xr} 
   &$0.1208~^{+~0.0044}_{-~0.0040}\phantom{0}$
                       &$0.1175 \pm 0.0055$&                     \\[0.5ex]
\hline
D0 II $p\bar{p}$ incl. jets \cite{Abazov:2008hua}
   &$0.1161~^{+~0.0041}_{-~0.0048}\phantom{0}$ 
                       &$0.1185 \pm 0.0055$&$0.1133 \pm 0.0063$  \\[0.5ex]
CDF II $p\bar{p}$ incl. jets \cite{Abulencia:2007ez}
   &                   &$0.1205 \pm 0.0045$&$0.1165 \pm 0.0025$  \\
D0 II $W \rightarrow l \nu$ asym. \cite{Abazov:2007pm}
   &                   &$-$                &$-$                  \\
CDF II $W \rightarrow l \nu$ asym. \cite{Acosta:2005ud} 
   &                   &$-$                &$-$                  \\
D0 II $Z$ rap. \cite{Abazov:2007jy}
   &                   &$0.1125 \pm 0.0100$&$0.1136 \pm 0.0084$  \\
CDF II $Z$ rap. \cite{CDFZ}
   &                   &$0.1160 \pm 0.0070$&$0.1157 \pm 0.0067$  \\
\hline 
MSTW 
   &                   &$0.1202~^{+~0.0012}_{-~0.0015}\phantom{0}$ 
                                           &$0.1171 \pm 0.0014$ \\[0.5ex]
\hline
\end{tabular} 
\renewcommand{\arraystretch}{1}  
\caption{\small
\label{tab:alphas5}
Comparison of the values of $\alpha_s(M_Z)$ obtained by BCDMS \cite{Benvenuti:1989fm},
NMC \cite{Arneodo:1993kz}, HERA-jet \cite{Aaron:2009vs,ZEUSas:2010} (see also \cite{Grindhammer:2011ur,Kogler:2011hw}),
and D0 \cite{Abazov:2009nc} at NLO with the results of the 
MSTW fits to DIS and other hard scattering data
at NLO and NNLO and the corresponding response of the different data sets
analysed, cf. Figs.~7a and 7b in~\cite{Martin:2009bu}.
Entries not given correspond to $\alpha_s(M_Z)$ central values below $0.110$ or above $0.130$; 
in case no errors are assigned these are larger than the bounds provided in form of the plots 
in~\cite{Martin:2009bu,THORNE}.}
\end{table}
\begin{table}[h]\centering
\renewcommand{\arraystretch}{1}
\begin{tabular}{|l|c|c|c|c|}
\hline
\multicolumn{1}{|c|}{Data Set } &
\multicolumn{1}{c|}{ABM11} &
\multicolumn{1}{c|}{BBG} &
\multicolumn{1}{c|}{NN21} &
\multicolumn{1}{c|}{MSTW} \\
\hline
BCDMS  & $0.1128 \pm 0.0020$ & $0.1126 \pm 0.0007$ & $0.1158 \pm 0.0015$ & $0.1101 \pm 0.0094$ \\
  & ($0.1084 \pm 0.0013$)& & & \\[1ex]
NMC    & $0.1055 \pm 0.0026$ & $0.1153 \pm 0.0039$ & $0.1150 \pm 0.0020$ & $0.1216 \pm 0.0074$ \\
  & ($0.1152 \pm 0.0007$)& & & \\[1ex]
SLAC   & $\begin{array}{c} 0.1184 \pm 0.0021 \\ (0.1128 \pm 0.0003) \end{array}$
& $0.1158 \pm 0.0034$ & $> 0.124$           & 
$~~\left\{{\small \begin{array}{c} 0.1140~ \pm~ 0.0060~{\rm ep}\\ 0.1220~ \pm~ 0.0060~{\rm ed} \end{array}}\right.$ \\[2ex]
HERA   & 
$\begin{array}{c} 0.1139 \pm 0.0014 \\ (0.1126 \pm 0.0002) \end{array}$
&                     & 
$\left\{{\small \begin{array}{c} 0.1199~ \pm~ 0.0019 \\ 0.1231~ \pm~ 0.0030 \end{array}}\right.~$                  
                                                                         & $ 0.1208 \pm 0.0058$ \\[2ex]
DY     &  ($0.101\phantom{0} \pm 0.025)\phantom{0}$  
                             & $-$                 & $-$                 & $0.1136 \pm 0.0100$ \\
\hline
       & $0.1134 \pm 0.0011$ &$0.1134 \pm 0.0020$  & $0.1173 \pm 0.0007$ & $0.1171 \pm 0.0014$\\ 
\hline
\end{tabular}
\renewcommand{\arraystretch}{1}   
\caption{
\small
\label{tab:alphas6}
Comparison of the pulls in $\alpha_s(M_Z)$ per data set between the
ABM11 as shown in Fig.~\ref{fig:scanas}, 
BBG~\cite{Blumlein:2006be}, 
NN21~\cite{Ball:2011us} and 
MSTW~\cite{Martin:2009bu} analyses at NNLO.
The values in parantheses of ABM11 correspond to Fig.~\ref{fig:scanfixed} where the shape
parameters are not refitted which is also the case for BBG.
}
\end{table}
\begin{table}[h!]\centering
\renewcommand{\arraystretch}{1}
\begin{tabular}{|l|l|l|}
\hline
\multicolumn{1}{|c|}{ } &
\multicolumn{1}{c|}{$\alpha_s({M_Z})$} &
\multicolumn{1}{c|}{  } \\
\hline
BBG      & $0.1134~^{+~0.0019}_{-~0.0021}$
         & {\rm valence~analysis, NNLO}  \cite{Blumlein:2006be}           
\\[0.5ex]
BB       & $0.1132  \pm 0.0022$
         & {\rm valence~analysis, NNLO}  \cite{BBprep}           
\\
GRS      & $0.112 $ & {\rm valence~analysis, NNLO}  \cite{Gluck:2006yz}           
\\
ABKM           & $0.1135 \pm 0.0014$ & {\rm HQ:~FFNS~$n_f=3$} \cite{Alekhin:2009ni}             
\\
ABKM           & $0.1129 \pm 0.0014$ & {\rm HQ:~BSMN-approach} 
\cite{Alekhin:2009ni}             
\\
JR       & $0.1124 \pm 0.0020$ & {\rm
dynamical~approach} \cite{JimenezDelgado:2008hf}   
\\
JR       & $0.1158 \pm 0.0035$ & {\rm
standard~fit}  \cite{JimenezDelgado:2008hf}    
\\
ABM11            & $0.1134\pm 0.0011$ &  \\
MSTW & $0.1171\pm 0.0014$ &  \cite{Martin:2009bu}     \\
NN21 & $0.1173\pm 0.0007$ &  \cite{Ball:2011us}     \\
CT10 & $0.118\phantom{0} \pm 0.005$  &  \cite{CTEQ12} \\
\hline
Gehrmann et al.& {{$0.1153 \pm 0.0017 \pm 0.0023$}} & {\rm
$e^+e^-$~thrust}~\cite{Gehrmann:2009eh}
\\
Abbate et al.& {{$0.1135 \pm 0.0011 \pm 0.0006$}} & {\rm
$e^+e^-$~thrust}~\cite{Abbate:2010xh}
\\
\hline
3 jet rate   & $0.1175 \pm 0.0025$ & Dissertori et al. 2009 \cite{arXiv:0910.4283}\\
Z-decay      & $0.1189 \pm 0.0026$ & BCK 2008/12  (N$^3$LO) \cite{arXiv:0801.1821,Baikov:2012er}\\
$\tau$ decay & $0.1212 \pm 0.0019$ & BCK 2008               \cite{arXiv:0801.1821}\\
$\tau$ decay & $0.1204 \pm 0.0016$ & Pich 2011              \cite{Bethke:2011tr}\\
$\tau$ decay & $0.1169 \pm 0.0025$ & Boito et al. 2011      \cite{arXiv:1110.1127}
\\
\hline
lattice      & $0.1205 \pm 0.0010$ & PACS-CS 2009 (2+1 fl.) \cite{Aoki:2009tf} \\
lattice      & $0.1184 \pm 0.0006$ & HPQCD 2010             \cite{arXiv:1004.4285} \\
lattice      & $0.1200 \pm 0.0014$ & ETM 2012 (2+1+1 fl.)   \cite{arXiv:1201.5770} 
\\
\hline
BBG & $0.1141~^{+~0.0020}_{-~0.0022}$
& {\rm valence~analysis, N$^3$LO$(^*)$}  \cite{Blumlein:2006be}            \\[0.5ex]
BB & $0.1137 \pm 0.0022$
& {\rm valence~analysis, N$^3$LO$(^*)$}  \cite{BBprep}            \\
\hline
{world average} & {$
0.1184 \pm 0.0007$  } & \cite{Bethke:2009jm} (2009)
\\
                & {$
0.1183 \pm 0.0010$  } & \cite{Bethke:2011tr} (2011)
\\
\hline
\end{tabular}
\renewcommand{\arraystretch}{1}   
\caption{
\small
\label{tab:alphas7}
Summary of recent NNLO QCD analyses of the DIS world data, supplemented by related measurements
using other processes.}
\end{table}

In Tab.~\ref{tab:alphas5} the $\alpha_s$ values determined by MSTW 
at NLO and NNLO \cite{Martin:2009bu} are compared for 
individual data sets in the fit. Here, the 1$\sigma$ errors, as defined by MSTW, are read off
the corresponding plots in~\cite{THORNE}.
This definition of `1$\sigma$' is obtained for values of $\Delta \chi^2$ much larger than one. 
Moreover, these values do even strongly vary between the different measurements used, 
which is unlike the case for the ABM and NN21 analyses. 
This procedure leads to an enlargement of errors, which, e.g., 
in the case of BCDMS translates into a NNLO value 
$\alpha_s(M_Z) = 0.1085 \pm 0.0095$ 
rather than the experimental one of $\Delta \alpha_s(M_Z) = \pm 0.0018$. 
The latter accuracy is reflected in other analyses, however, cf. 
$\Delta \alpha_s(M_Z)  = \pm 0.0007$~(BB), 
$\Delta \alpha_s(M_Z)  = \pm 0.0015$~(NN21), 
and
$\Delta \alpha_s(M_Z)  = \pm 0.0013$~(ABM11). 
Similar effects are present for various other data sets, 
as can be seen comparing the values given in Tabs.~\ref{tab:alphas1}--\ref{tab:alphas5}. 
In this way, almost all individual $\alpha_s(M_Z)$-errors, even at NNLO, 
are larger than the typical theory uncertainty of about $\pm 0.0050$ at NLO. 
We stress that the present analysis (ABM11) correctly accounts for 
the experimental systematic errors of all data sets used, cf. Tabs.~\ref{tab:alphas1} and~\ref{tab:alphas2}, 
and that an enlargement of errors has not been necessary. 
In view of this fact, it is somewhat surprising that the final error in the
MSTW analysis, i.e. $\Delta \alpha_s(M_Z)  = \pm 0.0014$~(MSTW),  
fully agrees with those obtained by BB, ABM11, and NN21, cf. Tab.~\ref{tab:alphas6}.
Unlike the case of NN21, the $\alpha_s(M_Z)$ values of MSTW become generally 
lower at NNLO if compared to NLO, with the exception of the D0 run II $Z$-boson rapidity data set, 
where the NNLO value is slightly higher.

It is also interesting to compare the $\alpha_s(M_Z)$ values obtained 
in the NN21~\cite{Lionetti:2011pw,Ball:2011us}, MSTW \cite{Martin:2009bu} and ABM11 analyses 
with respect to the individual data sets used in those fits.
At NLO NN21 obtains lower values for D0 R2CON than MSTW and ABM and a significantly higher value
for D0 ZRAP than MSTW, see Tabs.~\ref{tab:alphas4} and \ref{tab:alphas5}.
At NNLO, the individual $\alpha_s(M_Z)$ value for the data set CDF Zrap
moves upward with respect to the NLO value with a significantly larger error, 
while for MSTW the value remains the same as at NLO. 
The values of ABM11 given in Tab.~\ref{tab:alphas2} for Tevatron jet data 
are rather close to those of MSTW, both at NLO an NNLO$^*$. 
For the NuTeV data the NLO and NNLO $\alpha_s(M_Z)$ values show a bigger
difference for NN21 than for MSTW, while the NNLO values are rather similar.

In Tab.~\ref{tab:alphas6} we compare the $\alpha_s$ values of the ABM11, BBG, NN21 and MSTW 
analyses for those data sets which are commonly used at NNLO. 
An NLO comparison would still be subject to a scale error of $\sim 0.0050$, 
which is usually too large to differentiate between the various fit results. 
For ABM11 we present the $\alpha_s$ values extracted from Figs.~\ref{fig:scanfixed}
and~\ref{fig:scanas}. 
For the BCDMS data ABM11, BBG and MSTW obtain lower values, 
while NN21 differs, e.g., by $+2\sigma$ (or $\Delta \alpha_s(M_Z) = 0.0030$)
from MSTW and $+7\sigma$ (or $\Delta \alpha_s(M_Z) = 0.0110$) from ABM11. 
For the NMC data ABM, BBG, and NN21 do agree very well, 
while the value of MSTW shows an upward shift of 
$\Delta \alpha_s(M_Z) = 0.0066$ compared to NN21.  
The size of both these shifts is of the order of NLO scale uncertainty 
and should not be present at NNLO.
For the SLAC $ep$ data the values by ABM11, BBG and MSTW are consistent within errors. 
On the other hand, MSTW reports a much larger value for the SLAC $ed$ data than obtained 
by ABM11 and BBG.
NN21 obtains partial $\alpha_s(M_Z)$ values $> 0.124$ both at NLO and NNLO both in 
their global and DIS only analyses. 
This is in contrast to the present results, to BB, and to MSTW for the $ep$ data.
In the non-singlet BBG analysis the influence of the HERA data is strongly reduced, 
since most of these data are located within the quark-sea region. 
The fit results of NN21 and MSTW lead to values of $\alpha_s(M_Z) \sim 0.120$ 
while those of ABM11 yield a much lower value of 0.1139 (0.1126).
Note that the MSTW analysis does not yet include the HERA run~I combined data set~\cite{herapdf:2009wt}. 
The results for the DY data are consistent between ABM11 and MSTW, although 
the sensitivity of these data to $\alpha_s(M_Z)$ is comparably small.
In summary, despite the fact that NN21 and MSTW obtain nearly 
the same global fit values for $\alpha_s(M_Z)$, 
the above discussion shows that quite a series of individual pulls are different.
Both ABM11 and MSTW do not confirm the relatively large $\alpha_s(M_Z)$ values for the data sets 
of BCDMS and CDF R2KT and also the rise, at NNLO, of the NN21 value for CDF Zrap calls for further
clarification, cf. Tab.~\ref{tab:alphas4}.

Summarizing the comparison of the present results on $\alpha_s(M_Z)$ 
with the analyses~\cite{Blumlein:2006be,BBprep,Lionetti:2011pw,Ball:2011us,Martin:2009bu,THORNE}
we observe a good agreement with \cite{Blumlein:2006be,BBprep} and find differences in a series 
of data sets for~\cite{Lionetti:2011pw,Ball:2011us,Martin:2009bu,THORNE} both at NLO and NNLO, 
with partly different deviations in case of~\cite{Lionetti:2011pw,Ball:2011us} and~\cite{Martin:2009bu,THORNE}. 
NN21 does not agree with the BCDMS result. 
We find lower values of $\alpha_s(M_Z)$ both for the HERA and the SLAC data. 
We do not confirm part of the $\alpha_s(M_Z)$\ values found 
in~\cite{Lionetti:2011pw,Ball:2011us,Martin:2009bu,THORNE} for the jet data.
We would like to mention once more that in our analysis no rescaling of errors is performed, 
which varies for different data sets in the analysis~\cite{Martin:2009bu,THORNE}, 
but we have accounted for the systematic errors given by the experiments directly.
The procedure of\cite{Martin:2009bu,THORNE} naturally leads to re-weighting of the impact 
of different data sets on the value of $\alpha_s(M_Z)$.

Finally we would like to summarize the results of different determinations of 
$\alpha_s(M_Z)$ at NNLO and N$^3$LO (or N$^3$LO$^*$) in Tab.~\ref{tab:alphas7}. 
Some part of these results has been reported in~\cite{Bethke:2011tr}. 
Flavor non-singlet analyses of the DIS world data were performed 
in~\cite{Blumlein:2006be,Gluck:2006yz,BBprep}, with an accuracy of $\Delta \alpha_s(M_Z) \simeq 2\%$ at NNLO. 
The difference between the value at N$^3$LO$^*$ and NNLO amounts to $\sim 0.0007$, 
which provides an estimate for the size of a remaining uncertainty. 
The ABKM09 analysis~\cite{Alekhin:2009ni} 
is a combined flavor singlet/non-singlet fit of the DIS world data, DY and di-muon data.
Here a remaining difference of $\Delta \alpha_s(M_Z) = 0.0006$ due to the treatment of the
heavy-flavor corrections was observed. 
These uncertainties signal the typical theory errors remaining at the present level of description. 
The JR09 analysis obtained very similar results in combined flavor 
singlet/non-singlet fits~\cite{JimenezDelgado:2008hf} with a slightly larger value in the standard 
fit compared to the dynamical approach. 
The present analysis obtains the same values as those found 
in~\cite{Blumlein:2006be,BBprep,Alekhin:2009ni,Gluck:2006yz,JimenezDelgado:2008hf} before, 
while a slightly larger value $\alpha_s(M_Z) = 0.1147 \pm 0.0012$ was reported
in ABM10~\cite{Alekhin:2010iu} based on incorporating the combined HERA run~I data~\cite{herapdf:2009wt}. 
However, the improved treatment of the heavy-quark contributions finally led to the
present value $\alpha_s(M_Z) = 0.1134 \pm 0.0011$. 
The inclusion of Tevatron jet data, cf.~\cite{Alekhin:2011cf}, 
although only a NNLO$^*$ analysis, alters this values at most to $\alpha_s(M_Z) = 0.1149 \pm 0.0012$ and 
the effect of the complete, yet unknown NNLO corrections, remain to be seen.
Low values of $\alpha_s(M_Z)$ have not only been reported in analyses which
are predominantly based on DIS data, 
but also from those of thrust in $e^+ e^-$-annihilation 
in~\cite{Abbate:2010xh}, cf. also \cite{Gehrmann:2009eh}. 

Larger central values of $\alpha_s(M_Z)$ at NNLO (and similar in size) 
are reported by MSTW \cite{Martin:2009bu} and NN21 \cite{Ball:2011us}. 
These fits include a much broader set of hadronic scattering data in the
analysis and above we have pointed out various differences between these analyses, 
see Tabs.~\ref{tab:alphas4} and~\ref{tab:alphas5}. 
These differences manifest themselves in various cases 
in rather different pulls for the value of $\alpha_s(M_Z)$.
Note also that the 1$\sigma$ confidence level is very differently defined 
in the MSTW analysis compared to ABM11, NN21 and JR09. 
The (preliminary) central value of $\alpha_s(M_Z)$ reported by 
CT10~\cite{CTEQ12} is similar to MSTW and NN21 at NNLO, 
although accompanied by a rather large uncertainty of $\Delta \alpha_s(M_Z) = 0.0050$, 
which makes the CT10 result compatible with the lower values obtained 
in~\cite{Blumlein:2006be,BBprep,Alekhin:2009ni,Gluck:2006yz,JimenezDelgado:2008hf}. 
We have also discussed the reasons for the difference between the central values of
$\alpha_s(M_Z)$ in the NLO and NNLO 
analyses~\cite{Blumlein:2006be,BBprep,Alekhin:2009ni,Gluck:2006yz,JimenezDelgado:2008hf}
and~\cite{Martin:2009bu,Ball:2011us} and further comparisons may be performed. 
An earlier claim that this is caused by the Tevatron jet data is not confirmed,
cf.~\cite{Alekhin:2010iu}. 
Larger central values for $\alpha_s(M_Z)$ than 
in~\cite{Blumlein:2006be,BBprep,Alekhin:2009ni,Gluck:2006yz,JimenezDelgado:2008hf}
are obtained for the 3-jet rate in $e^+e^-$ annihilation~\cite{arXiv:0910.4283} at NNLO 
and for the $Z$-decay width at N$^3$LO \cite{arXiv:0801.1821}, see Tab.~\ref{tab:alphas7}.
The current $\alpha_s(M_Z)$ values at NNLO extracted from $\tau$-decays 
vary between 0.1212 and 0.1169 \cite{arXiv:0801.1821,Bethke:2011tr,arXiv:1110.1127}.
One lattice measurement \cite{arXiv:1004.4285} yields the same value as 
the current world average~\cite{Bethke:2011tr,Bethke:2009jm}. 
Other recent lattice results are compatible with this value and
more lattice studies are still underway aiming at improved 
systematics~\cite{SOMMER}.

%%
%% ---------------------------------------------------------------------
%%
\subsection{Heavy-quark masses}
\label{sec:hq-mass}
\begin{figure}[t!]
\centerline{
  \includegraphics[width=15.5cm]{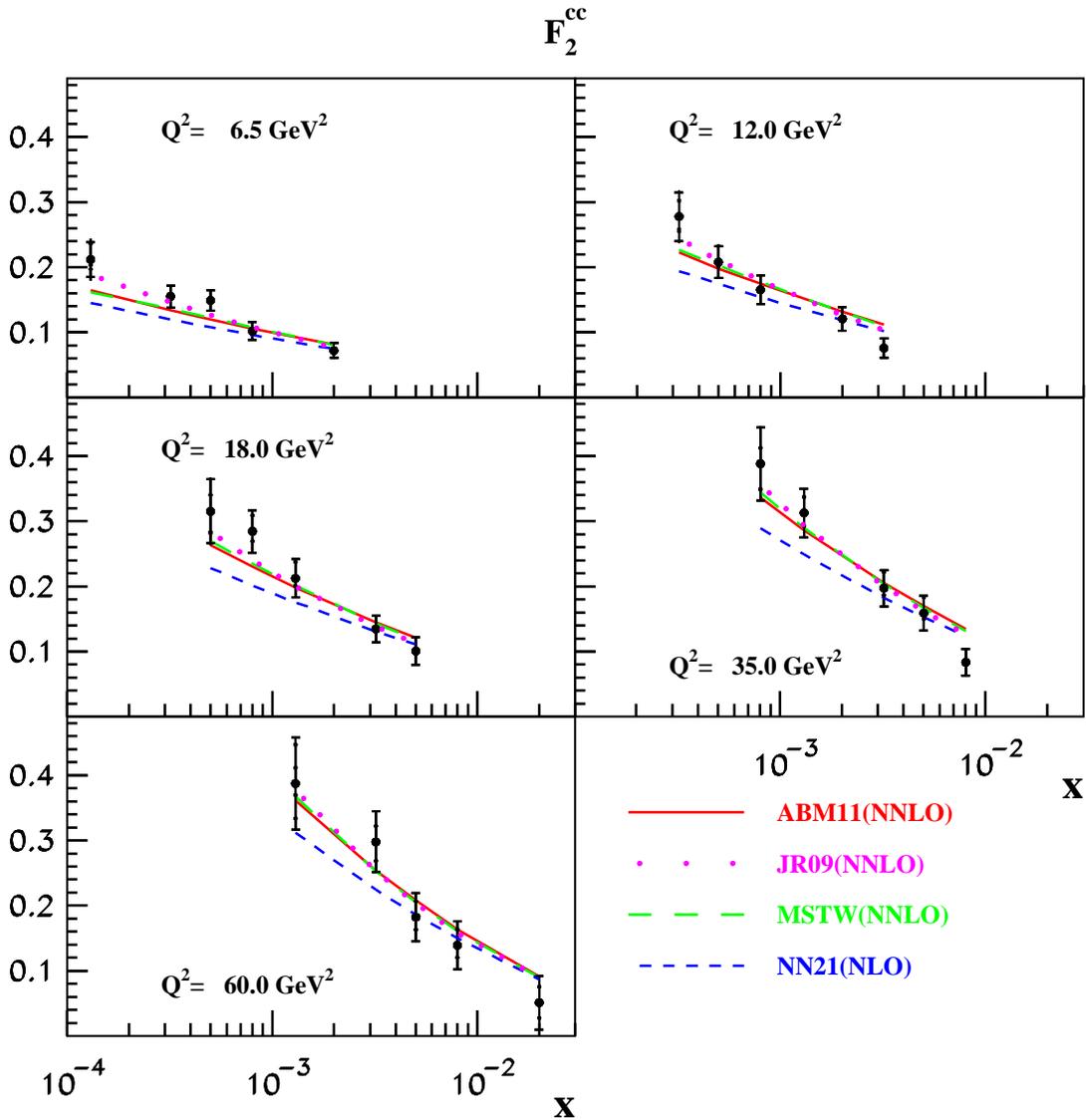}}
  \vspace*{-5mm}
  \caption{\small
    \label{fig:f2cc}
    Comparison of the data from~\cite{Aaron:2011gp} 
    for the semi-inclusive structure function $F_2^{cc}$ 
    at different values of the momentum transfer $Q^2$ versus $x$ 
    with predictions of various PDF sets at NLO and NNLO in QCD, 
    all taken in the FFNS with $n_f=3$ 
    and with a running-mass of $m_c=1.27~{\rm GeV}$~\cite{Nakamura:2010pdg},
    The NNLO predictions for $F_2^{cc}$ use 
    the ABM11 PDFs (solid curves), 
    the JR09 PDFs~\cite{JimenezDelgado:2008hf} (dots), and 
    the MSTW PDFs of~\cite{Martin:2009iq} (long dashes).
    The NLO calculations are based on the 
    NN21 PDFs of~\cite{Ball:2011mu} (short dashes).
  }
\end{figure}
\begin{figure}[th!]
  \includegraphics[width=8.0cm]{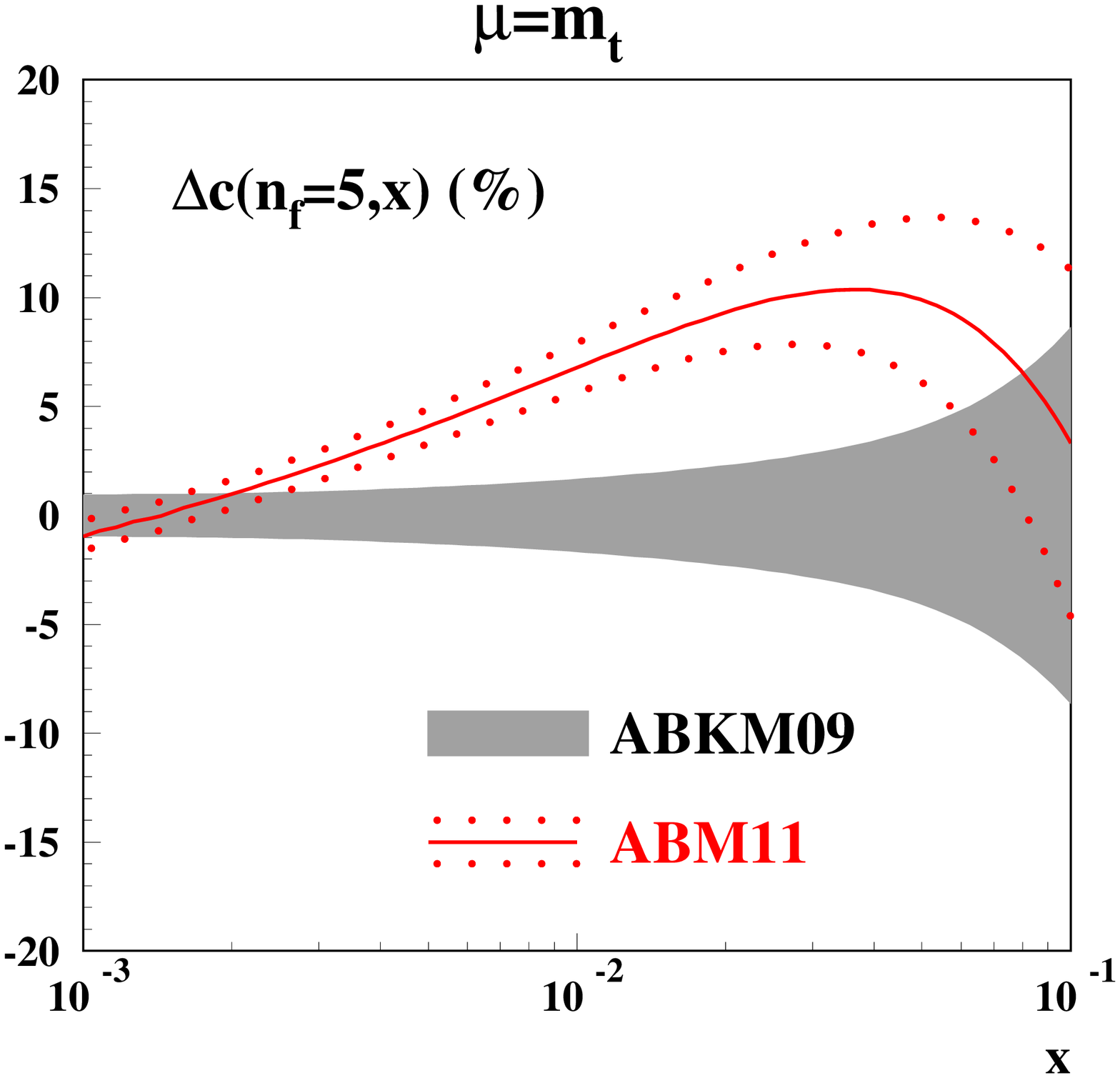}
  \includegraphics[width=8.0cm]{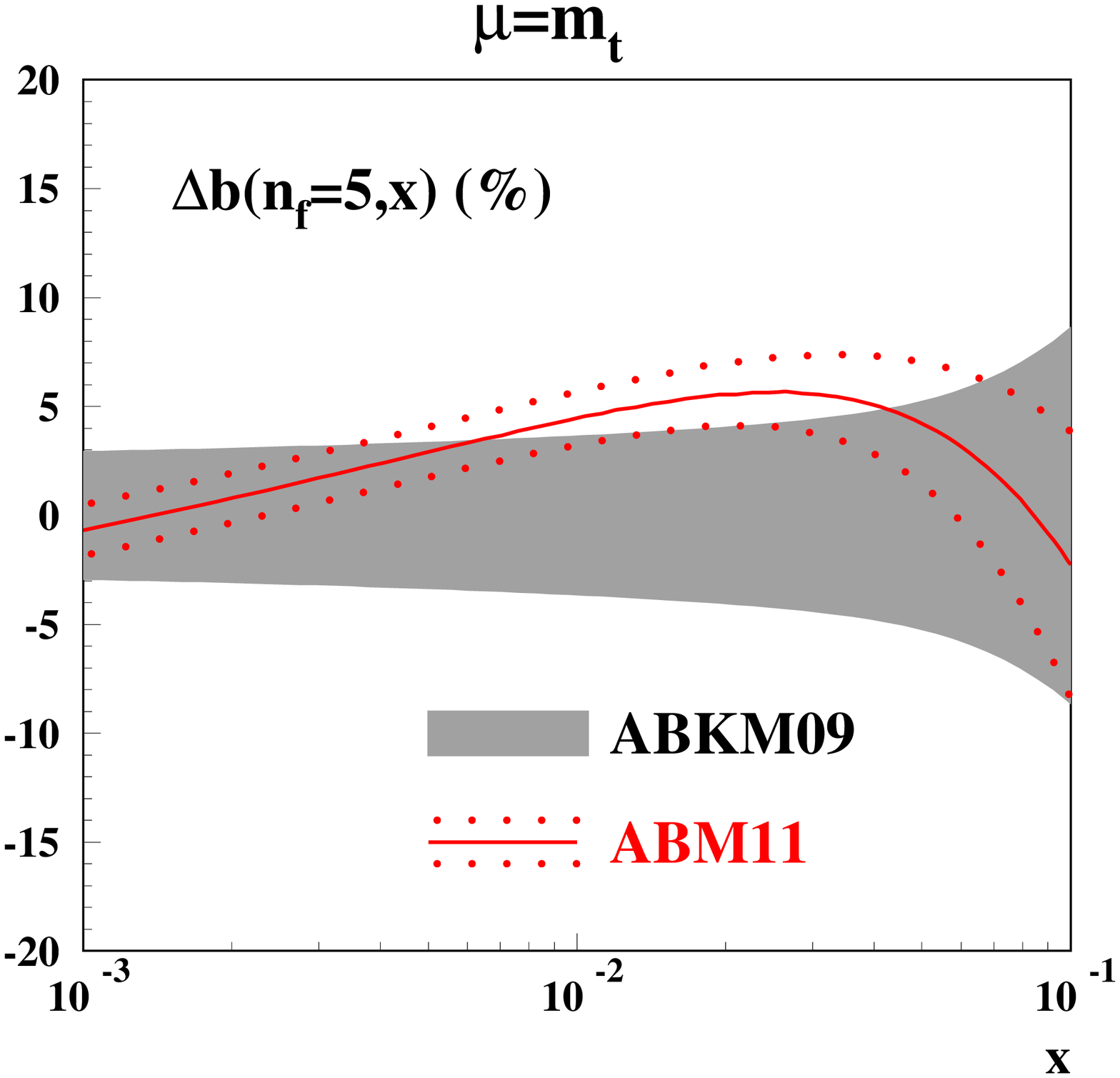}
  \caption{\small
    \label{fig:hq-pdf}
      The charm- (left) and the bottom-quark (right) PDFs obtained in the global fit:
      The dotted (red) lines denote the $\pm 1 \sigma$ band of relative
      uncertainties (in percent) and the solid (red) line indicates the
      central prediction resulting from the fit with
      the running masses of eq.~(\ref{eq:mcmbres}).
      For comparison the shaded (grey) area represents the results of ABKM09~\cite{Alekhin:2009ni}.
}
\end{figure}

The precise value of the heavy-quark masses is an important parameter in 
the description of DIS charm-quark production. 
In our fit we use the heavy-quark masses in the \MSbar-scheme and 
their implementation for heavy-quark DIS discussed in~\cite{Alekhin:2009ni} 
which allows us to relate the values for $m_c$ and $m_b$ directly 
to PDG results~\cite{Nakamura:2010pdg}, as done in eq.~(\ref{eq:mcmbinp}).

The current DIS data displays great sensitivity to the charm mass $m_c$ 
as we have demonstrated previously~\cite{Alekhin:2010sv,Alekhin:2011jq}.
Therefore, based on the pseudo-data input from eq.~(\ref{eq:mcmbinp}) 
we have released the uncertainty of the quark masses 
to obtain the following results 
\begin{eqnarray}
  \label{eq:mcmbres}
  m_c(m_c) \,=\, 1.27 \pm 0.06\,\, {\rm GeV}
  \, ,
  \qquad\qquad
  m_b(m_b) \,=\, 4.19 \pm 0.13\,\, {\rm GeV}
  \, ,
\end{eqnarray}
which shows that the error on $m_c$ from the inclusive DIS data used in the fit 
is comparable to the one quoted by the PDG~\cite{Nakamura:2010pdg} 
(although not comparable to the single most precise measurement listed therein).
Interestingly, we observe in the covariance matrix in Tabs.~\ref{tab:pdfco1}--\ref{tab:pdfco3} 
correlations of $m_c$ with $\alpha_s$ and some parameters of the gluon and the strange PDFs.
The precision of DIS data to the value of $m_c$ has previously also been exploited 
for the first direct determination of the running mass for charm quarks 
from hadronic processes with space-like kinematics as a variant of ABKM09 
yielding values consistent with but systematically somewhat lower than the PDG world 
average~\cite{Alekhin:2010sv,Alekhin:2011jq}.

It is interesting to compare the results of the fit to the most recent HERA data 
from the H1 collaboration~\cite{Aaron:2011gp} 
for the charm structure function $F_2^c$ in heavy-quark DIS 
extracted with the use of the {\tt HVQDIS} code~\cite{Harris:1995tu}.
This is done in Fig.~\ref{fig:f2cc} for our 3-flavor running-mass NNLO
predictions with $m_c$ of eq.~(\ref{eq:mcmbres}) using the
approximate NNLO QCD predictions of~\cite{Laenen:1998kp,Alekhin:2008hc,Presti:2010pd}.
The data are not used in our fit, however the agreement is quite good, as 
well as for the predictions based on the NNLO 3-flavor PDFs 
of JR09~\cite{JimenezDelgado:2008hf} and MSTW~\cite{Martin:2009iq}.
The predictions using the NLO 3-flavor NN21 PDFs of~\cite{Ball:2011mu} 
undershoot the data. 
The differences may be related to the peculiarities of the so-called general mass VFNS 
modeling employed in those fits~\cite{Martin:2009iq,Lai:2010vv,Ball:2011mu}. 
Also wrong analyses of the combined H1 and ZEUS data on $F_2^c$ exist~\cite{H1ZEUSmc:2010}.

Finally, we want to mention that the issue of heavy-quark masses also has consequences for heavy-quark PDFs, 
because the uncertainty on heavy-quark PDFs is directly related to the
accuracy of the numerical values for the quark masses $m_c$ or $m_b$.
The latter appear parametrically in the OMEs used to generate 
charm- and bottom-PDFs in schemes with $n_f=4$ or $n_f=5$ flavors.
With the results of eq.~(\ref{eq:mcmbres}) and the use of the \MSbar-scheme 
this uncertainty in the charm and bottom PDFs can be significantly reduced.

In Fig.~\ref{fig:hq-pdf} we display the PDFs generated in this way. 
We obtain a charm-PDF with comparable uncertainties 
to the one of~\cite{Alekhin:2009ni} (which then has used the pole mass definition for $m_c$),
while the resulting uncertainty of the bottom-PDF is greatly reduced, see also~\cite{Alekhin:2011jq}.
This improvement has a significant impact on LHC phenomenology, as it will, 
e.g., allow for precise predictions for the production of single-top-quarks and
other processes sensitive to bottom-PDFs.

%%
%% ---------------------------------------------------------------------
%%
\subsection{Moments of the PDFs}
\label{sec:pdfmoms}

\begin{table}[h]
\renewcommand{\arraystretch}{1.3}
\begin{center}
{\small
\begin{tabular}{|l|l|l|l|l|}
\hline
\multicolumn{1}{|c|}{ } &
\multicolumn{1}{c|}{$\langle x u_v(x)\rangle$} &
\multicolumn{1}{c|}{$\langle x d_v(x)\rangle$} &
\multicolumn{1}{c|}{$\langle x [u_v-d_v](x)\rangle$} &
\multicolumn{1}{c|}{$\langle x V(x)\rangle$} \\
\hline
ABM11
& $0.2971 \pm 0.0039$
& $0.1174 \pm 0.0050$
& $0.1797 \pm 0.0042$
& $0.1655 \pm 0.0039$
\\
ABKM09~\cite{Alekhin:2009ni}
& $0.2981 \pm 0.0025$ 
& $0.1191 \pm 0.0023$  
& $0.1790 \pm 0.0023$  
& $0.1647 \pm 0.0022$
\\
HERAPDF1.5~\cite{herapdf:2009wt,herapdfgrid:2011}
& $0.2938~^{+~0.0031}_{-~0.0052}$
& $0.1264~^{+~0.0054}_{-~0.0059}$
& $0.1674~^{+~0.0043}_{-~0.0052}$
& $0.1706~^{+~0.0071}_{-~0.0103}$
\\     
JR09~\cite{JimenezDelgado:2008hf,JimenezDelgado:2009tv}
& $0.2897 \pm 0.0035$ 
& $0.1253 \pm 0.0052$  
& $0.1645 \pm 0.0063$  
& $0.1513 \pm 0.0118$  
\\
MSTW~\cite{Martin:2009iq}
& $0.2816~^{+~0.0051}_{-~0.0042}$
& $0.1171~^{+~0.0027}_{-~0.0028}$
& $0.1645~^{+~0.0046}_{-~0.0034}$
& $0.1533~^{+~0.0041}_{-~0.0033}$
\\     
NN21~\cite{Ball:2011uy}
& $0.2913 \pm 0.0038$
& $0.1218 \pm 0.0042$
& $0.1695 \pm 0.0040$
& $0.1539 \pm 0.0030$
\\     
\hline
BBG~\cite{Blumlein:2006be}
& $0.2986 \pm 0.0029$ 
& $0.1239 \pm 0.0026$  
& $0.1747 \pm 0.0039$  
&
\\
BBG [N$^3$LO]~\cite{Blumlein:2006be}
& $0.3006 \pm 0.0031$  
& $0.1252 \pm 0.0027$  
& $0.1754 \pm 0.0041$ 
&
\\
\hline
\end{tabular}
}
\caption{\small 
  Comparison of the second moment of the valence quark distributions
  at NNLO and N$^3$LO obtained in different analyses at $Q^2 = 4~{\rm GeV}^2$.
}
\label{tab:2ndmom}
\end{center}
\end{table}

\begin{figure}[th!]
  \centerline{
    \includegraphics[width=10.0cm]{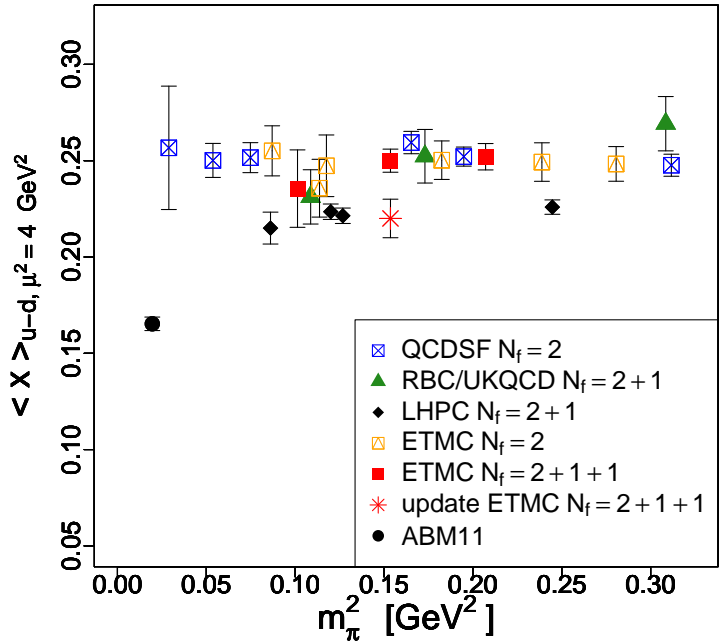}}
  \caption{\small
    \label{fig:2ndmom}
     Comparison of lattice computations for the second moment of the
     non-singlet distribution as a function of the pion mass $m_\pi$ 
     with the result of ABM11 given in Tab.~\ref{tab:2ndmom} 
     along with the uncertainties of the respective measurement. 
  }
\end{figure}

In Tab.~\ref{tab:2ndmom} we summarize different values of the second moment of the 
valence quark densities obtained in NNLO analyses at the scale $Q^2 = 4~{\rm GeV}^2$. 
It is evident, that these moments are rather stable quantities for all PDF sets considered 
as they are mostly influenced by the data normalization.
They are closely related to the moments which are being measured in lattice simulations. 
Of central importance is the quantity 
\begin{eqnarray}
  \label{eq:V2ndmom}
  \langle xV(Q^2) \rangle &=& 
  \int\limits_0^1 dx\, 
  x\left\{\left[u(x,Q^2) + \bar{u}_{s}(x,Q^2)\right]  
  -       
  \left[d(x,Q^2) + \bar{d}_{s}(x,Q^2)\right]\right\}
  \, , 
\end{eqnarray}
where $q \equiv q_{v} + q_{s}$ with $q=u,d$.

In Fig.~\ref{fig:2ndmom} the result for eq.~(\ref{eq:V2ndmom}) obtained in the
present analysis is compared with recent lattice computations using varying numbers of
flavors as a function of the pion mass $m_\pi$ employed on the lattice.
In detail, these are QCDSF ($n_f=2$)~\cite{Pleiter:2011gw}, RBC/UKQCD ($n_f=2+1$)~\cite{Aoki:2010xg}, 
LHPC ($n_f=2+1$)~\cite{Bratt:2010jn}, ETMC ($n_f=2$)~\cite{Alexandrou:2011nr} and 
ETMC ($n_f=2+1+1$)~\cite{Dinter:2011jt,Dinter:2011sg}.
It is apparent from Fig.~\ref{fig:2ndmom} that there are substantial
differences, even for low pion masses, 
between those lattice measurements and the experimental determinations of
Tab.~\ref{tab:2ndmom}. 
For very recent progress see~\cite{Bali:2012av}.

%%
%% ---------------------------------------------------------------------------
%%
\renewcommand{\theequation}{\thesection.\arabic{equation}}
\setcounter{equation}{0}
\renewcommand{\thefigure}{\thesection.\arabic{figure}}
\setcounter{figure}{0}
\renewcommand{\thetable}{\thesection.\arabic{table}}
\setcounter{table}{0}
\section{Benchmarks for cross sections }
\label{sec:crs}
In this section we quantify the impact of the new PDF set on predictions for
benchmark cross sections at the Tevatron and the LHC.
To that end, we confine ourselves to (mostly) inclusive cross sections which are known
to NNLO in QCD either completely or in very good approximation,
see~\cite{Alekhin:2010dd} for previous work along these lines.
NNLO accuracy is actually the first instance, where meaningful statements 
about the residual theoretical uncertainty are possible,
since at NLO the latter which is conventionally determined from a variation 
of the renormalization and factorization scale is generally still too large,
given the precision of present collider data.

In detail, we consider the following set of inclusive observables: 
hadronic $W$- and $Z$-boson
production~\cite{Hamberg:1990np,Harlander:2002wh}, the cross section
for Higgs boson production in the dominant channels,
ggF~\cite{Harlander:2002wh,Anastasiou:2002yz,Ravindran:2003um,Ravindran:2004mb},
vector-boson fusion (VBF) with {\tt VBFNNLO}~\cite{Bolzoni:2010xr,Bolzoni:2011cu}, 
and in Higgs-strahlung~\cite{Brein:2003wg}.
The cross section for top-quark pair production is approximately NNLO (based
on threshold resummation, see e.g.,~\cite{Langenfeld:2009wd}) 
and is computed with {\tt HATHOR} (version 1.2)~\cite{Aliev:2010zk}. 
We also consider the lepton ($l^\pm$) charge asymmetry in hadronic $W^\pm$-boson
production as a function of the rapidity~\cite{Anastasiou:2003yy,Anastasiou:2003ds,Catani:2009sm,Catani:2010en}.
Throughout the entire section we focus on the QCD corrections only. 
That is to say, we neglect all electroweak radiative effects at NLO, 
which often amount to corrections of ${\cal O}(\rm{few})\%$ at the LHC 
and, therefore, need to be considered in precision predictions.

The PDF uncertainties quoted here are calculated by summing over the $n_{PDF}$ sets 
provided by the various groups, where $n_{PDF}$ is the number of parameters used in the fit.
Typically, we quote the symmetric error according to 
\begin{equation}
  \label{eq:pdferr}
  \Delta \sigma_{PDF} \,=\, 
  \sqrt {\sum_{k=1,n_{PDF}} \, (\sigma_{0} - \sigma_{k} )^2}
  \, ,
\end{equation}
where $\sigma_{k}$ is obtained by using the $k$-th PDF $f^{k}_{i}$, 
which parametrizes the $\pm 1\sigma$-variation of the $k$-th fit parameter 
after diagonalization of the correlation matrix.
In some cases, e.g., for MSTW~\cite{Martin:2009iq}, asymmetric PDF errors are
provided, in which case the variation in the $k$-th fit parameter is given by a pair of PDFs $f^{k,\pm}_{i}$. 
The resulting asymmetric PDF error is then computed according to
\begin{eqnarray}
  \label{eq:pdfasyerr}
  \Delta \sigma_{PDF}^{+} &=& 
  \sqrt {\sum_{k=1,n_{PDF}} \, {\rm max}(0,+\sigma_{k,+}-\sigma_{0},+\sigma_{k,-}-\sigma_{0})^2}
  \, ,\\
  \Delta \sigma_{PDF}^{-} &=& 
  \sqrt {\sum_{k=1,n_{PDF}} \, {\rm min}(0,-\sigma_{k,+}+\sigma_{0},-\sigma_{k,-}+\sigma_{0})^2}
  \, .
\end{eqnarray}
For MSTW, we are using the set with 68\% confidence level error estimates throughout.

In a Monte Carlo approach like the one advocated by NN21~\cite{Ball:2011uy}, the PDF
uncertainty can be determined as the quadratic deviation from the central fit
as in Eq.~(\ref{eq:pdferr}), but with an additional factor $1/\sqrt{n_{PDF}}$.
For reasons of efficiency and run-times, we are using the NN21 PDF with 100
sets in our comparisons only, see also the discussion in Sec.~\ref{sec:lhapdf}.

\subsection{$W$- and $Z$-boson production}
\label{sec:wzprod}

We start by presenting results for $W$- and $Z$-boson production at the LHC at $\sqrt{s}=7~$TeV.
For the electroweak parameters, we follow~\cite{Alekhin:2010dd} and choose the scheme based 
on the set $(G_F, M_W, M_Z)$.
According to~\cite{Nakamura:2010pdg}, we have 
$G_F = 1.16637 \times 10^{-5}~{\rm GeV}^{-2}$, 
$M_W = 80.399  \pm 0.023$~GeV, 
$M_Z = 91.1876  \pm 0.0021$~GeV 
and the corresponding widths 
$\Gamma(W^\pm) = 2.085 \pm 0.042$~GeV and 
$\Gamma(Z) = 2.4952 \pm 0.0023$~GeV.
The weak mixing angle is then a dependent quantity, with  
\begin{equation}
  \label{eq:sintw}
  \hat{s}_Z^2 \,=\,  1 - \frac{M_W^2}{\hat{\rho} M_Z^2}  = 0.2307 \pm 0.0005
  \, ,
\end{equation}
and $\hat{\rho} = 1.01047 \pm 0.00015$. 
Finally, the Cabibbo angle $\theta_c$ yields the value $\sin^2 \theta_c = 0.051$.

At NNLO the theoretical uncertainty due to scale variation is small compared
to the PDF error, see Tabs.~\ref{tab:wz} and \ref{tab:wznf4}.
The change in the predictions between ABKM09 and ABM11 is small. 
NN21 and MSTW typically predict smaller cross sections with differences at the
level of 1$-$2$\sigma$, while the numbers of JR09 are significantly smaller,
see also the detailed discussion in~\cite{Alekhin:2010dd}.
Most importantly, there is the choice to consider in particular the $W$-boson
cross section in alternative schemes with $n_f=4$ or $n_f=5$ flavors, since
contributions of the initial bottom PDFs being proportional to the CKM
matrix element $V_{tb}$ are kinematically suppressed.

\begin{table}[ht!]
\renewcommand{\arraystretch}{1.3}
\begin{center}
{\small
\begin{tabular}{|l|l|l|l|l|l|}
\hline
&ABM11
&ABKM09~\cite{Alekhin:2009ni}
&JR09~\cite{JimenezDelgado:2008hf,JimenezDelgado:2009tv}
&MSTW~\cite{Martin:2009iq}
&NN21~\cite{Ball:2011uy}
\\     
\hline
$W^+$ &
${59.53}~^{+0.38}_{-0.23}~^{+0.88}_{-0.88}$
 &
${59.30}~^{+0.39}_{-0.24}~^{+0.93}_{-0.93}$
 &
${54.68}~^{+0.32}_{-0.19}~^{+1.30}_{-1.30}$
 &
${57.20}~^{+0.31}_{-0.14}~^{+1.02}_{-0.95}$
 &
${58.46}~^{+0.35}_{-0.21}~^{+0.91}_{-0.91}$
 \\
$W^-$ &
${39.97}~^{+0.28}_{-0.17}~^{+0.65}_{-0.65}$
 &
${39.70}~^{+0.28}_{-0.18}~^{+0.63}_{-0.63}$
 &
${37.22}~^{+0.24}_{-0.14}~^{+0.92}_{-0.92}$
 &
${39.89}~^{+0.24}_{-0.12}~^{+0.69}_{-0.67}$
 &
${39.75}~^{+0.27}_{-0.17}~^{+0.63}_{-0.63}$
 \\
$W^\pm$ &
${99.51}~^{+0.69}_{-0.41}~^{+1.43}_{-1.43}$
 &
${99.00}~^{+0.67}_{-0.41}~^{+1.53}_{-1.53}$
 &
${91.91}~^{+0.55}_{-0.34}~^{+2.14}_{-2.14}$
 &
${97.10}~^{+0.53}_{-0.27}~^{+1.66}_{-1.57}$
 &
${98.21}~^{+0.62}_{-0.38}~^{+1.40}_{-1.40}$
 \\
$Z$ &
${29.23}~^{+0.18}_{-0.10}~^{+0.42}_{-0.42}$
 &
${29.08}~^{+0.18}_{-0.10}~^{+0.46}_{-0.46}$
 &
${26.90}~^{+0.15}_{-0.08}~^{+0.58}_{-0.58}$
 &
${28.58}~^{+0.14}_{-0.07}~^{+0.49}_{-0.46}$
 &
${28.71}~^{+0.17}_{-0.09}~^{+0.38}_{-0.38}$
 \\
\hline
\end{tabular}
}
\caption{\small 
  \label{tab:wz}
  The total cross sections for gauge boson production 
  at the LHC ($\sqrt s = 7~{\rm TeV}$) for different PDF sets and to NNLO accuracy 
  The errors shown are the scale uncertainty based 
  on the shifts $\mu=M_{W/Z}/2$ and $\mu = 2M_{W/Z}$ 
  and, respectively, the 1$\sigma$ PDF uncertainty.  
  Numbers are in pb. 
}
\end{center}
\end{table}
\begin{table}[h]
\renewcommand{\arraystretch}{1.3}
\begin{center}
{\small
\begin{tabular}{|l|l|l|l|}
\hline
&ABM11
&ABKM09~\cite{Alekhin:2009ni}
&MSTW~\cite{Martin:2009iq}
\\     
\hline
$W^+$ &
${59.08}~^{+0.30}_{-0.14}~^{+0.87}_{-0.87}$
 &
${58.85}~^{+0.31}_{-0.15}~^{+0.92}_{-0.92}$
 &
${56.77}~^{+0.24}_{-0.08}~^{+1.01}_{-0.94}$
 \\
$W^-$ &
${39.70}~^{+0.22}_{-0.12}~^{+0.64}_{-0.64}$
 &
${39.43}~^{+0.22}_{-0.12}~^{+0.62}_{-0.62}$
 &
${39.61}~^{+0.19}_{-0.08}~^{+0.69}_{-0.66}$
 \\
$W^\pm$ &
${98.77}~^{+0.53}_{-0.25}~^{+1.41}_{-1.41}$
 &
${98.28}~^{+0.53}_{-0.27}~^{+1.51}_{-1.51}$
 &
${96.38}~^{+0.43}_{-0.16}~^{+1.65}_{-1.56}$
 \\
$Z$ &
${28.54}~^{+0.13}_{-0.05}~^{+0.42}_{-0.42}$
 &
${28.44}~^{+0.12}_{-0.06}~^{+0.45}_{-0.45}$
 &
${27.91}~^{+0.09}_{-0.03}~^{+0.50}_{-0.46}$
 \\
\hline
\end{tabular}
}
\caption{\small 
Same as Tab.~\ref{tab:wz} for the PDF sets with $n_f=4$.
}
\label{tab:wznf4}
\end{center}
\end{table}
Comparing the results in Tabs.~\ref{tab:wz} and \ref{tab:wznf4} 
for the PDF sets with $n_f=4$ and $n_f=5$ we observe that the 
numbers for the $n_f=5$ scheme are always larger, the differences being less than 
1$\sigma$ in the PDF uncertainty, though.
These differences, which become successively smaller at higher orders, i.e. as we go from
NLO to NNLO accuracy, originate from changes in the light flavor and the
gluon PDFs when the bottom PDF is generated perturbatively, recall Sec.~\ref{sec:hq-mass}.
In summary, the differences between the results in Tabs.~\ref{tab:wz} and \ref{tab:wznf4} 
for a given PDF set constitute an intrinsic uncertainty of the perturbative prediction.
Comparisons of heavy-flavor PDFs including mass effects have also been studied in~\cite{Gluck:2008gs}.

\begin{figure}[th!]
  \centerline{
    \includegraphics[width=8.0cm]{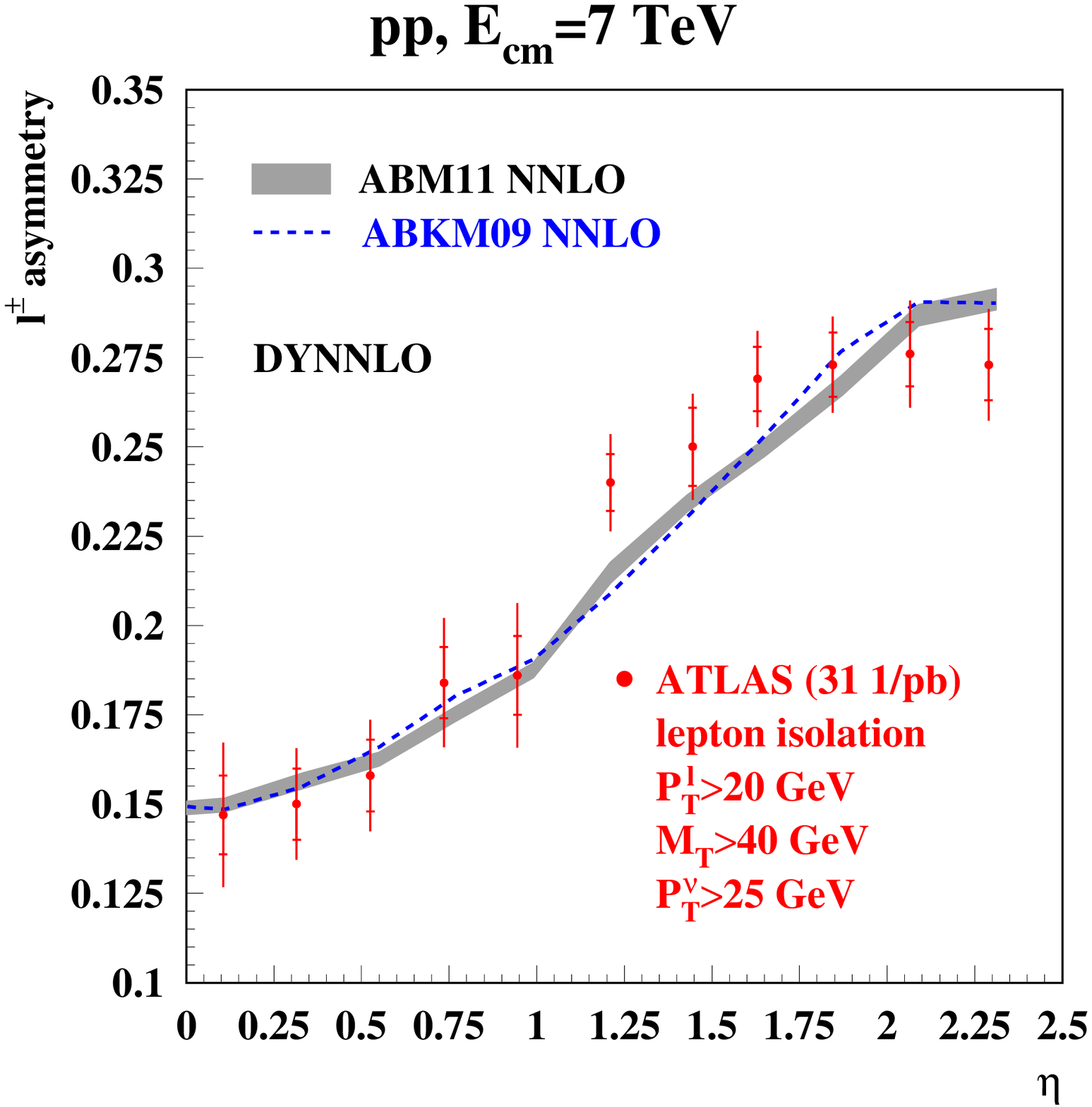}
    \includegraphics[width=8.0cm]{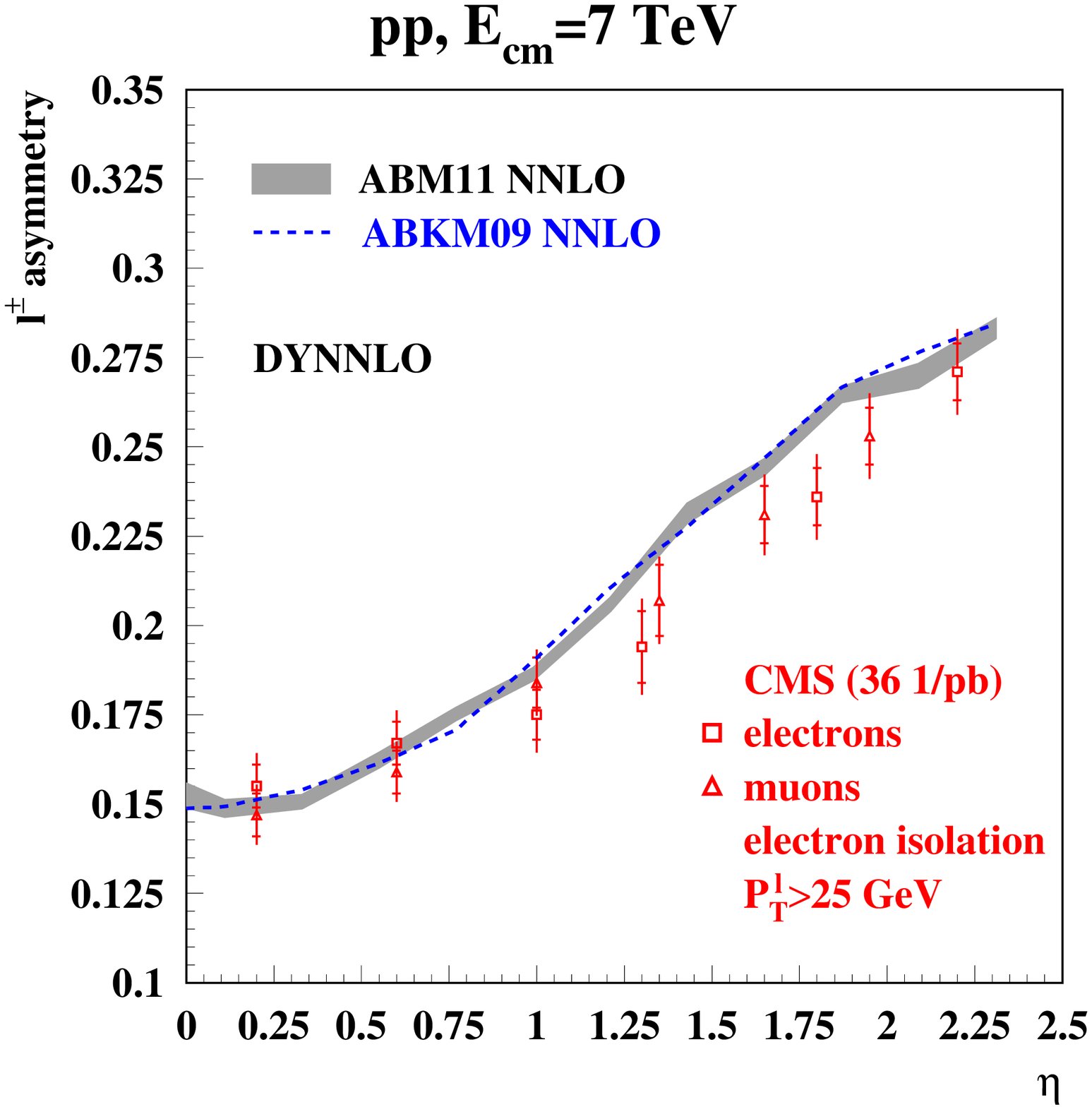}}
  \caption{\small
    \label{fig:lhc}
     The data on charged-lepton asymmetry versus the 
     lepton pseudo-rapidity $\eta$
     obtained by the ATLAS~\cite{Aad:2011yn} (left panel)
     and CMS~\cite{Chatrchyan:2011jz} (right panel) experiments 
     compared to the NNLO predictions based on the 
     {\tt DYNNLO} code~\cite{Catani:2009sm,Catani:2010en} and 
     the ABM11 NNLO PDFs with the shaded area showing the integration uncertainties. 
     The ABKM09 NNLO predictions are given for comparison by dashes, 
     without the integration uncertainties shown. 
  }
\end{figure}

Next, we address the charged-lepton asymmetry data~\cite{Aad:2011yn,Chatrchyan:2011jz} 
as obtained by the ATLAS and CMS experiments and compare it 
to the NNLO predictions based on the ABM11 PDFs in Fig.~\ref{fig:lhc}.  
All differential distributions for $W$- and $Z$-boson production are
computed with the fully exclusive NNLO program {\tt DYNNLO}~\cite{Catani:2009sm,Catani:2010en}, 
which allows to take into account the kinematical cuts imposed in the experiments (cf. Fig.~\ref{fig:lhc}), 
see also~\cite{Gavin:2010az} for an alternative code.

The overall agreement with both experiments is sufficiently good, however 
at values of $\eta\sim 1.5$ for the lepton pseudo-rapidity the data show 
a different trend with respect to the predictions.
Preliminary data on the charge-lepton asymmetry at large
rapidities obtained by the LHCb collaboration~\cite{Amhis:2012gj} are also in 
good agreement with the ABKM09 predictions. To check the impact of the 
LHC charged-lepton asymmetry data on our fit 
we have performed a variant of the ABM11 analysis which consists of 
adding the data of~\cite{Aad:2011yn,Chatrchyan:2011jz}. 
We have found, however, that the change in the PDF central 
values and their errors is only marginal in view of still big uncertainties in the data.  

\subsection{Higgs boson production}
\label{sec:higgsprod}

Let us now discuss the cross sections for the Standard Model Higgs boson
production, where all dominant channels are known to NNLO in QCD.

We start with ggF in Tabs.~\ref{tab:higgsggf}--\ref{tab:higgsggf14}, where 
the cross section is driven by the gluon luminosity and the value of $\alpha_s$ 
from the effective vertex.
The NNLO QCD corrections obtained
in~\cite{Harlander:2002wh,Anastasiou:2002yz,Ravindran:2003um,Ravindran:2004mb} 
still lead to a sizable increase in the cross section at nominal values of the scale, i.e. $\mu=m_H$. 
Further stabilization is achieved beyond NNLO on the basis of soft gluon
resummation, see e.g.,~\cite{Moch:2005ky}.

We observe in Tabs.~\ref{tab:higgsggf}--\ref{tab:higgsggf14} that the ABM11
predictions are rather stable with small changes only due to the gluon 
PDF discussed in Sec.~\ref{sec:results}. 
The values of MSTW are typically larger than those of ABM11 and of ABKM09, 
roughly by ${\cal O}(10\%)$ depending on the Higgs mass and the LHC collision energy,
which has direct consequences for the current Higgs searches at the LHC.
E.g., at $m_H=125~{\rm GeV}$ MSTW predicts an 8\% larger cross section of which 
6.5\% are due to the difference in $\alpha_s$.
In terms of PDF uncertainties, this discrepancy is significant at the level of 3$-$4$\sigma$. 
The reasons for the different gluon luminosities in the relevant $x$-range and
the value of $\alpha_s(M_Z)$ have been illustrated in Sec.~\ref{sec:results}.
The great sensitivity of the ggF rate to constraints from higher orders in QCD 
in the treatment of fixed-target DIS data has already been discussed
extensively in~\cite{Alekhin:2011ey}.

\begin{table}[th!]
\renewcommand{\arraystretch}{1.3}
\begin{center}
{\small
\begin{tabular}{|r|r|r|r|r|r|}
\hline
\multicolumn{1}{|c|}{$m_H$}
&\multicolumn{1}{|c|}{ABM11}
&\multicolumn{1}{|c|}{ABKM09~\cite{Alekhin:2009ni}}
&\multicolumn{1}{|c|}{JR09~\cite{JimenezDelgado:2008hf,JimenezDelgado:2009tv}}
&\multicolumn{1}{|c|}{MSTW~\cite{Martin:2009iq}}
&\multicolumn{1}{|c|}{NN21~\cite{Ball:2011uy}}
\\     
\hline
 100
 &$ 21.31~^{+ 2.20}_{- 2.10}~^{+ 0.46}_{- 0.46}$
 &$ 21.19~^{+ 2.21}_{- 2.11}~^{+ 0.60}_{- 0.60}$
 &$ 20.47~^{+ 1.99}_{- 1.87}~^{+ 0.70}_{- 0.70}$
 &$ 22.95~^{+ 2.50}_{- 2.34}~^{+ 0.25}_{- 0.35}$
 &$ 24.16~^{+ 2.73}_{- 2.49}~^{+ 0.31}_{- 0.31}$
 \\
 110
 &$ 17.43~^{+ 1.79}_{- 1.73}~^{+ 0.38}_{- 0.38}$
 &$ 17.31~^{+ 1.79}_{- 1.72}~^{+ 0.49}_{- 0.49}$
 &$ 16.91~^{+ 1.63}_{- 1.54}~^{+ 0.56}_{- 0.56}$
 &$ 18.84~^{+ 2.03}_{- 1.92}~^{+ 0.20}_{- 0.30}$
 &$ 19.83~^{+ 2.22}_{- 2.02}~^{+ 0.27}_{- 0.27}$
 \\
 115
 &$ 15.85~^{+ 1.62}_{- 1.57}~^{+ 0.34}_{- 0.34}$
 &$ 15.73~^{+ 1.62}_{- 1.57}~^{+ 0.45}_{- 0.45}$
 &$ 15.45~^{+ 1.48}_{- 1.40}~^{+ 0.50}_{- 0.50}$
 &$ 17.17~^{+ 1.85}_{- 1.75}~^{+ 0.20}_{- 0.26}$
 &$ 18.07~^{+ 2.01}_{- 1.84}~^{+ 0.25}_{- 0.25}$
 \\
 120
 &$ 14.46~^{+ 1.48}_{- 1.43}~^{+ 0.32}_{- 0.32}$
 &$ 14.34~^{+ 1.47}_{- 1.43}~^{+ 0.42}_{- 0.42}$
 &$ 14.18~^{+ 1.35}_{- 1.28}~^{+ 0.45}_{- 0.45}$
 &$ 15.70~^{+ 1.68}_{- 1.60}~^{+ 0.17}_{- 0.25}$
 &$ 16.51~^{+ 1.84}_{- 1.67}~^{+ 0.23}_{- 0.23}$
 \\
 125
 &$ 13.23~^{+ 1.35}_{- 1.31}~^{+ 0.30}_{- 0.30}$
 &$ 13.12~^{+ 1.34}_{- 1.31}~^{+ 0.38}_{- 0.38}$
 &$ 13.02~^{+ 1.24}_{- 1.17}~^{+ 0.41}_{- 0.41}$
 &$ 14.39~^{+ 1.54}_{- 1.47}~^{+ 0.17}_{- 0.22}$
 &$ 15.14~^{+ 1.68}_{- 1.53}~^{+ 0.21}_{- 0.21}$
 \\
 130
 &$ 12.14~^{+ 1.23}_{- 1.21}~^{+ 0.28}_{- 0.28}$
 &$ 12.04~^{+ 1.23}_{- 1.20}~^{+ 0.35}_{- 0.35}$
 &$ 12.01~^{+ 1.14}_{- 1.07}~^{+ 0.37}_{- 0.37}$
 &$ 13.24~^{+ 1.40}_{- 1.35}~^{+ 0.15}_{- 0.21}$
 &$ 13.92~^{+ 1.54}_{- 1.40}~^{+ 0.20}_{- 0.20}$
 \\
 140
 &$ 10.30~^{+ 1.04}_{- 1.03}~^{+ 0.24}_{- 0.24}$
 &$ 10.21~^{+ 1.03}_{- 1.02}~^{+ 0.31}_{- 0.31}$
 &$ 10.28~^{+ 0.97}_{- 0.92}~^{+ 0.32}_{- 0.32}$
 &$ 11.28~^{+ 1.19}_{- 1.16}~^{+ 0.14}_{- 0.18}$
 &$ 11.86~^{+ 1.30}_{- 1.19}~^{+ 0.18}_{- 0.18}$
 \\
 150
 &$  8.83~^{+ 0.89}_{- 0.88}~^{+ 0.21}_{- 0.21}$
 &$  8.75~^{+ 0.88}_{- 0.87}~^{+ 0.27}_{- 0.27}$
 &$  8.90~^{+ 0.83}_{- 0.79}~^{+ 0.27}_{- 0.27}$
 &$  9.71~^{+ 1.02}_{- 0.99}~^{+ 0.13}_{- 0.16}$
 &$ 10.21~^{+ 1.11}_{- 1.02}~^{+ 0.16}_{- 0.16}$
 \\
 160
 &$  7.63~^{+ 0.76}_{- 0.77}~^{+ 0.19}_{- 0.19}$
 &$  7.57~^{+ 0.76}_{- 0.76}~^{+ 0.24}_{- 0.24}$
 &$  7.76~^{+ 0.72}_{- 0.69}~^{+ 0.24}_{- 0.24}$
 &$  8.44~^{+ 0.88}_{- 0.86}~^{+ 0.11}_{- 0.14}$
 &$  8.86~^{+ 0.96}_{- 0.88}~^{+ 0.14}_{- 0.14}$
 \\
 180
 &$  5.84~^{+ 0.58}_{- 0.59}~^{+ 0.15}_{- 0.15}$
 &$  5.79~^{+ 0.57}_{- 0.58}~^{+ 0.19}_{- 0.19}$
 &$  6.04~^{+ 0.56}_{- 0.53}~^{+ 0.19}_{- 0.19}$
 &$  6.51~^{+ 0.67}_{- 0.67}~^{+ 0.10}_{- 0.12}$
 &$  6.83~^{+ 0.73}_{- 0.67}~^{+ 0.12}_{- 0.12}$
 \\
 200
 &$  4.58~^{+ 0.45}_{- 0.46}~^{+ 0.13}_{- 0.13}$
 &$  4.55~^{+ 0.45}_{- 0.46}~^{+ 0.16}_{- 0.16}$
 &$  4.83~^{+ 0.44}_{- 0.42}~^{+ 0.16}_{- 0.16}$
 &$  5.17~^{+ 0.53}_{- 0.53}~^{+ 0.09}_{- 0.10}$
 &$  5.42~^{+ 0.57}_{- 0.54}~^{+ 0.10}_{- 0.10}$
 \\
 220
 &$  3.69~^{+ 0.36}_{- 0.37}~^{+ 0.11}_{- 0.11}$
 &$  3.67~^{+ 0.36}_{- 0.37}~^{+ 0.14}_{- 0.14}$
 &$  3.97~^{+ 0.36}_{- 0.35}~^{+ 0.14}_{- 0.14}$
 &$  4.20~^{+ 0.43}_{- 0.43}~^{+ 0.07}_{- 0.08}$
 &$  4.41~^{+ 0.45}_{- 0.44}~^{+ 0.08}_{- 0.08}$
 \\
 260
 &$  2.55~^{+ 0.25}_{- 0.26}~^{+ 0.08}_{- 0.08}$
 &$  2.55~^{+ 0.25}_{- 0.26}~^{+ 0.10}_{- 0.10}$
 &$  2.85~^{+ 0.25}_{- 0.25}~^{+ 0.12}_{- 0.12}$
 &$  2.97~^{+ 0.30}_{- 0.31}~^{+ 0.06}_{- 0.07}$
 &$  3.11~^{+ 0.31}_{- 0.31}~^{+ 0.07}_{- 0.07}$
 \\
 300
 &$  1.93~^{+ 0.19}_{- 0.20}~^{+ 0.07}_{- 0.07}$
 &$  1.95~^{+ 0.19}_{- 0.20}~^{+ 0.09}_{- 0.09}$
 &$  2.23~^{+ 0.19}_{- 0.20}~^{+ 0.11}_{- 0.11}$
 &$  2.30~^{+ 0.23}_{- 0.24}~^{+ 0.05}_{- 0.06}$
 &$  2.41~^{+ 0.23}_{- 0.24}~^{+ 0.06}_{- 0.06}$
 \\
\hline
\end{tabular}
}
\caption{\small 
The total cross sections for Higgs production in ggF 
at the LHC ($\sqrt s = 7~{\rm TeV}$) for different PDF sets and to NNLO accuracy. 
The errors shown are the scale uncertainty based 
on the shifts $\mu=m_H/2$ and $\mu = 2m_H$ 
and, respectively, the 1$\sigma$ PDF uncertainty.  
Numbers are in pb. 
}
\label{tab:higgsggf}
\end{center}
\end{table}

\begin{table}[h!]
\renewcommand{\arraystretch}{1.3}
\begin{center}
{\small
\begin{tabular}{|r|r|r|r|r|r|}
\hline
\multicolumn{1}{|c|}{$m_H$}
&\multicolumn{1}{|c|}{ABM11}
&\multicolumn{1}{|c|}{ABKM09~\cite{Alekhin:2009ni}}
&\multicolumn{1}{|c|}{JR09~\cite{JimenezDelgado:2008hf,JimenezDelgado:2009tv}}
&\multicolumn{1}{|c|}{MSTW~\cite{Martin:2009iq}}
&\multicolumn{1}{|c|}{NN21~\cite{Ball:2011uy}}
\\     
\hline
 100
 &$ 26.91~^{+ 2.72}_{- 2.57}~^{+ 0.56}_{- 0.56}$
 &$ 26.82~^{+ 2.73}_{- 2.59}~^{+ 0.75}_{- 0.75}$
 &$ 25.64~^{+ 2.44}_{- 2.28}~^{+ 0.91}_{- 0.91}$
 &$ 28.86~^{+ 3.08}_{- 2.85}~^{+ 0.31}_{- 0.44}$
 &$ 30.35~^{+ 3.36}_{- 3.03}~^{+ 0.37}_{- 0.37}$
 \\
 110
 &$ 22.17~^{+ 2.21}_{- 2.13}~^{+ 0.46}_{- 0.46}$
 &$ 22.05~^{+ 2.22}_{- 2.13}~^{+ 0.61}_{- 0.61}$
 &$ 21.31~^{+ 2.01}_{- 1.88}~^{+ 0.72}_{- 0.72}$
 &$ 23.84~^{+ 2.51}_{- 2.36}~^{+ 0.25}_{- 0.37}$
 &$ 25.07~^{+ 2.74}_{- 2.48}~^{+ 0.32}_{- 0.32}$
 \\
 115
 &$ 20.22~^{+ 2.02}_{- 1.94}~^{+ 0.42}_{- 0.42}$
 &$ 20.10~^{+ 2.02}_{- 1.94}~^{+ 0.56}_{- 0.56}$
 &$ 19.52~^{+ 1.83}_{- 1.71}~^{+ 0.65}_{- 0.65}$
 &$ 21.78~^{+ 2.29}_{- 2.15}~^{+ 0.24}_{- 0.33}$
 &$ 22.91~^{+ 2.49}_{- 2.26}~^{+ 0.29}_{- 0.29}$
 \\
 120
 &$ 18.51~^{+ 1.84}_{- 1.78}~^{+ 0.39}_{- 0.39}$
 &$ 18.39~^{+ 1.84}_{- 1.78}~^{+ 0.51}_{- 0.51}$
 &$ 17.96~^{+ 1.68}_{- 1.58}~^{+ 0.59}_{- 0.59}$
 &$ 19.97~^{+ 2.09}_{- 1.98}~^{+ 0.21}_{- 0.31}$
 &$ 21.00~^{+ 2.28}_{- 2.06}~^{+ 0.28}_{- 0.28}$
 \\
 125
 &$ 16.99~^{+ 1.69}_{- 1.63}~^{+ 0.37}_{- 0.37}$
 &$ 16.87~^{+ 1.68}_{- 1.63}~^{+ 0.47}_{- 0.47}$
 &$ 16.53~^{+ 1.54}_{- 1.44}~^{+ 0.53}_{- 0.53}$
 &$ 18.36~^{+ 1.92}_{- 1.82}~^{+ 0.21}_{- 0.28}$
 &$ 19.30~^{+ 2.09}_{- 1.89}~^{+ 0.26}_{- 0.26}$
 \\
 130
 &$ 15.64~^{+ 1.55}_{- 1.51}~^{+ 0.34}_{- 0.34}$
 &$ 15.52~^{+ 1.54}_{- 1.50}~^{+ 0.44}_{- 0.44}$
 &$ 15.29~^{+ 1.42}_{- 1.33}~^{+ 0.48}_{- 0.48}$
 &$ 16.94~^{+ 1.76}_{- 1.68}~^{+ 0.18}_{- 0.27}$
 &$ 17.80~^{+ 1.92}_{- 1.74}~^{+ 0.24}_{- 0.24}$
 \\
 140
 &$ 13.36~^{+ 1.31}_{- 1.29}~^{+ 0.30}_{- 0.30}$
 &$ 13.25~^{+ 1.31}_{- 1.29}~^{+ 0.38}_{- 0.38}$
 &$ 13.16~^{+ 1.21}_{- 1.14}~^{+ 0.41}_{- 0.41}$
 &$ 14.52~^{+ 1.49}_{- 1.44}~^{+ 0.16}_{- 0.23}$
 &$ 15.25~^{+ 1.63}_{- 1.48}~^{+ 0.21}_{- 0.21}$
 \\
 150
 &$ 11.51~^{+ 1.13}_{- 1.12}~^{+ 0.26}_{- 0.26}$
 &$ 11.42~^{+ 1.12}_{- 1.11}~^{+ 0.34}_{- 0.34}$
 &$ 11.45~^{+ 1.05}_{- 0.99}~^{+ 0.35}_{- 0.35}$
 &$ 12.56~^{+ 1.29}_{- 1.25}~^{+ 0.15}_{- 0.20}$
 &$ 13.19~^{+ 1.40}_{- 1.28}~^{+ 0.19}_{- 0.19}$
 \\
 160
 &$ 10.01~^{+ 0.98}_{- 0.97}~^{+ 0.23}_{- 0.23}$
 &$  9.92~^{+ 0.97}_{- 0.97}~^{+ 0.29}_{- 0.29}$
 &$ 10.03~^{+ 0.91}_{- 0.86}~^{+ 0.30}_{- 0.30}$
 &$ 10.96~^{+ 1.12}_{- 1.09}~^{+ 0.14}_{- 0.18}$
 &$ 11.51~^{+ 1.22}_{- 1.10}~^{+ 0.17}_{- 0.17}$
 \\
 180
 &$  7.74~^{+ 0.75}_{- 0.76}~^{+ 0.19}_{- 0.19}$
 &$  7.67~^{+ 0.74}_{- 0.75}~^{+ 0.24}_{- 0.24}$
 &$  7.88~^{+ 0.71}_{- 0.67}~^{+ 0.24}_{- 0.24}$
 &$  8.55~^{+ 0.86}_{- 0.85}~^{+ 0.11}_{- 0.14}$
 &$  8.96~^{+ 0.94}_{- 0.85}~^{+ 0.14}_{- 0.14}$
 \\
 200
 &$  6.15~^{+ 0.59}_{- 0.60}~^{+ 0.16}_{- 0.16}$
 &$  6.10~^{+ 0.59}_{- 0.60}~^{+ 0.20}_{- 0.20}$
 &$  6.36~^{+ 0.56}_{- 0.54}~^{+ 0.20}_{- 0.20}$
 &$  6.84~^{+ 0.68}_{- 0.68}~^{+ 0.10}_{- 0.12}$
 &$  7.18~^{+ 0.73}_{- 0.69}~^{+ 0.12}_{- 0.12}$
 \\
 220
 &$  5.00~^{+ 0.48}_{- 0.49}~^{+ 0.14}_{- 0.14}$
 &$  4.96~^{+ 0.47}_{- 0.48}~^{+ 0.17}_{- 0.17}$
 &$  5.26~^{+ 0.46}_{- 0.45}~^{+ 0.17}_{- 0.17}$
 &$  5.61~^{+ 0.56}_{- 0.56}~^{+ 0.09}_{- 0.10}$
 &$  5.89~^{+ 0.59}_{- 0.57}~^{+ 0.10}_{- 0.10}$
 \\
 260
 &$  3.53~^{+ 0.34}_{- 0.35}~^{+ 0.11}_{- 0.11}$
 &$  3.52~^{+ 0.33}_{- 0.35}~^{+ 0.13}_{- 0.13}$
 &$  3.84~^{+ 0.32}_{- 0.33}~^{+ 0.14}_{- 0.14}$
 &$  4.04~^{+ 0.40}_{- 0.41}~^{+ 0.07}_{- 0.08}$
 &$  4.23~^{+ 0.41}_{- 0.41}~^{+ 0.08}_{- 0.08}$
 \\
 300
 &$  2.72~^{+ 0.26}_{- 0.27}~^{+ 0.09}_{- 0.09}$
 &$  2.73~^{+ 0.26}_{- 0.27}~^{+ 0.11}_{- 0.11}$
 &$  3.06~^{+ 0.25}_{- 0.27}~^{+ 0.13}_{- 0.13}$
 &$  3.18~^{+ 0.31}_{- 0.32}~^{+ 0.07}_{- 0.07}$
 &$  3.33~^{+ 0.31}_{- 0.33}~^{+ 0.07}_{- 0.07}$
 \\
\hline
\end{tabular}
}
\caption{\small 
Same as Tab.~\ref{tab:higgsggf} for the LHC at $\sqrt s = 8~{\rm TeV}$.
}
\label{tab:higgsggf8}
\end{center}
\end{table}

\begin{table}[h!]
\renewcommand{\arraystretch}{1.3}
\begin{center}
{\small
\begin{tabular}{|r|r|r|r|r|r|}
\hline
\multicolumn{1}{|c|}{$m_H$}
&\multicolumn{1}{|c|}{ABM11}
&\multicolumn{1}{|c|}{ABKM09~\cite{Alekhin:2009ni}}
&\multicolumn{1}{|c|}{JR09~\cite{JimenezDelgado:2008hf,JimenezDelgado:2009tv}}
&\multicolumn{1}{|c|}{MSTW~\cite{Martin:2009iq}}
&\multicolumn{1}{|c|}{NN21~\cite{Ball:2011uy}}
\\     
\hline
 100
 &$ 66.79~^{+ 6.13}_{- 5.63}~^{+ 1.31}_{- 1.31}$
 &$ 67.29~^{+ 6.28}_{- 5.78}~^{+ 1.80}_{- 1.80}$
 &$ 62.23~^{+ 5.46}_{- 4.92}~^{+ 2.62}_{- 2.62}$
 &$ 70.76~^{+ 6.91}_{- 6.23}~^{+ 0.80}_{- 1.12}$
 &$ 74.18~^{+ 7.50}_{- 6.54}~^{+ 0.78}_{- 0.78}$
 \\
 110
 &$ 56.35~^{+ 5.11}_{- 4.77}~^{+ 1.06}_{- 1.06}$
 &$ 56.62~^{+ 5.21}_{- 4.88}~^{+ 1.47}_{- 1.47}$
 &$ 52.77~^{+ 4.58}_{- 4.14}~^{+ 2.11}_{- 2.11}$
 &$ 59.75~^{+ 5.76}_{- 5.27}~^{+ 0.62}_{- 0.95}$
 &$ 62.63~^{+ 6.26}_{- 5.47}~^{+ 0.66}_{- 0.66}$
 \\
 115
 &$ 52.01~^{+ 4.70}_{- 4.40}~^{+ 0.99}_{- 0.99}$
 &$ 52.20~^{+ 4.79}_{- 4.49}~^{+ 1.35}_{- 1.35}$
 &$ 48.82~^{+ 4.21}_{- 3.80}~^{+ 1.92}_{- 1.92}$
 &$ 55.17~^{+ 5.31}_{- 4.85}~^{+ 0.60}_{- 0.84}$
 &$ 57.86~^{+ 5.74}_{- 5.04}~^{+ 0.62}_{- 0.62}$
 \\
 120
 &$ 48.14~^{+ 4.35}_{- 4.07}~^{+ 0.93}_{- 0.93}$
 &$ 48.26~^{+ 4.42}_{- 4.14}~^{+ 1.25}_{- 1.25}$
 &$ 45.33~^{+ 3.89}_{- 3.53}~^{+ 1.75}_{- 1.75}$
 &$ 51.12~^{+ 4.89}_{- 4.50}~^{+ 0.50}_{- 0.80}$
 &$ 53.58~^{+ 5.31}_{- 4.64}~^{+ 0.58}_{- 0.58}$
 \\
 125
 &$ 44.68~^{+ 4.02}_{- 3.78}~^{+ 0.85}_{- 0.85}$
 &$ 44.75~^{+ 4.07}_{- 3.85}~^{+ 1.16}_{- 1.16}$
 &$ 42.13~^{+ 3.60}_{- 3.26}~^{+ 1.59}_{- 1.59}$
 &$ 47.47~^{+ 4.52}_{- 4.18}~^{+ 0.50}_{- 0.71}$
 &$ 49.77~^{+ 4.91}_{- 4.30}~^{+ 0.54}_{- 0.54}$
 \\
 130
 &$ 41.59~^{+ 3.72}_{- 3.53}~^{+ 0.80}_{- 0.80}$
 &$ 41.61~^{+ 3.77}_{- 3.58}~^{+ 1.07}_{- 1.07}$
 &$ 39.32~^{+ 3.35}_{- 3.02}~^{+ 1.45}_{- 1.45}$
 &$ 44.22~^{+ 4.18}_{- 3.91}~^{+ 0.42}_{- 0.70}$
 &$ 46.35~^{+ 4.55}_{- 3.99}~^{+ 0.51}_{- 0.51}$
 \\
 140
 &$ 36.28~^{+ 3.23}_{- 3.09}~^{+ 0.70}_{- 0.70}$
 &$ 36.24~^{+ 3.26}_{- 3.12}~^{+ 0.94}_{- 0.94}$
 &$ 34.46~^{+ 2.90}_{- 2.65}~^{+ 1.22}_{- 1.22}$
 &$ 38.63~^{+ 3.63}_{- 3.41}~^{+ 0.37}_{- 0.59}$
 &$ 40.49~^{+ 3.94}_{- 3.47}~^{+ 0.45}_{- 0.45}$
 \\
 150
 &$ 31.92~^{+ 2.82}_{- 2.72}~^{+ 0.61}_{- 0.61}$
 &$ 31.85~^{+ 2.85}_{- 2.74}~^{+ 0.82}_{- 0.82}$
 &$ 30.49~^{+ 2.56}_{- 2.32}~^{+ 1.04}_{- 1.04}$
 &$ 34.05~^{+ 3.18}_{- 3.01}~^{+ 0.33}_{- 0.51}$
 &$ 35.69~^{+ 3.45}_{- 3.04}~^{+ 0.41}_{- 0.41}$
 \\
 160
 &$ 28.32~^{+ 2.49}_{- 2.42}~^{+ 0.54}_{- 0.54}$
 &$ 28.22~^{+ 2.50}_{- 2.44}~^{+ 0.72}_{- 0.72}$
 &$ 27.16~^{+ 2.25}_{- 2.06}~^{+ 0.90}_{- 0.90}$
 &$ 30.25~^{+ 2.80}_{- 2.67}~^{+ 0.30}_{- 0.44}$
 &$ 31.69~^{+ 3.04}_{- 2.67}~^{+ 0.37}_{- 0.37}$
 \\
 180
 &$ 22.75~^{+ 1.98}_{- 1.96}~^{+ 0.44}_{- 0.44}$
 &$ 22.62~^{+ 1.99}_{- 1.95}~^{+ 0.59}_{- 0.59}$
 &$ 22.03~^{+ 1.80}_{- 1.65}~^{+ 0.69}_{- 0.69}$
 &$ 24.39~^{+ 2.23}_{- 2.16}~^{+ 0.24}_{- 0.36}$
 &$ 25.56~^{+ 2.41}_{- 2.15}~^{+ 0.31}_{- 0.31}$
 \\
 200
 &$ 18.74~^{+ 1.62}_{- 1.61}~^{+ 0.37}_{- 0.37}$
 &$ 18.61~^{+ 1.62}_{- 1.61}~^{+ 0.49}_{- 0.49}$
 &$ 18.33~^{+ 1.47}_{- 1.38}~^{+ 0.54}_{- 0.54}$
 &$ 20.17~^{+ 1.83}_{- 1.79}~^{+ 0.21}_{- 0.29}$
 &$ 21.16~^{+ 1.94}_{- 1.79}~^{+ 0.27}_{- 0.27}$
 \\
 220
 &$ 15.78~^{+ 1.36}_{- 1.36}~^{+ 0.32}_{- 0.32}$
 &$ 15.66~^{+ 1.35}_{- 1.35}~^{+ 0.42}_{- 0.42}$
 &$ 15.61~^{+ 1.23}_{- 1.17}~^{+ 0.45}_{- 0.45}$
 &$ 17.05~^{+ 1.53}_{- 1.51}~^{+ 0.18}_{- 0.25}$
 &$ 17.90~^{+ 1.61}_{- 1.52}~^{+ 0.24}_{- 0.24}$
 \\
 260
 &$ 11.91~^{+ 1.01}_{- 1.04}~^{+ 0.26}_{- 0.26}$
 &$ 11.81~^{+ 1.00}_{- 1.03}~^{+ 0.33}_{- 0.33}$
 &$ 12.03~^{+ 0.91}_{- 0.91}~^{+ 0.34}_{- 0.34}$
 &$ 12.98~^{+ 1.15}_{- 1.15}~^{+ 0.16}_{- 0.20}$
 &$ 13.64~^{+ 1.18}_{- 1.16}~^{+ 0.19}_{- 0.19}$
 \\
 300
 &$  9.80~^{+ 0.83}_{- 0.86}~^{+ 0.23}_{- 0.23}$
 &$  9.73~^{+ 0.82}_{- 0.85}~^{+ 0.28}_{- 0.28}$
 &$ 10.11~^{+ 0.75}_{- 0.77}~^{+ 0.29}_{- 0.29}$
 &$ 10.79~^{+ 0.95}_{- 0.96}~^{+ 0.14}_{- 0.17}$
 &$ 11.34~^{+ 0.95}_{- 0.97}~^{+ 0.17}_{- 0.17}$
 \\
\hline
\end{tabular}
}
\caption{\small 
Same as Tab.~\ref{tab:higgsggf} for the LHC at $\sqrt s = 14~{\rm TeV}$.
}
\label{tab:higgsggf14}
\end{center}
\end{table}

\begin{table}[h!]
\renewcommand{\arraystretch}{1.3}
\begin{center}
{\small
\begin{tabular}{|r|r|r|r|r|r|}
\hline
\multicolumn{1}{|c|}{$m_H$}
&\multicolumn{1}{|c|}{ABM11}
&\multicolumn{1}{|c|}{ABKM09~\cite{Alekhin:2009ni}}
&\multicolumn{1}{|c|}{JR09~\cite{JimenezDelgado:2008hf,JimenezDelgado:2009tv}}
&\multicolumn{1}{|c|}{MSTW~\cite{Martin:2009iq}}
&\multicolumn{1}{|c|}{NN21~\cite{Ball:2011uy}}
\\     
\hline
 100
 &$\!1.673~^{+0.024}_{-0.022}~^{+0.020}_{-0.020}\!$
 &$\!1.643~^{+0.022}_{-0.026}~^{+0.012}_{-0.012}\!$
 &$\!1.599~^{+0.023}_{-0.017}~^{+0.022}_{-0.022}\!$
 &$\!1.616~^{+0.025}_{-0.034}~^{+0.029}_{-0.029}\!$
 &$\!1.603~^{+0.028}_{-0.028}~^{+0.021}_{-0.021}\!$
 \\
 110
 &$\!1.513~^{+0.025}_{-0.019}~^{+0.018}_{-0.018}\!$
 &$\!1.483~^{+0.026}_{-0.022}~^{+0.011}_{-0.011}\!$
 &$\!1.453~^{+0.014}_{-0.020}~^{+0.021}_{-0.021}\!$
 &$\!1.460~^{+0.025}_{-0.027}~^{+0.026}_{-0.026}\!$
 &$\!1.448~^{+0.027}_{-0.023}~^{+0.019}_{-0.019}\!$
 \\
 115
 &$\!1.440~^{+0.021}_{-0.021}~^{+0.017}_{-0.017}\!$
 &$\!1.411~^{+0.022}_{-0.014}~^{+0.011}_{-0.011}\!$
 &$\!1.388~^{+0.016}_{-0.022}~^{+0.020}_{-0.020}\!$
 &$\!1.391~^{+0.026}_{-0.028}~^{+0.025}_{-0.025}\!$
 &$\!1.378~^{+0.029}_{-0.024}~^{+0.018}_{-0.018}\!$
 \\
 120
 &$\!1.373~^{+0.022}_{-0.018}~^{+0.016}_{-0.016}\!$
 &$\!1.345~^{+0.022}_{-0.016}~^{+0.010}_{-0.010}\!$
 &$\!1.321~^{+0.018}_{-0.018}~^{+0.019}_{-0.019}\!$
 &$\!1.324~^{+0.025}_{-0.023}~^{+0.024}_{-0.024}\!$
 &$\!1.318~^{+0.021}_{-0.028}~^{+0.017}_{-0.017}\!$
 \\
 125
 &$\!1.307~^{+0.023}_{-0.018}~^{+0.016}_{-0.016}\!$
 &$\!1.285~^{+0.020}_{-0.022}~^{+0.010}_{-0.010}\!$
 &$\!1.260~^{+0.017}_{-0.020}~^{+0.019}_{-0.019}\!$
 &$\!1.264~^{+0.023}_{-0.023}~^{+0.023}_{-0.023}\!$
 &$\!1.252~^{+0.026}_{-0.019}~^{+0.016}_{-0.016}\!$
 \\
 130
 &$\!1.244~^{+0.025}_{-0.014}~^{+0.015}_{-0.015}\!$
 &$\!1.223~^{+0.022}_{-0.013}~^{+0.009}_{-0.009}\!$
 &$\!1.203~^{+0.017}_{-0.017}~^{+0.018}_{-0.018}\!$
 &$\!1.203~^{+0.026}_{-0.021}~^{+0.022}_{-0.022}\!$
 &$\!1.195~^{+0.027}_{-0.018}~^{+0.016}_{-0.016}\!$
 \\
 140
 &$\!1.137~^{+0.016}_{-0.015}~^{+0.014}_{-0.014}\!$
 &$\!1.116~^{+0.021}_{-0.013}~^{+0.008}_{-0.008}\!$
 &$\!1.098~^{+0.018}_{-0.016}~^{+0.017}_{-0.017}\!$
 &$\!1.099~^{+0.024}_{-0.019}~^{+0.020}_{-0.020}\!$
 &$\!1.093~^{+0.021}_{-0.020}~^{+0.015}_{-0.015}\!$
 \\
 150
 &$\!1.038~^{+0.019}_{-0.013}~^{+0.013}_{-0.013}\!$
 &$\!1.018~^{+0.021}_{-0.011}~^{+0.008}_{-0.008}\!$
 &$\!1.006~^{+0.015}_{-0.014}~^{+0.016}_{-0.016}\!$
 &$\!1.003~^{+0.021}_{-0.016}~^{+0.019}_{-0.019}\!$
 &$\!0.999~^{+0.020}_{-0.019}~^{+0.013}_{-0.013}\!$
 \\
 160
 &$\!0.950~^{+0.018}_{-0.011}~^{+0.012}_{-0.012}\!$
 &$\!0.934~^{+0.017}_{-0.012}~^{+0.007}_{-0.007}\!$
 &$\!0.923~^{+0.016}_{-0.014}~^{+0.015}_{-0.015}\!$
 &$\!0.918~^{+0.018}_{-0.013}~^{+0.017}_{-0.017}\!$
 &$\!0.915~^{+0.017}_{-0.014}~^{+0.012}_{-0.012}\!$
 \\
 180
 &$\!0.800~^{+0.016}_{-0.009}~^{+0.010}_{-0.010}\!$
 &$\!0.785~^{+0.016}_{-0.007}~^{+0.006}_{-0.006}\!$
 &$\!0.781~^{+0.013}_{-0.010}~^{+0.014}_{-0.014}\!$
 &$\!0.773~^{+0.019}_{-0.011}~^{+0.015}_{-0.015}\!$
 &$\!0.771~^{+0.018}_{-0.011}~^{+0.011}_{-0.011}\!$
 \\
 200
 &$\!0.679~^{+0.013}_{-0.008}~^{+0.009}_{-0.009}\!$
 &$\!0.669~^{+0.012}_{-0.007}~^{+0.005}_{-0.005}\!$
 &$\!0.665~^{+0.011}_{-0.008}~^{+0.012}_{-0.012}\!$
 &$\!0.658~^{+0.014}_{-0.010}~^{+0.013}_{-0.013}\!$
 &$\!0.656~^{+0.016}_{-0.009}~^{+0.009}_{-0.009}\!$
 \\
 220
 &$\!0.580~^{+0.011}_{-0.006}~^{+0.008}_{-0.008}\!$
 &$\!0.571~^{+0.012}_{-0.005}~^{+0.005}_{-0.005}\!$
 &$\!0.570~^{+0.010}_{-0.006}~^{+0.011}_{-0.011}\!$
 &$\!0.562~^{+0.013}_{-0.007}~^{+0.011}_{-0.011}\!$
 &$\!0.561~^{+0.012}_{-0.007}~^{+0.008}_{-0.008}\!$
 \\
 260
 &$\!0.429~^{+0.010}_{-0.003}~^{+0.006}_{-0.006}\!$
 &$\!0.424~^{+0.009}_{-0.003}~^{+0.004}_{-0.004}\!$
 &$\!0.425~^{+0.008}_{-0.004}~^{+0.009}_{-0.009}\!$
 &$\!0.417~^{+0.010}_{-0.005}~^{+0.008}_{-0.008}\!$
 &$\!0.418~^{+0.010}_{-0.005}~^{+0.006}_{-0.006}\!$
 \\
 300
 &$\!0.324~^{+0.008}_{-0.002}~^{+0.004}_{-0.004}\!$
 &$\!0.321~^{+0.007}_{-0.003}~^{+0.003}_{-0.003}\!$
 &$\!0.323~^{+0.006}_{-0.003}~^{+0.007}_{-0.007}\!$
 &$\!0.316~^{+0.008}_{-0.003}~^{+0.007}_{-0.007}\!$
 &$\!0.316~^{+0.007}_{-0.003}~^{+0.005}_{-0.005}\!$
 \\
\hline
\end{tabular}
}
\caption{\small 
The total VBF cross sections 
at the LHC ($\sqrt s = 7~{\rm TeV}$)
for different PDF sets and to NNLO accuracy 
as computed with {\tt VBFNNLO}~\cite{Bolzoni:2010xr,Bolzoni:2011cu}.
Errors shown are the scale uncertainties evaluated by varying $\mu_r$ and $\mu_f$ in the
interval $\mu_r,\mu_f \in [Q/4,4Q]$ and, respectively, the PDF uncertainties.
Numbers are in pb.
}
\label{tab:higgsvbf}
\end{center}
\end{table}

Next in size comes the VBF channel. All numbers in Tab.~\ref{tab:higgsvbf} are 
computed with the {\tt VBFNNLO} program~\cite{Bolzoni:2010xr,Bolzoni:2011cu} 
in the structure function approach, which describes VBF as a double DIS process, 
where two (virtual) vector-bosons $V_i$ (independently) emitted from
the hadronic initial states fuse into a Higgs boson. 
Although the structure function approach to VBF is not truly exact at NNLO it includes the bulk 
of the radiative corrections so that the remaining contributions, 
are both, parametrically small and kinematically suppressed, see~\cite{Bolzoni:2011cu}. 
The residual theory uncertainty based on the scale variation is rather small
and the cross sections for all PDF sets considered in Tab.~\ref{tab:higgsvbf}
agree well, typically within 1$-$2$\sigma$ of the PDF uncertainties.

\begin{table}[th!]
\renewcommand{\arraystretch}{1.3}
\begin{center}
{\small
\begin{tabular}{|l|l|l|l|l|l|}
\hline
\multicolumn{1}{|c|}{$m_H$}
&\multicolumn{1}{|c|}{ABM11}
&\multicolumn{1}{|c|}{ABKM09~\cite{Alekhin:2009ni}}
&\multicolumn{1}{|c|}{JR09~\cite{JimenezDelgado:2008hf,JimenezDelgado:2009tv}}
&\multicolumn{1}{|c|}{MSTW~\cite{Martin:2009iq}}
&\multicolumn{1}{|c|}{NN21~\cite{Ball:2011uy}}
\\     
\hline
100
 &
${1298}~^{+14}_{-37}~^{+17}_{-17}$
 &
${1270}~^{+14}_{-37}~^{+14}_{-14}$
 &
${1229}~^{+17}_{-35}~^{+17}_{-17}$
 &
${1256}~^{+14}_{-36}~^{+22}_{-18}$
 &
${1273}~^{+19}_{-40}~^{+20}_{-20}$
 \\
110
 &\phantom{1}
${960}~^{+10}_{-27}~^{+13}_{-13}$
 &\phantom{1}
${938}~^{+10}_{-27}~^{+10}_{-10}$
 &\phantom{1}
${911}~^{+12}_{-26}~^{+12}_{-12}$
 &\phantom{1}
${930}~^{+10}_{-27}~^{+16}_{-13}$
 &\phantom{1}
${942}~^{+14}_{-30}~^{+15}_{-15}$
 \\
115
 &\phantom{1}
${832}~^{+9}_{-24}~^{+11}_{-11}$
 &\phantom{1}
${812}~^{+9}_{-23}~^{+9}_{-9}$
 &\phantom{1}
${790}~^{+11}_{-23}~^{+10}_{-10}$
 &\phantom{1}
${805}~^{+8}_{-23}~^{+14}_{-12}$
 &\phantom{1}
${816}~^{+12}_{-26}~^{+13}_{-13}$
 \\
120
 &\phantom{1}
${723}~^{+7}_{-20}~^{+10}_{-10}$
 &\phantom{1}
${706}~^{+7}_{-20}~^{+8}_{-8}$
 &\phantom{1}
${687}~^{+9}_{-20}~^{+9}_{-9}$
 &\phantom{1}
${700}~^{+7}_{-20}~^{+12}_{-10}$
 &\phantom{1}
${709}~^{+10}_{-22}~^{+11}_{-11}$
 \\
125
 &\phantom{1}
${631}~^{+6}_{-18}~^{+9}_{-9}$
 &\phantom{1}
${616}~^{+6}_{-18}~^{+7}_{-7}$
 &\phantom{1}
${600}~^{+8}_{-17}~^{+8}_{-8}$
 &\phantom{1}
${611}~^{+6}_{-18}~^{+11}_{-9}$
 &\phantom{1}
${619}~^{+9}_{-19}~^{+10}_{-10}$
 \\
130
 &\phantom{1}
${553}~^{+5}_{-16}~^{+8}_{-8}$
 &\phantom{1}
${539}~^{+5}_{-15}~^{+6}_{-6}$
 &\phantom{1}
${527}~^{+6}_{-16}~^{+5}_{-5}$
 &\phantom{1}
${536}~^{+5}_{-16}~^{+9}_{-8}$
 &\phantom{1}
${543}~^{+7}_{-18}~^{+9}_{-9}$
 \\
140
 &\phantom{1}
${428}~^{+5}_{-12}~^{+6}_{-6}$
 &\phantom{1}
${417}~^{+5}_{-11}~^{+5}_{-5}$
 &\phantom{1}
${409}~^{+5}_{-12}~^{+4}_{-4}$
 &\phantom{1}
${415}~^{+4}_{-12}~^{+7}_{-6}$
 &\phantom{1}
${421}~^{+6}_{-14}~^{+7}_{-7}$
 \\
150
 &\phantom{1}
${336}~^{+3}_{-9}~^{+5}_{-5}$
 &\phantom{1}
${327}~^{+3}_{-9}~^{+4}_{-4}$
 &\phantom{1}
${321}~^{+5}_{-9}~^{+3}_{-3}$
 &\phantom{1}
${326}~^{+3}_{-9}~^{+6}_{-5}$
 &\phantom{1}
${330}~^{+5}_{-10}~^{+6}_{-6}$
 \\
160
 &\phantom{1}
${267}~^{+2}_{-8}~^{+4}_{-4}$
 &\phantom{1}
${259}~^{+3}_{-7}~^{+3}_{-3}$
 &\phantom{1}
${256}~^{+3}_{-8}~^{+2}_{-2}$
 &\phantom{1}
${259}~^{+2}_{-8}~^{+5}_{-4}$
 &\phantom{1}
${262}~^{+4}_{-8}~^{+5}_{-5}$
\\
\hline
\end{tabular}
}
\caption{\small 
The total cross sections $\sigma(WH)$ for associated Higgs production in the $WH$ mode
at the LHC ($\sqrt s = 7~{\rm TeV}$) for different PDF sets and to NNLO accuracy.
The errors shown are the scale uncertainty based 
on the shifts $\mu=(m_H+M_{W})/2$ and $\mu = 2(m_H+M_{W})$ 
and, respectively, the 1$\sigma$ PDF uncertainty.  
Numbers are in fb. 
}
\label{tab:higgsWH}
\end{center}
\end{table}

\begin{table}[th!]
\renewcommand{\arraystretch}{1.3}
\begin{center}
{\small
\begin{tabular}{|l|l|l|l|l|l|}
\hline
\multicolumn{1}{|c|}{$m_H$}
&\multicolumn{1}{|c|}{ABM11}
&\multicolumn{1}{|c|}{ABKM09~\cite{Alekhin:2009ni}}
&\multicolumn{1}{|c|}{JR09~\cite{JimenezDelgado:2008hf,JimenezDelgado:2009tv}}
&\multicolumn{1}{|c|}{MSTW~\cite{Martin:2009iq}}
&\multicolumn{1}{|c|}{NN21~\cite{Ball:2011uy}}
\\     
\hline
100
 &
${669}~^{+7}_{-20}~^{+9}_{-9}$
 &
${656}~^{+7}_{-20}~^{+7}_{-7}$
 &
${635}~^{+9}_{-19}~^{+5}_{-5}$
 &
${651}~^{+7}_{-19}~^{+10}_{-9}$
 &
${655}~^{+10}_{-22}~^{+10}_{-10}$
 \\
110
 &
${499}~^{+5}_{-15}~^{+7}_{-7}$
 &
${488}~^{+6}_{-14}~^{+5}_{-5}$
 &
${474}~^{+7}_{-14}~^{+4}_{-4}$
 &
${486}~^{+5}_{-15}~^{+9}_{-7}$
 &
${488}~^{+7}_{-16}~^{+8}_{-8}$
 \\
115
 &
${433}~^{+5}_{-13}~^{+6}_{-6}$
 &
${424}~^{+5}_{-12}~^{+5}_{-5}$
 &
${413}~^{+5}_{-13}~^{+4}_{-4}$
 &
${422}~^{+4}_{-13}~^{+7}_{-6}$
 &
${424}~^{+6}_{-14}~^{+7}_{-7}$
 \\
120
 &
${378}~^{+4}_{-11}~^{+5}_{-5}$
 &
${370}~^{+4}_{-11}~^{+4}_{-4}$
 &
${360}~^{+5}_{-11}~^{+3}_{-3}$
 &
${368}~^{+4}_{-11}~^{+7}_{-5}$
 &
${370}~^{+5}_{-12}~^{+6}_{-6}$
 \\
125
 &
${331}~^{+3}_{-10}~^{+5}_{-5}$
 &
${323}~^{+4}_{-9}~^{+4}_{-4}$
 &
${316}~^{+4}_{-10}~^{+3}_{-3}$
 &
${322}~^{+4}_{-9}~^{+6}_{-5}$
 &
${324}~^{+5}_{-11}~^{+5}_{-5}$
 \\
130
 &
${290}~^{+3}_{-8}~^{+4}_{-4}$
 &
${284}~^{+3}_{-8}~^{+3}_{-3}$
 &
${277}~^{+4}_{-8}~^{+2}_{-2}$
 &
${283}~^{+3}_{-8}~^{+5}_{-4}$
 &
${285}~^{+4}_{-10}~^{+5}_{-5}$
 \\
140
 &
${226}~^{+2}_{-7}~^{+3}_{-3}$
 &
${221}~^{+2}_{-7}~^{+2}_{-2}$
 &
${216}~^{+3}_{-6}~^{+2}_{-2}$
 &
${221}~^{+2}_{-7}~^{+4}_{-3}$
 &
${222}~^{+3}_{-8}~^{+4}_{-4}$
 \\
150
 &
${178}~^{+2}_{-5}~^{+3}_{-3}$
 &
${174}~^{+2}_{-5}~^{+2}_{-2}$
 &
${171}~^{+2}_{-5}~^{+2}_{-2}$
 &
${174}~^{+2}_{-5}~^{+3}_{-3}$
 &
${175}~^{+2}_{-6}~^{+3}_{-3}$
 \\
160
 &
${142}~^{+1}_{-4}~^{+2}_{-2}$
 &
${138}~^{+2}_{-4}~^{+2}_{-2}$
 &
${136}~^{+2}_{-4}~^{+1}_{-1}$
 &
${139}~^{+1}_{-5}~^{+3}_{-2}$
 &
${139}~^{+2}_{-4}~^{+2}_{-2}$
 \\
\hline
\end{tabular}
}
\caption{\small 
Same as Tab.~\ref{tab:higgsWH} for $\sigma(ZH)$ and the scale uncertainty based 
on the shifts $\mu=(m_H+M_{Z})/2$ and $\mu = 2(m_H+M_{Z})$.
}
\label{tab:higgsZH}
\end{center}
\end{table}

Last in line we consider the Higgs-strahlung process, that is the  
associated $WH$ and $ZH$ production using the NNLO QCD corrections of~\cite{Brein:2003wg}. 
See also~\cite{Ferrera:2011bk} for fully exclusive QCD calculations at NNLO.
The dominant part of the hard partonic cross section is the same as for 
$W$- and $Z$-boson production discussed above in Sec.~\ref{sec:wzprod}, so that essentially
the same PDFs are probed, although at slightly larger values of $x$.
The numbers in Tabs.~\ref{tab:higgsWH} and \ref{tab:higgsZH} 
do not contain the gluon induced contribution for $\sigma(ZH)$~\cite{Kniehl:1990iva}. 
Top-quark mediated effects, which we neglect here, 
yield small perturbative corrections which are 
largely independent of the production model, 
that is to say of the parton luminosity, see also~\cite{Brein:2011vx} for a recent discussion. 
For the Higgs-strahlung the scale uncertainty is typically small and 
the predictions of the various PDF sets in Tabs.~\ref{tab:higgsWH} and \ref{tab:higgsZH} 
agree well within the quoted PDF uncertainties.

\begin{table}[th!]
\renewcommand{\arraystretch}{1.3}
\begin{center}
{\small
\begin{tabular}{|l|l|l|l|l|l|}
\hline
\multicolumn{1}{|c|}{$m_t$}
&\multicolumn{1}{|c|}{ABM11}
&\multicolumn{1}{|c|}{ABKM09~\cite{Alekhin:2009ni}}
&\multicolumn{1}{|c|}{JR09~\cite{JimenezDelgado:2008hf,JimenezDelgado:2009tv}}
&\multicolumn{1}{|c|}{MSTW~\cite{Martin:2009iq}}
&\multicolumn{1}{|c|}{NN21~\cite{Ball:2011uy}}
\\     
\hline
165&$167.9~^{+3.6}_{-9.3}~^{+7.5}_{-7.5}$&$171.4~^{+3.5}_{-9.3}~^{+9.6}_{-9.6}$&$204.3~^{+3.1}_{-9.1}~^{+14.7}_{-14.7}$&$209.9~^{+4.0}_{-11.8}~^{+5.5}_{-5.6}$&$216.1~^{+4.6}_{-10.8}~^{+5.8}_{-5.8}$\\
166&$162.6~^{+3.5}_{-9.0}~^{+7.3}_{-7.3}$&$166.1~^{+3.4}_{-9.0}~^{+9.3}_{-9.3}$&$197.2~^{+3.9}_{-7.9}~^{+14.4}_{-14.4}$&$203.5~^{+3.9}_{-11.4}~^{+5.3}_{-5.4}$&$209.4~^{+4.4}_{-10.5}~^{+5.7}_{-5.7}$\\
167&$157.5~^{+3.4}_{-8.8}~^{+7.0}_{-7.0}$&$160.9~^{+3.3}_{-8.7}~^{+9.0}_{-9.0}$&$191.2~^{+3.8}_{-7.6}~^{+14.0}_{-14.0}$&$197.3~^{+3.8}_{-11.1}~^{+5.2}_{-5.3}$&$202.9~^{+4.3}_{-10.1}~^{+5.5}_{-5.5}$\\
168&$152.5~^{+3.3}_{-8.5}~^{+6.8}_{-6.8}$&$155.9~^{+3.2}_{-8.5}~^{+8.8}_{-8.8}$&$185.5~^{+3.7}_{-7.4}~^{+13.7}_{-13.7}$&$191.3~^{+3.7}_{-10.8}~^{+5.1}_{-5.1}$&$196.7~^{+4.2}_{-9.8}~^{+5.4}_{-5.4}$\\
169&$147.8~^{+3.2}_{-8.2}~^{+6.7}_{-6.7}$&$151.1~^{+3.1}_{-8.2}~^{+8.5}_{-8.5}$&$179.8~^{+3.7}_{-7.1}~^{+13.3}_{-13.3}$&$185.5~^{+3.7}_{-10.4}~^{+4.9}_{-5.0}$&$190.6~^{+4.2}_{-9.4}~^{+5.2}_{-5.2}$\\
170&$143.2~^{+3.2}_{-8.0}~^{+6.5}_{-6.5}$&$146.4~^{+3.0}_{-8.0}~^{+8.3}_{-8.3}$&$174.5~^{+3.6}_{-6.9}~^{+13.0}_{-13.0}$&$179.9~^{+3.5}_{-10.1}~^{+4.8}_{-4.9}$&$184.9~^{+4.1}_{-9.1}~^{+5.1}_{-5.1}$\\
171&$138.8~^{+3.1}_{-7.7}~^{+6.3}_{-6.3}$&$141.9~^{+3.0}_{-7.7}~^{+8.1}_{-8.1}$&$169.2~^{+3.6}_{-6.6}~^{+12.7}_{-12.7}$&$174.5~^{+3.5}_{-9.7}~^{+4.6}_{-4.7}$&$179.2~^{+4.0}_{-8.7}~^{+4.9}_{-4.9}$\\
172&$134.5~^{+3.0}_{-7.5}~^{+6.1}_{-6.1}$&$137.6~^{+2.9}_{-7.5}~^{+7.9}_{-7.9}$&$164.2~^{+3.5}_{-6.5}~^{+12.4}_{-12.4}$&$169.3~^{+3.3}_{-9.5}~^{+4.5}_{-4.6}$&$173.9~^{+3.9}_{-8.5}~^{+4.8}_{-4.8}$\\
173&$130.4~^{+2.9}_{-7.2}~^{+5.9}_{-5.9}$&$133.4~^{+2.8}_{-7.3}~^{+7.6}_{-7.6}$&$159.3~^{+3.4}_{-6.3}~^{+12.0}_{-12.0}$&$164.3~^{+3.3}_{-9.2}~^{+4.4}_{-4.5}$&$168.7~^{+3.8}_{-8.2}~^{+4.7}_{-4.7}$\\
174&$126.4~^{+2.8}_{-7.0}~^{+5.8}_{-5.8}$&$129.4~^{+2.7}_{-7.1}~^{+7.4}_{-7.4}$&$154.6~^{+3.4}_{-6.0}~^{+11.8}_{-11.8}$&$159.4~^{+3.2}_{-8.9}~^{+4.3}_{-4.4}$&$163.7~^{+3.8}_{-7.9}~^{+4.6}_{-4.6}$\\
175&$122.6~^{+2.8}_{-6.8}~^{+5.6}_{-5.6}$&$125.5~^{+2.7}_{-6.8}~^{+7.2}_{-7.2}$&$150.0~^{+3.3}_{-5.8}~^{+11.5}_{-11.5}$&$154.7~^{+3.1}_{-8.7}~^{+4.2}_{-4.2}$&$158.9~^{+3.6}_{-7.8}~^{+4.4}_{-4.4}$\\
176&$118.9~^{+2.7}_{-6.6}~^{+5.5}_{-5.5}$&$121.8~^{+2.6}_{-6.6}~^{+7.0}_{-7.0}$&$145.6~^{+3.3}_{-5.6}~^{+11.3}_{-11.3}$&$150.2~^{+3.1}_{-8.4}~^{+4.1}_{-4.1}$&$154.2~^{+3.5}_{-7.5}~^{+4.3}_{-4.3}$\\
177&$115.3~^{+2.6}_{-6.4}~^{+5.3}_{-5.3}$&$118.1~^{+2.5}_{-6.5}~^{+6.9}_{-6.9}$&$141.3~^{+3.1}_{-5.5}~^{+11.0}_{-11.0}$&$145.8~^{+2.9}_{-8.2}~^{+4.0}_{-4.0}$&$149.8~^{+3.3}_{-7.4}~^{+4.2}_{-4.2}$\\
178&$111.9~^{+2.5}_{-6.2}~^{+5.2}_{-5.2}$&$114.6~^{+2.5}_{-6.3}~^{+6.7}_{-6.7}$&$137.2~^{+3.1}_{-5.2}~^{+10.7}_{-10.7}$&$141.6~^{+2.9}_{-7.9}~^{+3.9}_{-3.9}$&$145.4~^{+3.3}_{-7.1}~^{+4.1}_{-4.1}$\\
179&$108.5~^{+2.5}_{-6.0}~^{+5.0}_{-5.0}$&$111.3~^{+2.4}_{-6.1}~^{+6.5}_{-6.5}$&$133.4~^{+2.8}_{-5.2}~^{+10.4}_{-10.4}$&$137.6~^{+2.7}_{-7.8}~^{+3.8}_{-3.8}$&$141.2~^{+3.1}_{-7.0}~^{+4.0}_{-4.0}$\\
180&$105.3~^{+2.4}_{-5.9}~^{+4.9}_{-4.9}$&$108.0~^{+2.3}_{-5.9}~^{+6.3}_{-6.3}$&$129.6~^{+2.8}_{-5.0}~^{+10.2}_{-10.2}$&$133.6~^{+2.7}_{-7.6}~^{+3.7}_{-3.7}$&$137.1~^{+3.0}_{-6.8}~^{+3.9}_{-3.9}$\\
\hline
\end{tabular}
}
\caption{\small 
The total cross section for top-quark pair production at the LHC ($\sqrt s = 7~{\rm TeV}$)
for different PDF sets and to (approximate) NNLO accuracy 
as computed with {\tt HATHOR} (version 1.2)~\cite{Aliev:2010zk}  
as a function of the pole mass $m_t$.
The errors shown are the scale uncertainty based 
on the shifts $\mu=m_t/2$ and $\mu = 2m_t$ 
and, respectively, the 1$\sigma$ PDF uncertainty.  
All rates are in pb. 
}
\label{tab:ttbar}
\end{center}
\end{table}

\begin{table}[ht!]
\renewcommand{\arraystretch}{1.3}
\begin{center}
{\small
\begin{tabular}{|l|l|l|l|l|l|}
\hline
\multicolumn{1}{|c|}{$m_t$}
&\multicolumn{1}{|c|}{ABM11}
&\multicolumn{1}{|c|}{ABKM09~\cite{Alekhin:2009ni}}
&\multicolumn{1}{|c|}{JR09~\cite{JimenezDelgado:2008hf,JimenezDelgado:2009tv}}
&\multicolumn{1}{|c|}{MSTW~\cite{Martin:2009iq}}
&\multicolumn{1}{|c|}{NN21~\cite{Ball:2011uy}}
\\     
\hline
165&$243.8~^{+4.6}_{-12.9}~^{+10.0}_{-10.0}$&$247.4~^{+4.5}_{-12.9}~^{+12.7}_{-12.7}$&$289.3~^{+3.9}_{-12.2}~^{+18.3}_{-18.3}$&$298.1~^{+5.0}_{-16.0}~^{+7.1}_{-7.4}$&$307.5~^{+5.8}_{-14.7}~^{+7.6}_{-7.6}$\\
166&$236.2~^{+4.5}_{-12.6}~^{+9.7}_{-9.7}$&$239.9~^{+4.3}_{-12.5}~^{+12.3}_{-12.3}$&$279.5~^{+5.0}_{-10.6}~^{+17.9}_{-17.9}$&$289.2~^{+4.9}_{-15.5}~^{+7.0}_{-7.2}$&$298.3~^{+5.6}_{-14.2}~^{+7.4}_{-7.4}$\\
167&$229.0~^{+4.4}_{-12.2}~^{+9.4}_{-9.4}$&$232.6~^{+4.2}_{-12.1}~^{+12.0}_{-12.0}$&$271.2~^{+4.9}_{-10.2}~^{+17.4}_{-17.4}$&$280.5~^{+4.7}_{-15.1}~^{+6.8}_{-7.0}$&$289.2~^{+5.5}_{-13.7}~^{+7.2}_{-7.2}$\\
168&$222.0~^{+4.3}_{-11.8}~^{+9.2}_{-9.2}$&$225.5~^{+4.1}_{-11.8}~^{+11.6}_{-11.6}$&$263.2~^{+4.8}_{-10.0}~^{+17.1}_{-17.1}$&$272.2~^{+4.6}_{-14.7}~^{+6.6}_{-6.8}$&$280.6~^{+5.4}_{-13.3}~^{+7.0}_{-7.0}$\\
169&$215.2~^{+4.1}_{-11.5}~^{+8.9}_{-8.9}$&$218.7~^{+4.0}_{-11.4}~^{+11.3}_{-11.3}$&$255.4~^{+4.8}_{-9.6}~^{+16.6}_{-16.6}$&$264.1~^{+4.6}_{-14.2}~^{+6.4}_{-6.6}$&$272.2~^{+5.4}_{-12.8}~^{+6.8}_{-6.8}$\\
170&$208.7~^{+4.0}_{-11.1}~^{+8.7}_{-8.7}$&$212.1~^{+3.9}_{-11.1}~^{+11.0}_{-11.0}$&$248.0~^{+4.7}_{-9.3}~^{+16.2}_{-16.2}$&$256.4~^{+4.5}_{-13.8}~^{+6.2}_{-6.4}$&$264.1~^{+5.2}_{-12.4}~^{+6.6}_{-6.6}$\\
171&$202.4~^{+3.9}_{-10.8}~^{+8.4}_{-8.4}$&$205.8~^{+3.8}_{-10.7}~^{+10.7}_{-10.7}$&$240.7~^{+4.6}_{-8.9}~^{+15.8}_{-15.8}$&$248.8~^{+4.4}_{-13.3}~^{+6.1}_{-6.2}$&$256.3~^{+5.2}_{-11.9}~^{+6.5}_{-6.5}$\\
172&$196.4~^{+3.8}_{-10.5}~^{+8.2}_{-8.2}$&$199.7~^{+3.7}_{-10.4}~^{+10.4}_{-10.4}$&$233.7~^{+4.5}_{-8.7}~^{+15.5}_{-15.5}$&$241.6~^{+4.2}_{-13.0}~^{+5.9}_{-6.1}$&$248.8~^{+5.0}_{-11.6}~^{+6.3}_{-6.3}$\\
173&$190.5~^{+3.7}_{-10.2}~^{+8.0}_{-8.0}$&$193.8~^{+3.6}_{-10.1}~^{+10.2}_{-10.2}$&$227.0~^{+4.4}_{-8.4}~^{+15.1}_{-15.1}$&$234.6~^{+4.1}_{-12.6}~^{+5.8}_{-5.9}$&$241.5~^{+4.9}_{-11.2}~^{+6.1}_{-6.1}$\\
174&$184.8~^{+3.6}_{-9.9}~^{+7.8}_{-7.8}$&$188.1~^{+3.5}_{-9.8}~^{+9.9}_{-9.9}$&$220.4~^{+4.3}_{-8.1}~^{+14.7}_{-14.7}$&$227.8~^{+4.0}_{-12.2}~^{+5.6}_{-5.8}$&$234.5~^{+4.8}_{-10.8}~^{+6.0}_{-6.0}$\\
175&$179.3~^{+3.5}_{-9.6}~^{+7.6}_{-7.6}$&$182.5~^{+3.4}_{-9.5}~^{+9.7}_{-9.7}$&$214.0~^{+4.3}_{-7.9}~^{+14.4}_{-14.4}$&$221.3~^{+4.0}_{-11.8}~^{+5.5}_{-5.6}$&$227.9~^{+4.6}_{-10.6}~^{+5.8}_{-5.8}$\\
176&$174.1~^{+3.4}_{-9.3}~^{+7.4}_{-7.4}$&$177.2~^{+3.3}_{-9.3}~^{+9.4}_{-9.4}$&$207.9~^{+4.2}_{-7.6}~^{+14.2}_{-14.2}$&$214.9~^{+3.9}_{-11.5}~^{+5.4}_{-5.5}$&$221.3~^{+4.5}_{-10.3}~^{+5.7}_{-5.7}$\\
177&$168.9~^{+3.4}_{-9.0}~^{+7.2}_{-7.2}$&$172.1~^{+3.2}_{-9.0}~^{+9.2}_{-9.2}$&$202.0~^{+4.0}_{-7.4}~^{+13.8}_{-13.8}$&$208.8~^{+3.7}_{-11.2}~^{+5.2}_{-5.4}$&$215.0~^{+4.2}_{-10.1}~^{+5.6}_{-5.6}$\\
178&$164.0~^{+3.3}_{-8.8}~^{+7.0}_{-7.0}$&$167.1~^{+3.2}_{-8.7}~^{+8.9}_{-8.9}$&$196.2~^{+4.0}_{-7.1}~^{+13.5}_{-13.5}$&$202.9~^{+3.7}_{-10.9}~^{+5.1}_{-5.2}$&$208.9~^{+4.2}_{-9.8}~^{+5.4}_{-5.4}$\\
179&$159.3~^{+3.2}_{-8.5}~^{+6.8}_{-6.8}$&$162.3~^{+3.1}_{-8.5}~^{+8.7}_{-8.7}$&$190.9~^{+3.6}_{-7.0}~^{+13.1}_{-13.1}$&$197.2~^{+3.5}_{-10.7}~^{+5.0}_{-5.1}$&$203.1~^{+3.9}_{-9.7}~^{+5.3}_{-5.3}$\\
180&$154.7~^{+3.1}_{-8.3}~^{+6.6}_{-6.6}$&$157.6~^{+3.0}_{-8.3}~^{+8.5}_{-8.5}$&$185.6~^{+3.6}_{-6.8}~^{+12.8}_{-12.8}$&$191.7~^{+3.5}_{-10.4}~^{+4.8}_{-4.9}$&$197.3~^{+3.9}_{-9.3}~^{+5.1}_{-5.1}$\\
\hline
\end{tabular}
}
\caption{\small 
Same as Tab.~\ref{tab:ttbar} for the LHC at $\sqrt s = 8~{\rm TeV}$.
}
\label{tab:ttbar8}
\end{center}
\end{table}

\begin{table}[h]
\renewcommand{\arraystretch}{1.3}
\begin{center}
{\small
\begin{tabular}{|r|r|r|r|r|r|}
\hline
\multicolumn{1}{|c|}{$m_t$}
&\multicolumn{1}{|c|}{ABM11}
&\multicolumn{1}{|c|}{ABKM09~\cite{Alekhin:2009ni}}
&\multicolumn{1}{|c|}{JR09~\cite{JimenezDelgado:2008hf,JimenezDelgado:2009tv}}
&\multicolumn{1}{|c|}{MSTW~\cite{Martin:2009iq}}
&\multicolumn{1}{|c|}{NN21~\cite{Ball:2011uy}}
\\     
\hline
165&$995.2~^{+11.0}_{-43.6}~^{+28.5}_{-28.5}$&$993.3~^{+10.7}_{-42.9}~^{+35.4}_{-35.4}$&$1076.9~^{+8.3}_{-37.2}~^{+39.7}_{-39.7}$&$1131.4~^{+11.6}_{-50.5}~^{+18.4}_{-20.3}$&$1174.6~^{+14.0}_{-46.0}~^{+20.8}_{-20.8}$\\
166&$967.3~^{+10.7}_{-42.5}~^{+27.7}_{-27.7}$&$965.6~^{+10.3}_{-41.9}~^{+34.5}_{-34.5}$&$1043.1~^{+12.5}_{-31.6}~^{+38.9}_{-38.9}$&$1100.2~^{+11.1}_{-49.1}~^{+18.0}_{-19.9}$&$1141.9~^{+13.4}_{-44.3}~^{+20.3}_{-20.3}$\\
167&$940.2~^{+10.5}_{-41.3}~^{+27.0}_{-27.0}$&$938.6~^{+10.2}_{-40.6}~^{+33.6}_{-33.6}$&$1014.8~^{+12.3}_{-30.6}~^{+37.8}_{-37.8}$&$1070.0~^{+10.8}_{-47.7}~^{+17.6}_{-19.3}$&$1110.3~^{+13.7}_{-42.8}~^{+19.8}_{-19.8}$\\
168&$914.1~^{+10.2}_{-40.2}~^{+26.2}_{-26.2}$&$912.7~^{+9.8}_{-39.6}~^{+32.8}_{-32.8}$&$987.6~^{+12.2}_{-29.8}~^{+37.1}_{-37.1}$&$1041.0~^{+10.6}_{-46.6}~^{+17.1}_{-18.9}$&$1080.0~^{+13.1}_{-41.6}~^{+19.3}_{-19.3}$\\
169&$888.7~^{+9.9}_{-39.1}~^{+25.6}_{-25.6}$&$887.4~^{+9.7}_{-38.5}~^{+32.1}_{-32.1}$&$960.8~^{+12.4}_{-28.8}~^{+36.2}_{-36.2}$&$1012.4~^{+10.8}_{-45.1}~^{+16.7}_{-18.4}$&$1050.5~^{+13.0}_{-40.5}~^{+19.0}_{-19.0}$\\
170&$864.2~^{+9.7}_{-38.0}~^{+25.1}_{-25.1}$&$863.1~^{+9.4}_{-37.5}~^{+31.3}_{-31.3}$&$935.3~^{+12.1}_{-28.0}~^{+35.5}_{-35.5}$&$985.3~^{+10.5}_{-44.0}~^{+16.3}_{-17.9}$&$1022.0~^{+12.9}_{-39.4}~^{+18.5}_{-18.5}$\\
171&$840.5~^{+9.5}_{-37.0}~^{+24.4}_{-24.4}$&$839.6~^{+9.1}_{-36.5}~^{+30.4}_{-30.4}$&$910.0~^{+11.9}_{-26.8}~^{+34.6}_{-34.6}$&$958.4~^{+10.4}_{-42.5}~^{+15.9}_{-17.5}$&$994.1~^{+12.8}_{-37.8}~^{+18.0}_{-18.0}$\\
172&$817.6~^{+9.2}_{-36.0}~^{+23.8}_{-23.8}$&$816.7~^{+8.9}_{-35.4}~^{+29.7}_{-29.7}$&$886.1~^{+11.5}_{-26.5}~^{+33.9}_{-33.9}$&$933.1~^{+9.9}_{-41.8}~^{+15.6}_{-17.1}$&$967.7~^{+12.3}_{-37.0}~^{+17.6}_{-17.6}$\\
173&$795.3~^{+9.0}_{-35.0}~^{+23.3}_{-23.3}$&$794.6~^{+8.8}_{-34.5}~^{+29.1}_{-29.1}$&$862.7~^{+11.4}_{-25.6}~^{+33.1}_{-33.1}$&$908.3~^{+9.8}_{-40.5}~^{+15.2}_{-16.7}$&$941.8~^{+11.8}_{-35.8}~^{+17.1}_{-17.1}$\\
174&$773.8~^{+8.8}_{-34.1}~^{+22.7}_{-22.7}$&$773.2~^{+8.6}_{-33.6}~^{+28.4}_{-28.4}$&$840.0~^{+11.2}_{-24.7}~^{+32.4}_{-32.4}$&$884.2~^{+9.6}_{-39.3}~^{+14.9}_{-16.3}$&$916.7~^{+11.8}_{-34.6}~^{+16.6}_{-16.6}$\\
175&$752.9~^{+8.6}_{-33.2}~^{+22.2}_{-22.2}$&$752.5~^{+8.4}_{-32.7}~^{+27.7}_{-27.7}$&$817.9~^{+11.2}_{-23.9}~^{+31.9}_{-31.9}$&$860.9~^{+9.5}_{-38.2}~^{+14.6}_{-16.0}$&$893.0~^{+11.2}_{-34.1}~^{+16.3}_{-16.3}$\\
176&$732.7~^{+8.4}_{-32.3}~^{+21.7}_{-21.7}$&$732.5~^{+8.1}_{-31.8}~^{+26.9}_{-26.9}$&$796.4~^{+11.1}_{-23.0}~^{+31.5}_{-31.5}$&$838.2~^{+9.5}_{-37.0}~^{+14.3}_{-15.6}$&$869.9~^{+10.4}_{-33.8}~^{+16.0}_{-16.0}$\\
177&$713.2~^{+8.3}_{-31.5}~^{+21.2}_{-21.2}$&$713.0~^{+8.0}_{-31.0}~^{+26.4}_{-26.4}$&$775.8~^{+10.5}_{-22.6}~^{+30.7}_{-30.7}$&$816.5~^{+8.8}_{-36.5}~^{+14.0}_{-15.2}$&$847.0~^{+10.2}_{-32.8}~^{+15.6}_{-15.6}$\\
178&$694.3~^{+8.1}_{-30.7}~^{+20.7}_{-20.7}$&$694.2~^{+7.8}_{-30.2}~^{+25.7}_{-25.7}$&$755.6~^{+10.4}_{-21.6}~^{+30.0}_{-30.0}$&$795.1~^{+8.8}_{-35.4}~^{+13.7}_{-14.9}$&$824.9~^{+10.0}_{-31.8}~^{+15.2}_{-15.2}$\\
179&$675.9~^{+7.8}_{-29.9}~^{+20.2}_{-20.2}$&$676.0~^{+7.6}_{-29.5}~^{+25.1}_{-25.1}$&$736.9~^{+9.4}_{-21.5}~^{+29.3}_{-29.3}$&$774.9~^{+8.2}_{-35.0}~^{+13.3}_{-14.5}$&$804.0~^{+9.2}_{-31.4}~^{+14.9}_{-14.9}$\\
180&$658.2~^{+7.6}_{-29.1}~^{+19.6}_{-19.6}$&$658.3~^{+7.4}_{-28.7}~^{+24.5}_{-24.5}$&$718.2~^{+9.4}_{-20.8}~^{+28.7}_{-28.7}$&$754.9~^{+8.3}_{-33.9}~^{+13.0}_{-14.2}$&$783.1~^{+9.3}_{-30.6}~^{+14.6}_{-14.6}$\\
\hline
\end{tabular}
}
\caption{\small 
Same as Tab.~\ref{tab:ttbar} for the LHC at $\sqrt s = 14~{\rm TeV}$.
}
\label{tab:ttbar14}
\end{center}
\end{table}

\subsection{Top-quark pair production}
\label{sec:ttbar}

Here we present results for top-quark pair production at the LHC at $\sqrt{s}=7~$TeV.
The cross section is driven predominantly by the gluon luminosity
and the value of $\alpha_s$, much like in the case of Higgs production in ggF.
By far the largest parametric dependence of the cross section resides, 
however, in the value of top-quark mass. 
This is currently quoted with an experimental uncertainty of less than 1\% 
as $m_t = 172.9 \pm 1.1 \GeV$ based on the kinematic reconstruction 
from the decay products and comparison to Monte Carlo simulations~\cite{Nakamura:2010pdg}.
This value is commonly considered to be pole mass of the top-quark, although 
one should keep in mind, that the reconstruction of the
top-quark momenta from the observed (color-neutral) hadron momenta 
carries a further intrinsic uncertainty of ${\cal O}(\Lambda_{\rm QCD})$.
Top quark mass measurements 
in a well defined scheme for $m_t$ from the inclusive cross section at the Tevatron 
give lower masses, with a larger uncertainty though~\cite{Langenfeld:2009wd,Abazov:2011pta}.

In Tab.~\ref{tab:ttbar} we summarize the cross section values 
for a range of top-quark masses $165 \le m_t \le 180$ in the pole mass scheme.
The theoretical uncertainty at (approximate) NNLO is quantified 
by a variation of scale in the range $m_t/2 \le \mu \le 2m_t$.
We see that the predictions of ABKM09 and ABM11 are largely the same, the latter
being slightly smaller due to a smaller gluon PDF in the relevant $x$-range, cf. Fig.~\ref{fig:pdf1}.
In comparison to other PDFs, the predictions based on ABM11 in Tab.~\ref{tab:ttbar} are significantly
smaller and it seems, that precision measurements of the cross section at the LHC can, 
potentially constrain the gluon PDF.
This discriminating power, however, relies 
critically on the accurate knowledge 
of all other non-perturbative parameters, in particular $m_t$.

Cross section predictions using the running top-quark mass in the
\MSbar-scheme instead of the pole mass definition for $m_t$ in Tab.~\ref{tab:ttbar} 
generally display improved stability of the perturbative expansion 
and good properties of apparent convergence which is reflected in smaller 
uncertainties due the scale variation, see e.g.~\cite{Langenfeld:2009wd}.
The PDF uncertainties given in Tab.~\ref{tab:ttbar} and the differences between the various PDF sets 
are, of course, largely unaffected.

\subsection{Hadronic jet production}
\label{sec:jets}
\begin{figure}[th]
\centering
    {
    \includegraphics[angle=0,width=10.5cm]{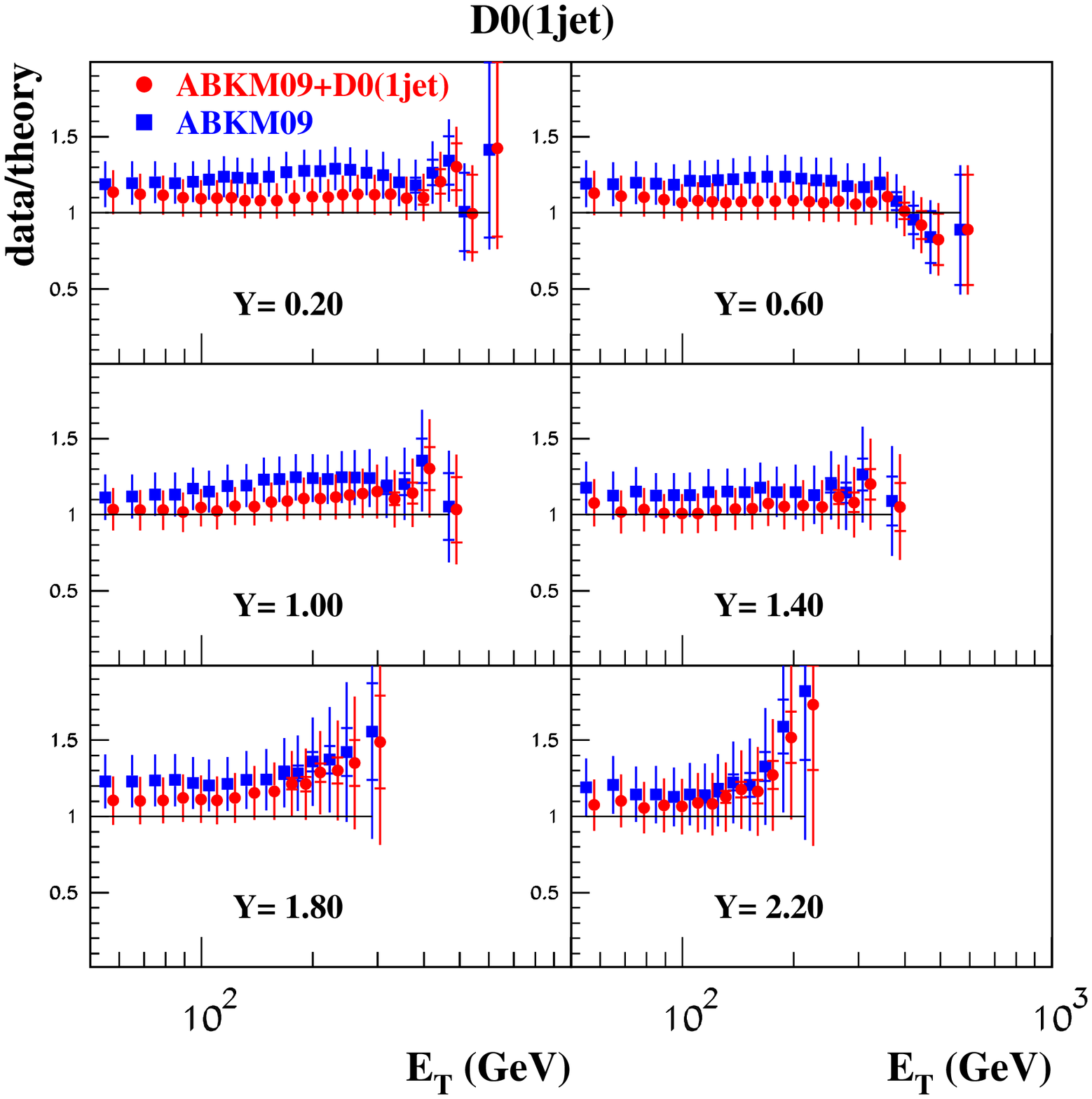}
    }
    \caption{\small
      \label{fig:jets1}
      Cross section data for 1-jet inclusive production from the D0 collaboration~\cite{Abazov:2008hua}
      as a function of the jet's transverse energy $E_T$  
      for the renormalization and factorization scales equal to $E_T$
      compared to the result of~\cite{Alekhin:2009ni} (circles) 
      and a re-fit including this data (squares) including 
      the NNLO threshold resummation corrections to the jet production~\cite{Kidonakis:2000gi}. 
    }
\end{figure}
\begin{figure}[ht!]
\centering
    {
    \includegraphics[angle=0,width=10.5cm]{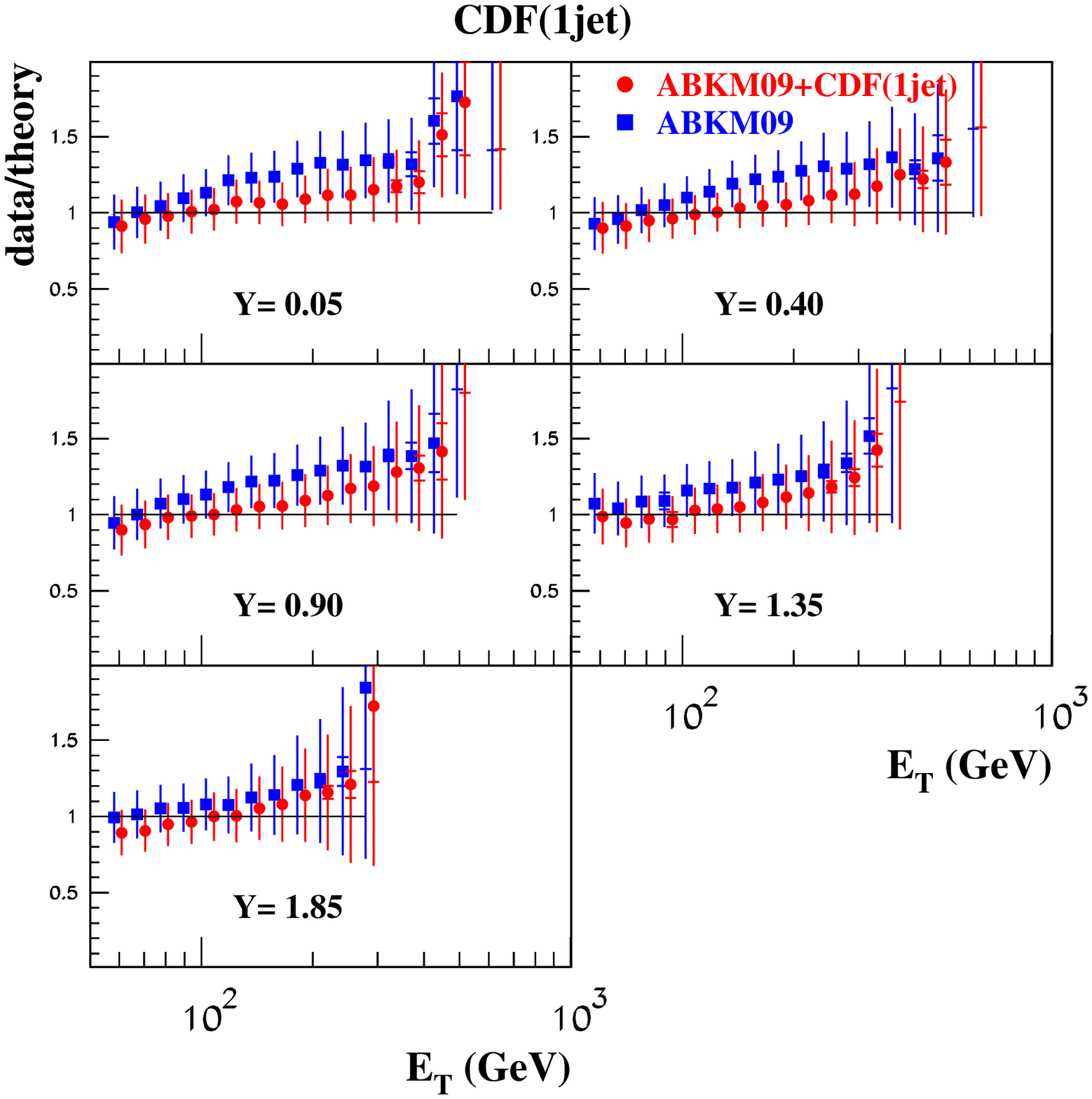}
    }
    \caption{\small
      \label{fig:jets2}
      Same as Fig.~\ref{fig:jets1} 
      for the cross section data for 1-jet inclusive production from 
      the CDF collaboration
      using a $k_T$ jet algorithm~\cite{Abulencia:2007ez}.
    }
\end{figure}

The hadronic jet production is sensitive to all nucleon PDFs, 
in particular to the large-$x$ gluon distribution, which is not sufficiently sensitive 
to other processes employed in PDF fits. 
Therefore the inclusive jet production data obtained by the Tevatron collider experiments~\cite{Abazov:2008hua,Abulencia:2007ez}
are used in the PDF fits of~\cite{Martin:2009iq,Lai:2010vv,Ball:2011mu}
in order to provide better constraints on the large-$x$ gluon. 
The calculation of the full NNLO QCD corrections to this process is still in
progress (see~\cite{GehrmannDeRidder:2011aa,Bolzoni:2010bt} and references therein). 
This precludes a consistent use of the Tevatron jet data in our NNLO PDF fit. 
Nevertheless in order to check any potential impact of the jet Tevatron data on our PDFs we 
have performed trial variants of the NNLO ABKM09 fit with the Tevatron jet data added~\cite{Alekhin:2011cf}.
The NLO QCD corrections~\cite{Nagy:2001fj,Nagy:2003tz} and the partial (soft gluon enhanced) 
NNLO corrections due to threshold resummation~\cite{Kidonakis:2000gi} have been computed with the {\tt FastNLO}
tool~\cite{Kluge:2006xs,Wobisch:2011ij}.

In order to allow more flexibility of the large-$x$ gluon distribution we have added the term $\gamma_{2,g}x^2$
to the polynomial of eq.~(\ref{eq:Ppdf5}) with an additional fitted parameter $\gamma_{2,g}$ making sure that further expansions   
of this polynomial do not improve the quality of those fits. 
In general, the Tevatron jet data overshoot the ABKM09 predictions, nevertheless they can be smoothly accommodated in the fit. 
The typical value of $\chi^2/NDP\approx 1$ is achieved with account of the error correlations for the jet data sets 
of~\cite{Abazov:2008hua,Abazov:2010fr,Abulencia:2007ez,Aaltonen:2008eq} 
once they are included into the NNLO ABKM09 fit. 
Meanwhile the various data sets demonstrate a somewhat different trend with respect to the ABKM09 predictions. 
E.g., the off-set of the D0 inclusive jet data~\cite{Abazov:2008hua} does not depend on the jet energy $E_T$ 
and therefore may be attributed to the impact of the currently missing full
NNLO corrections, cf. Figs.~\ref{fig:jets1} and \ref{fig:jets2}.
In contrast, for the CDF data of~\cite{Abulencia:2007ez} obtained with the $k_T$ jet algorithm 
the pulls rise with $E_T$ and can be reduced only by means of a modification of the PDF shapes.

\begin{table}[ht!]
\centering
\begin{tabular}{|c|c|c|c|c|c|}
\hline
  $\sigma(H) [pb]$
  & ABKM09 
  & D0 1-jet inc. 
  & D0 di-jet 
  & CDF 1-jet inc.
  & CDF 1-jet inc.
\\
  & 
  & 
  & 
  & (cone)
  & ($k_T$)
\\[0.5ex]
\hline
% mh = 115
%Tevatron(1.96) & 
%    {\bf 0.885(55)} & 0.981(33) &  0.954(30) & 0.932(28)  & 0.962(28) 
%\\[0.5ex]
%LHC(7)
%    & 
%    {\bf 15.72(45)} & 16.08(32) &  16.10(29) & 15.45(30) & 15.81(30) 
%\\
% mh = 120
Tevatron(1.96) & 
    {\bf 0.770(50)} & 0.859(29) &  0.833(27) & 0.815(25)  & 0.842(25) 
\\
LHC(7)
    & 
    {\bf 14.34(41)} & 14.68(29) &  14.69(27) & 14.11(28) & 14.44(27) 
\\
\hline
\end{tabular}
\caption{\small
\label{tab:hxsvalues}
The predicted cross sections for Higgs boson production in ggF with 
$m_H = 120$~GeV at Tevatron ($\sqrt{s}=1.96$~TeV) and at LHC ($\sqrt{s}=7$~TeV) from 
NNLO variants of the ABKM09 fit~\cite{Alekhin:2009ni} corresponding to Tab.~\ref{tab:alphas2}.
The uncertainty in brackets refers to the $1\sigma$ standard deviation for the
combined uncertainty on the PDFs and the value of $\alpha_s(M_Z)$.
The values in bold correspond to the published result~\cite{Alekhin:2010dd}.
}
\end{table}

The values of $\alpha_s$ extracted from the trial ABKM09 fits with the Tevatron jet data included 
are compared with the nominal ABKM09 value in Tab.~\ref{tab:alphas2}.
At most, they are bigger by 1$\sigma$, while for the CDF cone jet algorithm data~\cite{Aaltonen:2008eq} 
the central value of $\alpha_s$ is even the same. 
A recent evaluation of $\alpha_s$ using the ATLAS inclusive jet cross section data
yields $\alpha_s(M_Z) = 0.1151$ at NLO~\cite{Malaescu:2012ts}.
The predictions for the light Higgs production cross section, 
which are defined by the gluon distribution at $x\lesssim 0.1$, are also not very 
sensitive to the constraints coming from the Tevatron data, cf. Tab.~\ref{tab:hxsvalues}.
The impact of the Tevatron jet data on the large-$x$ gluon distribution is more significant. 
However, in this context we note that the Tevatron dijet and 3-jet production data are in good 
agreement with the ABKM09 predictions~\cite{Wobisch:2012iu}, in contrast to the  
case of inclusive jet production at Tevatron. 
The analysis of Tevatron data 3-jet production has also shown, 
that the predictions of MSTW~\cite{Martin:2009iq} agree even better 
with the data of~\cite{Abazov:2011ub} than ABKM09. 
However, for the case of CT10~\cite{Lai:2010vv} this is opposite. 

\begin{figure}[ht!]
\centering
    {
    \includegraphics[angle=0,width=10.5cm]{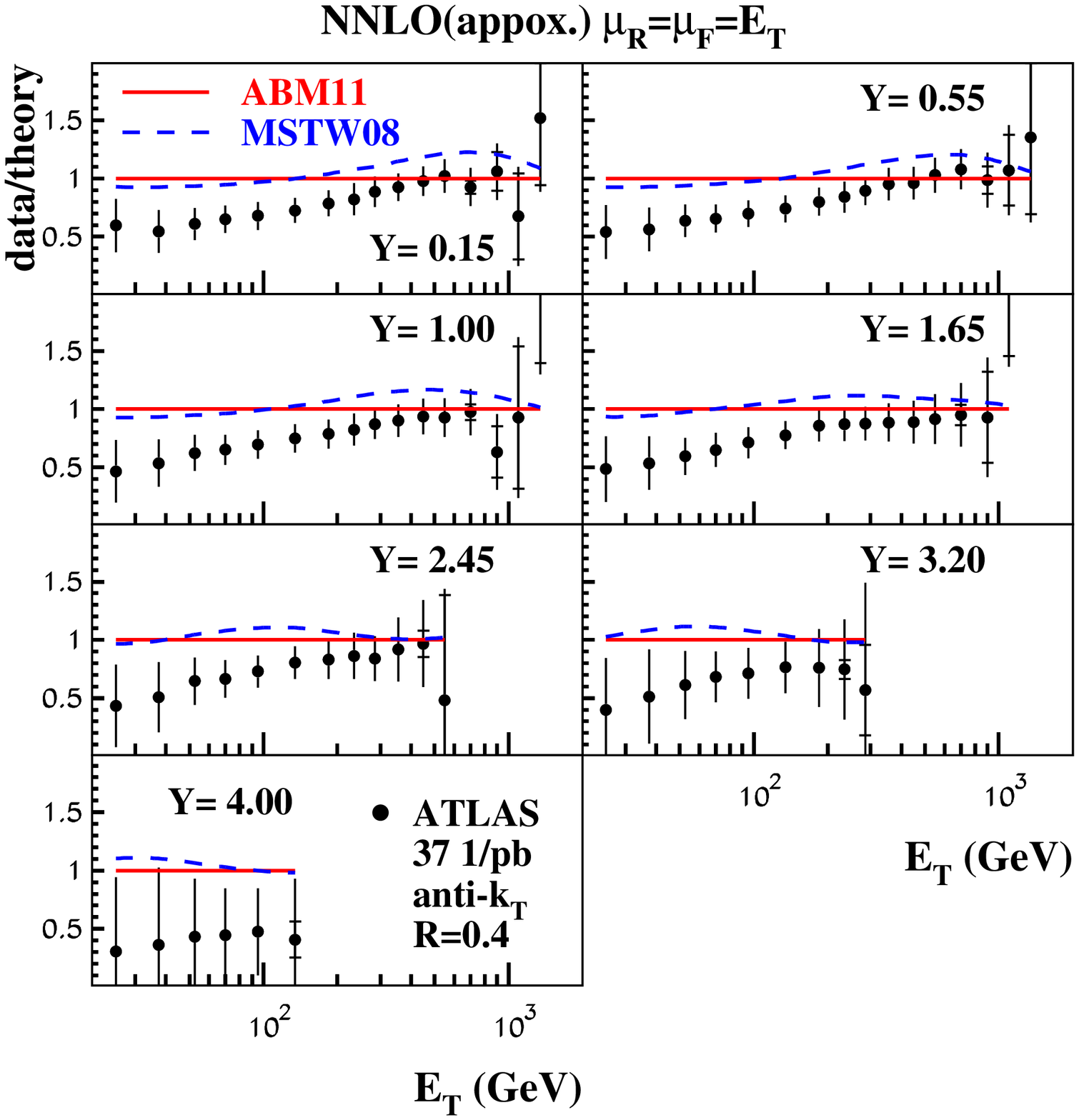}
    }
    \caption{\small
      \label{fig:jets-atlas}
      Cross section data for 1-jet inclusive production from the 
      ATLAS collaboration~\cite{Abazov:2008hua}
      as a function of the jet's transverse energy $E_T$ for $\mu_R=\mu_F=E_T$
      compared to the result of the present analysis (solid)
      and to MSTW~\cite{Martin:2009iq} (dashed). The theory predictions include 
      the NNLO threshold resummation corrections to the jet production~\cite{Kidonakis:2000gi}. 
    }
\end{figure}

\begin{figure}[ht!]
\centering
    {
    \includegraphics[angle=0,width=10.5cm]{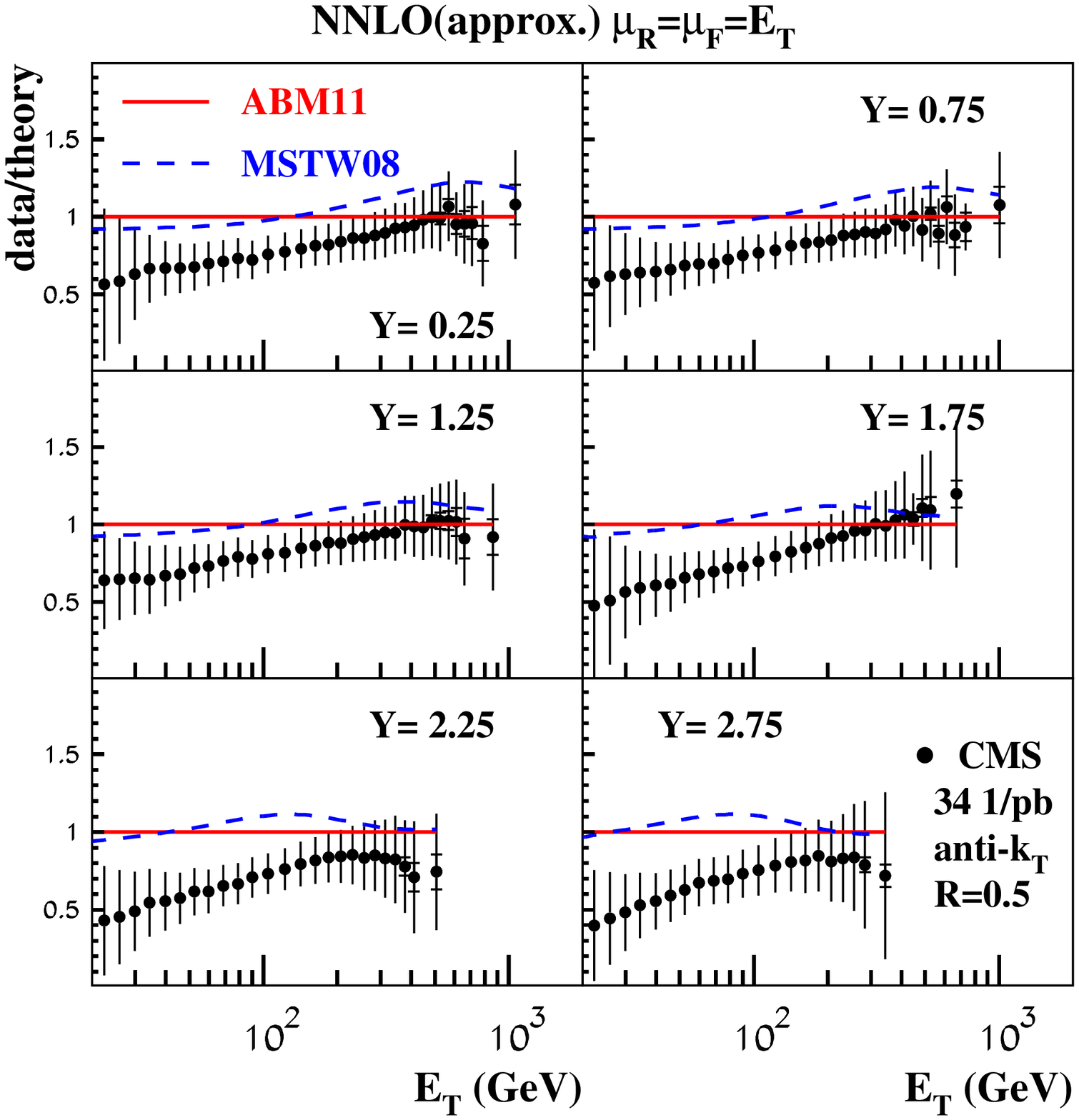}
    }
    \caption{\small
      \label{fig:jets-cms}
      Same as Fig.~\ref{fig:jets-atlas} for data of the CMS collaboration~\cite{CMS:2011ab}.
    }
\end{figure}

The trend of the first LHC data on the jet production 
with respect to the various PDF predictions is different from the Tevatron measurements. 
The ABKM09 predictions are in better agreement with the CMS and ATLAS inclusive data 
of~\cite{Rabbertz:1368241,Aad:2011fc} than the predictions based on the PDFs 
of~\cite{Martin:2009iq,Lai:2010vv,Ball:2011mu}, 
which were tuned to the Tevatron inclusive jet data. 
In Figs.~\ref{fig:jets-atlas} and \ref{fig:jets-cms} we show a comparison of
the result of the present fit and of MSTW~\cite{Martin:2009iq} with 
LHC jet data from ATLAS~\cite{Aad:2011fc} and from CMS~\cite{CMS:2011ab} 
calculated with the updated version of the {\tt FastNLO} code~\cite{Wobisch:2011ij}.
The trend of these data in comparison the predictions is the same for both experiments.
We find good agreement at large transverse energy $E_T$ and small rapidity $Y$.
The predictions overshoot the data for $E_T < 100~{\rm GeV}$, i.e., 
in a range where the theory description is evidently incomplete.
The predictions also overshoot the data (within the errors) at large $Y$.
In summary this is the opposite trend as compared to the Tevatron data.
If included in the PDF analysis and taking only data with $E_T > 100-150~{\rm GeV}$ say, 
the LHC jet data may potentially lead to a decrease of the large-$x$ gluons.
This said, it should be kept in mind that jet data from the LHC is still subject 
to large systematic errors, though.
In summary, these ambiguities in the data as well as the limitations in the 
current theoretical treatment prevent the use of hadronic jet data in our fit. 

Finally let us also comment on~\cite{Thorne:2011kq} in this context    
which has studied the compatibility of the Tevatron jet data with the 
predictions based on the PDFs of ABKM09~\cite{Alekhin:2009ni} 
(see also~\cite{Alekhin:2011cf}).
The particular focus of~\cite{Thorne:2011kq} has been on shifts 
in the normalization uncertainty of the experimental data.
Unfortunately, the study of~\cite{Thorne:2011kq} overestimates 
the statistical significance of such shifts in the normalization errors, 
since the PDF errors in the ABKM09 PDF sets have not been taken into account. 
The latter however should have a significant impact on the comparison because 
the large-$x$ gluon PDF of ABKM09, which is most relevant for such 
comparisons to Tevatron jet data, carries quite a large uncertainty in itself. 
This is corroborated by the fact that the value of $\chi^2$ obtained 
in the variants of the ABKM09 fit with the Tevatron data included~\cite{Alekhin:2011cf} 
are quite good, illustrating in an indirect manner the irrelevance of~\cite{Thorne:2011kq}.

\subsection{LHAPDF library}
\label{sec:lhapdf}

For the cross section computations presented here we have used the 
{\tt LHAPDF} library~\cite{Whalley:2005nh,lhapdf:2011} to interface to our
PDFs and to those of other groups.
For wider public use and to facilitate cross section computations by the interested reader, 
we provide the results of the current analysis in the form of data grids 
accessible with the most recent version {\tt lhapdf-5.8.7} of the {\tt LHAPDF} library, 
which can be obtained from {\tt http://projects.hepforge.org/lhapdf}. 
Please follow the instructions in the {\tt LHAPDF} package to install the {\tt LHAPDF} library.  

In detail, we provide three NLO grids for $n_f=3,4,5$ flavors,
\begin{verbatim}
      abm11_3n_nlo.LHgrid (0+28),
      abm11_4n_nlo.LHgrid (0+28),
      abm11_5n_nlo.LHgrid (0+28),
\end{verbatim}
with the central fit and 28 additional sets for the combined symmetric
uncertainty on the PDFs 
and on $\alpha_s$ from eqs.~(\ref{eq:abm11as-nlo}), (\ref{eq:abm11as-nnlo}), 
the heavy-quark masses from eq.~(\ref{eq:mcmbres}), and the deuteron
correction, cf. Fig.~\ref{fig:deut}.
Likewise at NNLO, we have
\begin{verbatim}
      abm11_3n_nnlo.LHgrid (0+28),
      abm11_4n_nnlo.LHgrid (0+28),
      abm11_5n_nnlo.LHgrid (0+28).
\end{verbatim}

For future measurements of the strong coupling constant, 
e.g., from data on hadronic jet production 
as well as for detailed phenomenological studies of the parametric dependence 
of observables on $\alpha_s$, we also provide grids with fixed values 
of $\alpha_s$, see the discussion in Sec.~\ref{sec:alphas}.
At NLO these are 20 sets in the range $\alpha_s=0.11\dots0.13$ 
and at NNLO 16 sets for $\alpha_s=0.105\dots0.12$, 
cf. also Fig.~\ref{fig:aschi2}. 
The sets for the $\alpha_s$ scan are denoted
\begin{verbatim}
  abm11_5n_as_nlo.LHgrid (0+20),
  abm11_5n_as_nnlo.LHgrid (0+16). 
\end{verbatim}

During the computation of the cross sections with the {\tt LHAPDF} library 
we have noticed significant differences in the run-time of our programs, 
when determining the cross section value and its uncertainty for a particular PDF set.
Usually, this requires sampling of the error PDFs which describe the error of
the corresponding parameter in a given set.
The run-time, of course, depends on the number of parameters in a given set, 
but also on the parametrization of the PDF grid and details of the look-up algorithm.
Efficiency for phenomenological studies requires sufficiently precise and at same time fast cross section computations.

Interestingly, for all observables discussed above, we have found the same pattern for the required run-times.
The grids of ABKM09~\cite{Alekhin:2009ni} and ABM11 
are both equally fast, i.e., $t_{\rm ABKM09} \simeq t_{\rm ABM11}$. 
Next come the JR09~\cite{JimenezDelgado:2008hf,JimenezDelgado:2009tv} grids which are
slightly slower by roughly a factor $t_{\rm JR09} \simeq 1.3\, \cdot\, t_{\rm ABM11}$ 
followed by the grids of MSTW~\cite{Martin:2009iq} with $t_{\rm MSTW} \simeq 2\,\cdot\, t_{\rm ABM11}$.
The grids of NN21~\cite{Ball:2011uy} are last in this row with run-times 
up by roughly a factor of five, $t_{\rm NN21} \simeq 5\, \cdot\, t_{\rm ABM11}$. 
This is correlated with the enormous size of the latter grids of ${\cal O}(100)$~MByte for NN21 with 100 PDF sets,  
as compared, e.g., to the size of ${\cal O}(5)$~MByte for ABM11 or to ${\cal O}(15)$~MByte for MSTW. 
Note that the NN21 grids with 1000 PDF sets have a size of about ${\cal O}(1)$~GByte 
and typical run times up by another order of magnitude, making PDF error computations even more inefficient.

%%
%% ---------------------------------------------------------------------------
%%
\section{Conclusions}
\label{sec:conclusions}

We have presented the PDF set ABM11 which determines the parton content of the
nucleon and measures $\alpha_s(M_Z)$ at NNLO accuracy in QCD.
In order to achieve a description of our analysis which is as complete and 
as transparent possible, we have provided an extensive discussion of all ingredients. 
This begins with the theoretical foundations which are most advanced and (almost)
fully consistent at NNLO in QCD using well-defined renormalization schemes for
all parameters involved.
We have given detailed information on our treatment of the additional higher twist terms and nuclear
corrections as well as on all data sets involved and on our fit ansatz.
The results presented have been exposed to numerous consistency checks to
test stability and the statistical quality.

We have found good agreement with our previous results in ABKM09 and the PDF
sets of ABM11 are readily available with the {\tt LHAPDF} library 
for precision phenomenology at the LHC and other hadron colliders.
We have studied the differences with respect to other PDF groups using 
the code {\tt OPENQCDRAD} for standardized precision comparison.
In this way some of the differences with respect to other PDF sets could be explained. 
The observed differences in the gluon PDFs are clearly outstanding in this context. 
This has important implications for the Higgs boson searches at the LHC 
and requires urgently further studies.
In our present fit, thanks to {\tt OPENQCDRAD}, we have provided a framework in which
such questions can be addressed in the future.

Despite the fact that various extractions of $\alpha_s(M_Z)$ have reached an impressive level of precision of 1\%, 
there exist still larger systematic differences which depend on both, the particular observable 
under consideration as well as on the specific analysis carried out for a
given observables (if compared to other analyses of the same observable). 
We have tried to analyse the status of the measurement of $\alpha_s(M_Z)$ from 
world DIS data and the effect of including other hard scattering data. 
At present we feel, that not all data used in global analyses are of
sufficient quality, considering the precision currently envisaged for $\alpha_s(M_Z)$. 

The differences of the present analysis with respect to other groups have to be clarified in the future by 
performing dedicated mutual comparisons addressing the data sets used in the analyses, 
details of the theoretical description as well as the data analysis. 
Here an important issue concerns the systematics of different data sets. 
In some cases the analyses show tensions and 
it is even known from the experimental analyses themselves that significant differences 
exist which are of systematic nature. 
It is a rather delicate matter how to deal with those
data in rather refined analyses at the precision of the NNLO level in QCD. 
These effects are partly responsible for the different results presently
obtained in PDF fits of various groups.
In the future, this situation can be significantly improved by 
{\it (i)} the availability of NNLO QCD corrections to hard jet cross sections; 
{\it (ii)} hard multi-jet data from the LHC with well-controlled systematics and  
within a wider energy range than available at Tevatron; 
{\it (iii)} precise $W^\pm, Z$ and Drell-Yan data from
the LHC. 
Gradually, also other hard scattering cross sections will be understood to NNLO
and allow to constrain the PDFs much further.

It is therefore clear that the simplest way to reduce PDF uncertainties is {\it not} 
to discard non-global sets. Rather, it consists of very detailed studies of data sets, 
using the most complete theoretical descriptions, following D. Hilbert's request: 
%%{\sc Statt des t\"orichten Ignorabimus hei\ss{}e im Gegenteil unsere Losung:} 
{\sc ``Wir m\"ussen wissen - wir werden wissen!''}~\cite{HILBERT}. 
Precision QCD analyses will always refer to compatible sets of precise data 
with a detailed account of the systematic errors and all known theoretical corrections, 
which allows to measure all non-perturbative PDF parameters 
and the strong coupling constant $\alpha_s(M_Z)$ as precisely as possible.

The QCD corrections are of vital importance to understand the production mechanism of 
{\it new states} as detailed as possible.
In particular, the anticipated signal of a Higgs boson at a mass $m_H \sim 126 \GeV$ \cite{HIGGS1} has
to be measured to highest precision. 
The aim of further PDF analyses has to consist in extracting both
the gluon PDF and $\alpha_s(M_Z)$ at even higher statistical and systematic precision 
because the dominant Higgs production in ggF behaves as 
$\propto \alpha_s^2 G(x,Q^2) \otimes G(x,Q^2)$ 
and it is well-known that even the NNLO QCD corrections in ggF 
contribute a numerically significant portion to the total cross section.
Also various other discoveries at hadron 
colliders may crucially depend on the thorough and precise understanding of the
PDFs and QCD at high-energy scales.

\subsection*{Note added:}
In the present analysis the comparison to NN21 has been carried out with the
help of 
the grid {\tt NNPDF21\_FFN3\_100} in the {\tt LHAPDF} repository dating from Aug 02, 2011. 
After publication of this article the grids of NN21 have been replaced 
and Figs.~\ref{fig:slacw} and \ref{fig:f2cc} have been updated 
with the grid {\tt NNPDF21\_FFN3\_100} now dating from Feb 17, 2012. 
With this new grid our conclusions remain unchanged.

%%
%% ---------------------------------------------------------------------
%%
\subsection*{Acknowledgments}
We thank J.~Baglio, H.~B\"ottcher, S.~Dinter, A.~Djouadi, V.~Drach, P.~Jimenez-Delgado, E.~Reya, and R.~Thorne 
for discussions.
We are thankful to V.~Drach for preparing the plot in Fig.~\ref{fig:2ndmom}, 
to R.~Thorne for providing us with the corresponding NLO plots not contained in~\cite{Martin:2009bu} 
and to M.~Zaro for computing the cross sections in Tab.~\ref{tab:higgsvbf}.
We also gratefully acknowledge the support of M.~Whalley to
integrate the results of the ABM11 fit into the {\tt LHAPDF} library~\cite{Whalley:2005nh,lhapdf:2011}. 

J.B. acknowledges support from Technische Universit\"at Dortmund.
This work has been supported by Helmholtz Gemeinschaft under contract VH-HA-101 ({\it Alliance Physics at the Terascale}),
by the Deutsche Forschungsgemeinschaft in Sonderforschungs\-be\-reich/Transregio~9
and by the European Commission through contract PITN-GA-2010-264564 ({\it LHCPhenoNet}).

\appendix
%%
%% ---------------------------------------------------------------------
%%
\renewcommand{\theequation}{\ref{sec:appA}.\arabic{equation}}
\setcounter{equation}{0}
\renewcommand{\thefigure}{\ref{sec:appA}.\arabic{figure}}
\setcounter{figure}{0}
\renewcommand{\thetable}{\ref{sec:appA}.\arabic{table}}
\setcounter{table}{0}
\section{Statistics}
\label{sec:appA}

\subsection{Statistical procedures}
\label{sec:statistics}

In our analysis we infer the vector of fitted parameters $\vec{\theta}$ 
from the experimental data minimizing the $\chi^2$ functional 
\begin{equation}
\chi^2(\vec{\theta}) \,=\, 
\sum_{i,j=1}^{N} (f_i(\vec{\theta})-y_i) E_{ij} (f_j(\vec{\theta})-y_j)
\, ,
\label{eq:EQCHI}
\end{equation}
where $f_i(\vec{\theta})$ is the fitted model, $y_i$ are 
the measurements, and $E_{ij}$ is the measurement error matrix with 
the indexes $i,j$ running through all $N$
data points included into the fit. The error matrix $E_{ij}$ is the inverse of 
the covariance matrix $C_{ij}$. If the data are uncorrelated 
the covariance matrix is diagonal and 
\begin{equation}
C_{ij} \,=\, 
\delta_{ij}\sigma_i\sigma_j
\, , 
\label{eq:covunc}
\end{equation}
where $\sigma_i$ are uncorrelated errors in the measurements 
$y_i$. 
If, in addition, 
the data are subject to the point-to-point correlated systematic 
fluctuations, the off-diagonal 
terms appear in the covariance matrix as well. 
For counting experiments the systematic errors are commonly 
multiplicative. Therefore the general form of the 
covariance matrix employed in our analysis reads
\begin{equation}
C_{ij} \,=\, 
\eta_i\eta_j f_i(\vec{\theta}) f_j(\vec{\theta}) + \delta_{ij}\sigma_i\sigma_j
\, , 
\label{eq:covcorr}
\end{equation}
where $\eta_i$ is the relative correlated systematic error in the 
measurement $y_i$. The uncorrelated errors may stem both from the 
statistical and systematic uncertainties.
In the later case they are combined 
with the statistical ones in quadrature to obtain $\sigma_j$.
The estimator based on the 
functional of eq.~(\ref{eq:EQCHI}) is statistically  
efficient and asymptotically 
unbiased in the limit of $N\rightarrow \infty$~.
In case of the analysis of correlated data 
the fitted parameters may be biased~\cite{D'Agostini:1993uj}, 
however for the definition of eq.~(\ref{eq:covcorr}) the 
estimator is nevertheless asymptotically unbiased~\cite{Alekhin:2000es}. 
The errors in the fitted parameters given in Tab.~\ref{tab:fitvalues}
and the correlations coefficients of Tabs.~\ref{tab:pdfco1}--\ref{tab:pdfco3}
are propagated from the uncertainties 
in the data with account of the correlations in the latter and 
correspond to the standard statistical criterion 
$\Delta\chi^2=1$. The errors in the data normalization factors
for the selected experiments, cf. Tabs.~\ref{tab:NMC} and~\ref{tab:SLAC},
which were fitted simultaneously with other parameters, 
are calculated in the same way. 

\bigskip

The detailed information on the experimental uncertainities in all data sets considered 
in the fit and listed in Sec.~\ref{sec:data} are available from {\tt http://arxiv.org}
as an attachment to the arXiv version of our paper.
This includes in particular the systematic errors in the SLAC data~\cite{WHITLOW} discussed in Sec.~\ref{sec:incldis} 
and, likewise, the systematic errors in the NuTeV data and the corrections on the unmeasured phase space 
for the NuTeV and the CCFR data~\cite{MASON} discussed in Sec.~\ref{sec:di-muon}.

\subsection{Parameter correlation matrix}
\label{sec:covmat}
Here, we finally present the covariance matrix for the correlations of the fit parameters of ABM11 
discussed in Sec.~\ref{sec:results}, cf. Tab.~\ref{tab:fitvalues} and eq.~(\ref{eq:mcmbinp}) for $m_c$ and $m_b$.
The strong coupling $\alpha_s(M_Z)$ is given in eq.~(\ref{eq:abm11as-nnlo}), 
while in Tabs.~\ref{tab:pdfco1}--\ref{tab:pdfco3} we quote the correlations for 
$\alpha_s(\mu_0)$ with $\mu_0^2=1.5~{\rm GeV}^2$.

The correlations matrices for all other variants of the PDF fit discussed 
in the paper, e.g. those fits in Sec.~\ref{sec:jets} including Tevatron jet data 
are omitted for the brevity. 
However they are available from the authors upon request.

%
% co1
\begin{center}
\begin{table}
\renewcommand{\arraystretch}{1.5}
\begin{center}
\footnotesize
\begin{sideways}
\begin{tabular}{|c|c|c|c|c| c|c|c|c|c| c|c|c|c|c|}
\hline
&
$a_u$ & 
$b_u$ &
$\gamma_{1,u}$ &
$\gamma_{2,u}$ &
$a_d$ &
$b_d$ &
$A_\Delta$ &
$b_\Delta$ &
$A_{us}$ &
$a_{us}$ &
$b_{us}$ &
$a_g$ &
$b_g$ &
$\gamma_{1,g}$ \\
\hline
$a_u$ 
 & 1.0000 & 0.9692 & 0.9787 & -0.7929 & 0.7194 & 0.5279 & -0.1460 & -0.1007 & 0.7481 & 0.6835 & -0.4236 & -0.2963 & 0.3391 & 0.3761 \\ 
$b_u$ 
 & & 1.0000 & 0.9396 & -0.7244 & 0.6792 & 0.4939 & -0.1146 & -0.1099 & 0.7404 & 0.6840 & -0.4146 & -0.3138 & 0.3464 & 0.3738  \\ 
$\gamma_{1,u}$ 
 & & & 1.0000 & -0.8940 & 0.6506 & 0.4646 & -0.1865 & -0.0539 & 0.6728 & 0.6093 & -0.4799 & -0.2755 & 0.3441 & 0.3717 \\ 
$\gamma_{2,u}$ 
 & & & & 1.0000 & -0.4102 & -0.2267 & 0.2357 & -0.0182 & -0.4075 & -0.3495 & 0.4543 & 0.1713 & -0.3156 & -0.3149 \\ 
$a_d$ 
 & & & & & 1.0000 & 0.8827 & -0.2155 & -0.1964 & 0.6875 & 0.6435 & -0.3030 & -0.3354 & 0.2635 & 0.3500 \\ 
$b_d$ 
 & & & & & & 1.0000 & -0.2462 & -0.0979 & 0.5359 & 0.5099 & -0.2957 & -0.3443 & 0.3157 & 0.3763 \\ 
$A_\Delta$ 
 & & & & & & & 1.0000 & -0.2068 & -0.0689 & -0.0698 & 0.2381 & -0.0168 & 0.0384 & 0.0453 \\ 
$b_\Delta$ 
 & & & & & & & & 1.0000 & 0.1015 & 0.1279 & -0.4146 & -0.0852 & -0.1185 & -0.0892 \\ 
$A_{us}$ 
 & & & & & & & & & 1.0000 & 0.9884 & -0.4678 & -0.4679 & 0.1961 & 0.2504 \\ 
$a_{us}$ 
 & & & & & & & & & & 1.0000 & -0.4520 & -0.5195 & 0.1982 & 0.2596 \\ 
$b_{us}$ 
 & & & & & & & & & & & 1.0000 & 0.1436 & 0.0444 & -0.0180  \\ 
$a_g$ 
 & & & & & & & & & & & & 1.0000 & -0.6289 & -0.7662 \\ 
$b_g$ 
 & & & & & & & & & & & & & 1.0000 & 0.9392 \\ 
$\gamma_{1,g}$ 
 & & & & & & & & & & & & & & 1.0000  \\ 
\hline
\end{tabular}
\end{sideways}
\caption{ \small
The covariance matrix for the PDF parameters in
Tab.~\ref{tab:fitvalues}, $\alpha_s$, $m_c$ and $m_b$.}
\label{tab:pdfco1}
\end{center}
\end{table}
\end{center}

% co2
\begin{center}
\begin{table}
\renewcommand{\arraystretch}{1.5}
\begin{center}
\footnotesize
\begin{sideways}
\begin{tabular}{|c|c|c|c|c| c|c|c|c|c| c|c|c|c|c}
\hline
&
$\alpha_s(\mu_0)$ &
$\gamma_{1,\Delta}$ &
$\gamma_{1,us}$ &
$\gamma_{1,d}$ &
$\gamma_{2,d}$ &
$A_{s}$ &
$b_s$ &
$a_s$ &
$\gamma_{3,u}$ &
$m_c(m_c)$ &
$\gamma_{3,us}$ &
$m_b(m_b)$ &
$a_\Delta$ \\
\hline
$a_u$ 
 & -0.0435 & 0.0000 & -0.8480 & 0.6008 & 0.1535 & -0.0034 & -0.0437 & -0.0355 & 0.8111 & 0.0796 & -0.4797 & 0.0044 & -0.1718 \\ 
$b_u$ 
 & -0.1251 & 0.0316 & -0.8375 & 0.5537 & 0.1806 & 0.0008 & -0.0345 & -0.0276 & 0.7001 & 0.0625 & -0.4889 & -0.0005 & -0.1452 \\ 
$\gamma_{1,u}$ 
 & -0.0849 & -0.0637 & -0.8133 & 0.5422 & 0.1667 & -0.0324 & -0.0671 & -0.0638 & 0.8948 & 0.0726 & -0.4033 & 0.0075 & -0.2028 \\ 
$\gamma_{2,u}$ 
 & 0.0920 & 0.1659 & 0.5760 & -0.3308 & -0.2276 & 0.0799 & 0.0966 & 0.1098 & -0.9749 & -0.0631 & 0.1728 & -0.0142 & 0.2353 \\ 
$a_d$ 
 & -0.0321 & -0.0137 & -0.7618 & 0.9630 & -0.1842 & 0.0007 & -0.0414 & -0.0167 & 0.4878 & 0.0227 & -0.4735 & -0.0078 & -0.2088 \\ 
$b_d$ 
 & -0.1666 & -0.1167 & -0.6060 & 0.9351 & -0.5969 & -0.0064 & -0.0249 & -0.0203 & 0.3007 & -0.0045 & -0.3782 & -0.0132 & -0.2121 \\ 
$A_\Delta$ 
 & 0.0206 & 0.8718 & 0.1649 & -0.2544 & 0.1916 & -0.0232 & -0.0212 & -0.0294 & -0.2398 & 0.0202 & 0.0667 & 0.0034 & 0.9721 \\ 
$b_\Delta$ 
 & 0.0086 & -0.6291 & -0.1067 & -0.1834 & -0.1103 & 0.0594 & 0.0577 & 0.0711 & 0.0052 & -0.0063 & -0.1768 & -0.0083 & -0.0662 \\ 
$A_{us}$ 
 & 0.0043 & -0.0481 & -0.8662 & 0.5862 & 0.0768 & -0.0341 & -0.0659 & -0.0493 & 0.4485 & 0.1559 & -0.8164 & -0.0008 & -0.0417 \\ 
$a_{us}$ 
 & -0.0459 & -0.0650 & -0.8255 & 0.5493 & 0.0606 & -0.0119 & -0.0441 & -0.0255 & 0.3870 & 0.0940 & -0.8628 & -0.0055 & -0.0375 \\ 
$b_{us}$ 
 & -0.0382 & 0.3783 & 0.7032 & -0.3288 & 0.1278 & -0.0734 & -0.0445 & -0.0807 & -0.4262 & -0.0100 & 0.3911 & 0.0040 & 0.1782 \\ 
$a_g$ 
 & 0.3785 & 0.0061 & 0.3050 & -0.3280 & 0.1338 & 0.0936 & 0.0718 & 0.1165 & -0.1744 & -0.0137 & 0.4886 & 0.0323 & -0.0360 \\ 
$b_g$ 
 & -0.6085 & 0.1017 & -0.0873 & 0.2827 & -0.2104 & -0.0543 & -0.0114 & -0.1223 & 0.2973 & 0.1560 & -0.1337 & 0.0141 & 0.0066 \\ 
$\gamma_{1,g}$ 
 & -0.4642 & 0.1021 & -0.1778 & 0.3605 & -0.1962 & -0.0708 & -0.0396 & -0.1230 & 0.3132 & 0.0425 & -0.1977 & 0.0071 & 0.0201 \\ 
\hline
\end{tabular}
\end{sideways}
\caption{ \small
The covariance matrix for the PDF parameters in
Tab.~\ref{tab:fitvalues}, $\alpha_s$, $m_c$ and $m_b$.}
\label{tab:pdfco2}
\end{center}
\end{table}
\end{center}

% co3
\begin{center}
\begin{table}
\renewcommand{\arraystretch}{1.5}
\begin{center}
\footnotesize
\begin{sideways}
\begin{tabular}{|c|c|c|c|c| c|c|c|c|c| c|c|c|c|c}
\hline
&
$\alpha_s(\mu_0)$ &
$\gamma_{1,\Delta}$ &
$\gamma_{1,us}$ &
$\gamma_{1,d}$ &
$\gamma_{2,d}$ &
$A_{s}$ &
$b_s$ &
$a_s$ &
$\gamma_{3,u}$ &
$m_c(m_c)$ &
$\gamma_{3,us}$ &
$m_b(m_b)$ &
$a_\Delta$ \\
\hline
$\alpha_s(\mu_0)$ 
 & 1.0000 & 0.0176 & -0.0394 & -0.0798 & 0.2357 & -0.0018 & -0.0982 & -0.0075 & -0.0291 & 0.1904 & 0.0676 & 0.0562 & 0.0136 \\ 
$\gamma_{1,\Delta}$ 
 & & 1.0000 & 0.1183 & -0.0802 & 0.2640 & -0.0427 & -0.0489 & -0.0550 & -0.1595 & 0.0193 & 0.0985 & 0.0069 & 0.7657 \\ 
$\gamma_{1,us}$ 
 & & & 1.0000 & -0.6753 & -0.0493 & -0.0525 & 0.0158 & -0.0445 & -0.6039 & -0.0656 & 0.6590 & 0.0017 & 0.1487 \\ 
$\gamma_{1,d}$ 
 & & & & 1.0000 & -0.4041 & -0.0213 & -0.0513 & -0.0366 & 0.4145 & 0.0148 & -0.3931 & -0.0086 & -0.2284 \\ 
$\gamma_{2,d}$ 
 & & & & & 1.0000 & 0.0308 & -0.0016 & 0.0326 & 0.1801 & 0.0276 & -0.0510 & 0.0111 & 0.1212 \\ 
$A_{s}$ 
 & & & & & & 1.0000 & 0.8570 & 0.9749 & -0.0664 & -0.0206 & -0.4355 & 0.0017 & -0.0139 \\ 
$b_s$ 
 & & & & & & & 1.0000 & 0.8730 & -0.0894 & -0.0706 & -0.3708 & 0.0005 & -0.0127 \\ 
$a_s$ 
 & & & & & & & & 1.0000 & -0.0967 & -0.1234 & -0.4403 & -0.0050 & -0.0172 \\ 
$\gamma_{3,u}$ 
 & & & & & & & & & 1.0000 & 0.0674 & -0.2082 & 0.0153 & -0.2378 \\ 
$m_c(m_c)$ 
 & & & & & & & & & & 1.0000 & -0.0010 & 0.0505 & 0.0141 \\ 
$\gamma_{3,us}$ 
 & & & & & & & & & & & 1.0000 & 0.0083 & 0.0276 \\ 
$m_b(m_b)$ 
 & & & & & & & & & & & & 1.0000 & 0.0006 \\ 
$a_\Delta$ 
 & & & & & & & & & & & & & 1.0000 \\ 
\hline
\end{tabular}
\end{sideways}
\caption{ \small
The covariance matrix for the PDF parameters in
Tab.~\ref{tab:fitvalues}, $\alpha_s$, $m_c$ and $m_b$.}
\label{tab:pdfco3}
\end{center}
\end{table}
\end{center}

\cleardoublepage
\newpage

%%
%% ---------------------------------------------------------------------
%%
{\footnotesize
%\bibliography{abmbib}
%\bibliographystyle{h-physrev5.bst}

}

\end{document}